%% file: Concave_normal.tex
\documentclass[a4paper,12pt]{article}
\usepackage{latexsym,amsmath}
\usepackage{amssymb}
\usepackage[dvips]{graphicx}
\usepackage{lscape}
\usepackage{color}
\usepackage{undertilde}
\usepackage{accents}
\usepackage[T1]{fontenc}
\usepackage[utf8]{inputenc}
\usepackage{authblk}
\usepackage[authoryear,round,longnamesfirst]{natbib}
\usepackage{booktabs}
\usepackage{multirow}
\usepackage{float}
\usepackage{setspace}
\usepackage{subcaption}
\usepackage{tcolorbox}
\usepackage{mathtools}
\usepackage{epstopdf}
\usepackage{geometry}
\usepackage{multicol}
\usepackage{csquotes}
\usepackage{hyperref}
\usepackage{etoolbox}
\usepackage{rotating}
\usepackage{enumitem}
\usepackage[useregional]{datetime2}
\usepackage[toc,page]{appendix}
\usepackage{amsbsy}
\usepackage{adjustbox}
\usepackage[flushleft]{threeparttable}

\graphicspath{{./P3Images/}}

\newtheorem{lemma}{Lemma}

\numberwithin{equation}{section}
\patchcmd{\abstract}{\small}{}{}{}
\allowdisplaybreaks

\hypersetup{ colorlinks = true, citecolor= blue}

\newcommand{\z}{{\mbox{\boldmath $z$}}}

\newcommand{\w}{{\mbox{\boldmath $w$}}}

\newcommand{\bdelta}{{\mbox{\boldmath $\delta$}}}

\def \z {{\mbox{\boldmath$z$}}}

\newcommand{\uiota}             {\mbox{\boldmath$\uiota$}}

\def\beq{\begin{equation}}
\def\eeq{\end{equation}}
\def\beqa{\begin{eqnarray}}
\def\eeqa{\end{eqnarray}}
\def\beqan{\begin{eqnarray*}}
\def\eeqan{\end{eqnarray*}}

\def\bc{\begin{center}}
\def\ec{\end{center}}
\def\btable{\begin{table}[htbp]}
\def\etable{\end{table}}
\def\bfig{\begin{figure}[htbp]}
\def\efig{\end{figure}}

\def\bi{\begin{itemize}}
\def\ei{\end{itemize}}

\newlength{\tempheight}
\newlength{\tempwidth}
\newcommand{\rowname}[1]% #1 = text
{\rotatebox{90}{\makebox[\tempheight][c]{\textbf{#1}}}}
\newcommand{\columnname}[1]% #1 = text
{\makebox[\tempwidth][c]{\textbf{#1}}}

\let\oldabstract\abstract
\let\oldendabstract\endabstract
\makeatletter
\renewenvironment{abstract}
{%
               {\list{}{\addtolength{\leftmargin}{1em} % change this value to add or remove length to the the default
                        \listparindent 1.5em%
                        \itemindent    \listparindent%
                        \rightmargin   \leftmargin%
                        \parsep        \z@ \@plus\p@}%
                \item\relax}%
               {\endlist}%
\oldabstract}
{\oldendabstract}
\renewcommand{\@makefnmark}{\hbox{\textsuperscript{\tiny{\@thefnmark}}}}
\makeatother

\title{A placement-value based approach to concave ROC analysis}
\author[1]{Soutik Ghosal}
\author[2]{Zhen Chen\thanks{zhen.chen@nih.gov}}
\affil[1]{Division of Biostatistics, Public Health Sciences Department, School of Medicine, University of Virginia, Charlottesville, VA 22903}
\affil[2]{Biostatistics and Bioinformatics Branch, Division of Intramural Population Health Research, {\it Eunice Kennedy Shriver} National Institute of Child Health and Human Development, Bethesda, MD 20892}

\date{}

\begin{document}
\maketitle
%\begin{document} 
\begin{abstract}
The receiver operating characteristic (ROC) curve is an important graphic tool for evaluating a test in a wide range of disciplines. While useful, an ROC curve can cross the chance line, either by having an S-shape or a hook at the extreme specificity. These non-concave ROC curves are sub-optimal according to decision theory, as there are points that are superior than those corresponding to the portions below the chance line with either the same sensitivity or specificity. We extend the literature by proposing a novel placement value-based approach to ensure concave curvature of the ROC curve, and utilize Bayesian paradigm to make estimations under both a parametric and a semiparametric framework. We conduct extensive simulation studies to assess the performance of the proposed methodology under various scenarios, and apply it to a pancreatic cancer dataset.

\end{abstract}

%\newpage
\section{Introduction} \label{Intro}

The receiver operating characteristic (ROC) curve is a popular two-dimensional graphic tool to assess how a test of interest differentiates an affected and a reference population \citep{birdsall1966theory,peterson1954theory}. It is constructed by plotting pairs of false positive and true positive rates across various thresholds of the test value. A useful summary measure based on the ROC curve analysis is the area under the ROC curve (AUC), which is a scalar between 0.5 and 1.0, with higher values for higher discrimination capacity of the test \citep{mcclish1989analyzing}. %Recent advances in epidemiological and biomedical research have generated increased interest in ROC methodologies that allow covariates adjustment \citep{pepe1997regression, de2013bayesian, alonzo2002distribution, gorsevski2006spatial, ishwaran2000general}, especially those capable of assessing covariate effects directly on ROC and AUC. %\citep{cai2004semi}. %For example, when using estimated fetal weight (EFW) from ultrasound examinations during pregnancy to predict large for gestation age (LGA), it is of interest to explore whether the diagnostic accuracy of EFW varies by the gestational age at which the ultrasound examination is taken, or whether this accuracy differs between male and female newborn. 
%Traditional ROC regressions take an indirect approach by considering covariates in both the affected and reference test score distributions \citep{tosteson1988general, toledano1996ordinal}. While straightforward to implement, this indirect approach can be difficult to interpret. As a result, Pepe and coauthors [REF] have suggested a novel approach that regress ROC directly as a function of covariates. This is achieved by the use of placement values (PV), a concept that represents the position of affected test scores relative to the reference population. Recently, extensions have been made along this PV-based approach that allow semi-parametric and non-parametric modeling of the ROC curve \citep{cai2002semiparametric,cai2004semi,lin2012direct}.

\begin{figure}[ht]
    \centering
    \begin{subfigure}[t]{0.5\textwidth}
        \centering
        \includegraphics[width=0.9\linewidth]{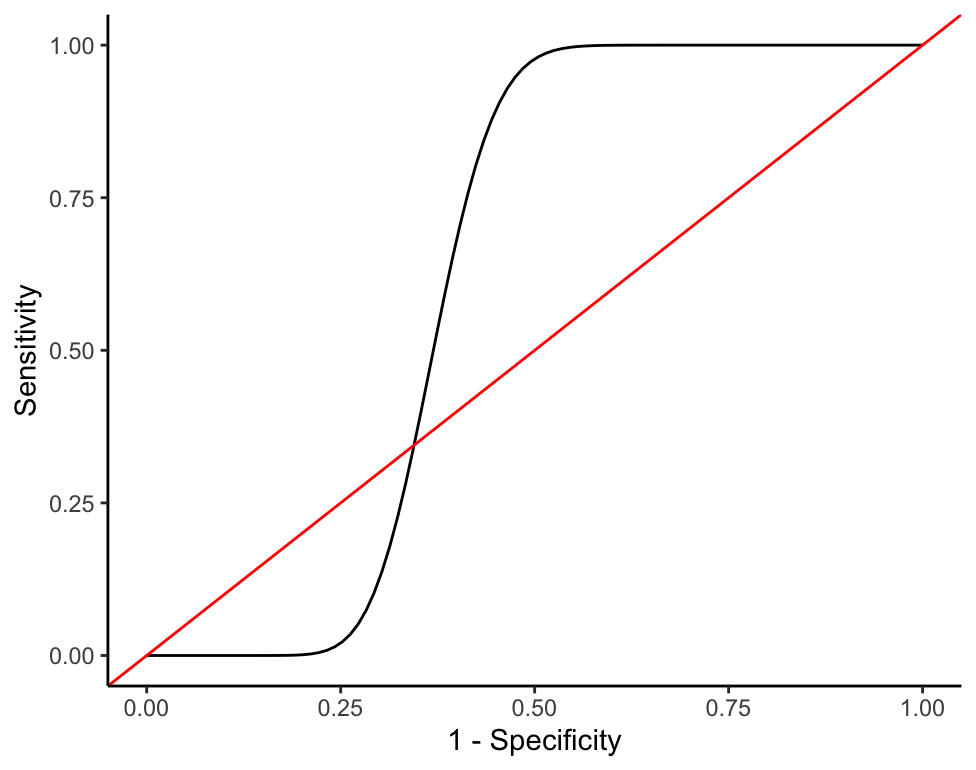} 
        \caption{} 
        \label{fig:ROCs}
    \end{subfigure}%
    %\hfill
    \begin{subfigure}[t]{0.5\textwidth}
        \centering
        \includegraphics[width=\linewidth]{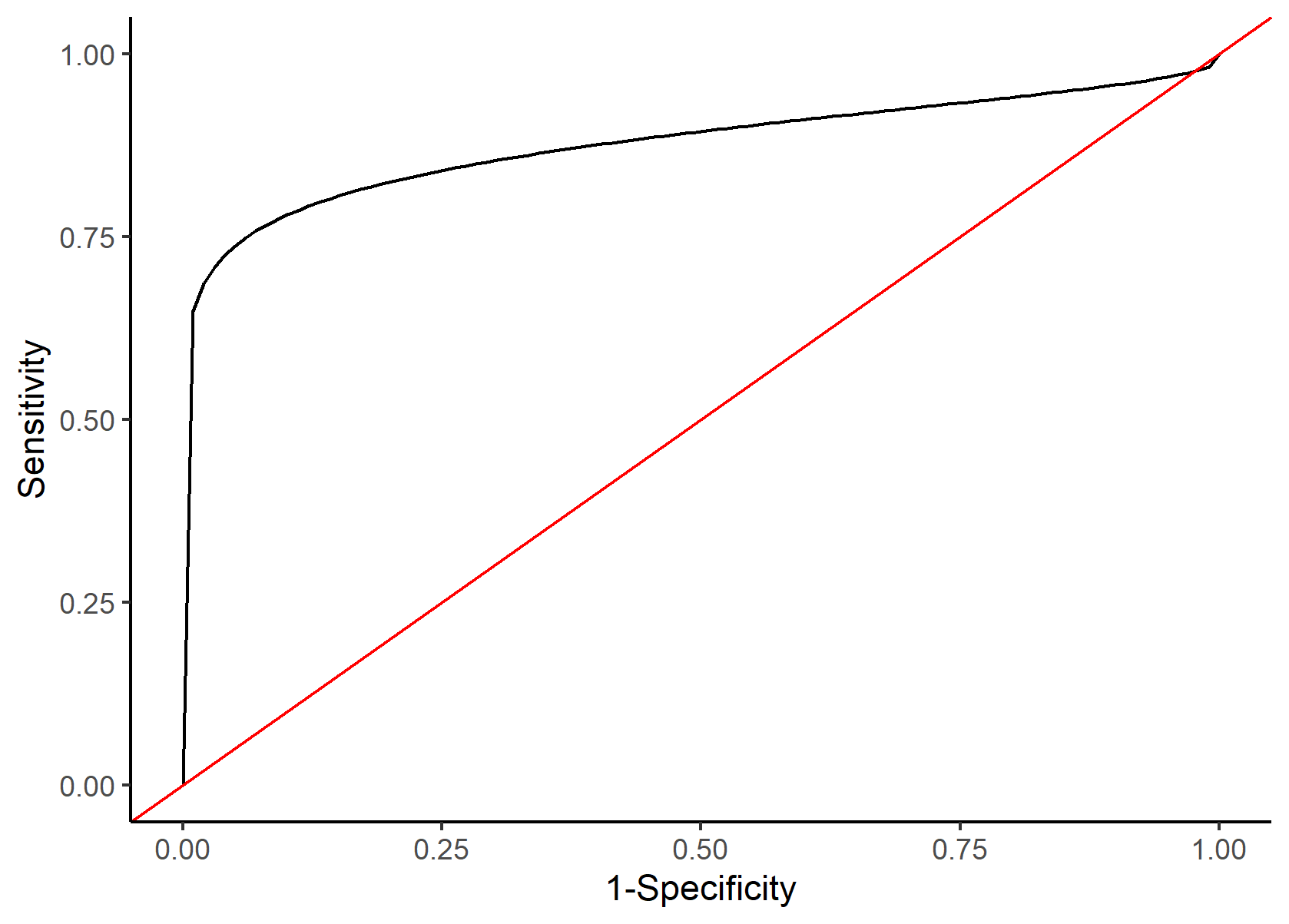} 
        \caption{} 
        \label{fig:Pancreas_BN}
    \end{subfigure}
    \caption{(a) An S-shaped ROC curve from a Bi-Normal model with unequal variances, (b) an ROC curve with a ``hook'', \enquote{naïvely} estimated using the pancreatic cancer data detailed in Section 4. The red line is the chance line.}
\end{figure}

There is no inherent shape constraint on a ROC curve based on its definition, with the exception that it is monotonely nondecreasing. A ROC curve can be \enquote{S}-shaped. This can arise under the well-known Bi-Normal model with a considerable discrepancy in standard deviations between the affected and reference groups \citep{green1966signal}, as illustrated in Figure~\ref{fig:ROCs}. A ROC curve may feature a discernible dip at higher specificity levels, exhibiting the so-called \enquote{hooks} phenomenon. Figure~\ref{fig:Pancreas_BN} depicts such an example that was obtained from a non-concave standard Bi-Normal model for assessing the diagnostic potential of the carbohydrate antigen CA199 for pancreatic cancer \citep{wieand1989family}. In both cases, the ROC curve features a portion below the chance line, suggesting that it is sub-optimal as there are always decision points that have better sensitivity or specificity. 

These nonconcave ROC curve shapes are not uncommon in medical research and have been extensively discussed in the literature \citep{bandos2017estimating, hillis2011using, hillis2012simulation}. These symptoms are indicative of non-proper ROC curves that are usually based on a non-optimal decision rule. Recent literature advocates for the use of proper ROC curves \citep{egan1975signal,metz1999proper} that ensure a concave shape curvature and interpretable sensitivity measures at any specificity level. Concave ROC curves have been proposed either by considering special distributions for the affected and reference test scores or by considering alternative decision variables. The former includes the Bi-Gamma model of \citet{dorfman1996proper}, the Bi-Lomax model of \citet{campbell1993application}, and the Bi-Beta model of \citet{mossman2016using}, among others. A common theme of these works is the use of some shared parameters in the affected and reference distributions to induce concavity. In the latter, the proper Bi-Normal model of \citet{metz1999proper} resorts to a likelihood ratio transformation of the test scores as a new decision variable; see also \citet{hillis2016equivalence} and \citet{sacchetto2018proper}.

While the literature has addressed concavity in ROC analysis, existing methodologies are predominantly parametric, making them susceptible to model mis-specifications. This paper aims to bridge this gap by proposing alternative concave ROC models that are less sensitive to such mis-specifications. Our proposed approach leverages placement values (PV), a concept used by \citet{pepe2003statistical} and others that makes use of standardization of affected test scores relative to the reference population. The advantage of the PV-based approach lies in its representation of the ROC curve as the distribution function of PVs. Recent extensions in this approach, pioneered by \citet{cai2002semiparametric,cai2004semi,lin2012direct}, enable semiparametric and nonparametric modeling of ROC curves. To achieve concavity in the ROC curve, we adopt the PV-based approach and incorporate the concept of concave distribution functions introduced by \citet{hansen2002nonparametric}. In essence, our approach expresses a ROC curve as the distribution function of PVs, and utilizes the concave distribution function concept to impose the desired concave shape of the distribution function through a mixture distribution. Furthermore, we present both parametric and semi-parametric frameworks of the proposed approach to accommodate complex scenarios effectively, and adopt a Baeysian perspective to facilitate the estimation and inference. %The benefit of this integrative framework is immediate: the constructed ROC curves are not only directly modeled as a function of covariates, they are also concave at every value of the covariates. %This can be achieved with a little loss of efficiency. 
%to obtain proper estimates of diagnostic accuracy at all plausible values of GAs. Following \citet{ghosal2019discriminatory}, we can see that low value of GA (17 weeks) results in improper ROC (see Figure \ref{fig:ROC_SCAN_PVBN_17}). Thus in this article, we propose a new framework that allows simultaneous concavity and direct covariates adjustment. This is achieved by using the placement value framework \citep{pepe2003statistical,cai2004semi,sullivan2004analysis} of ROCs and building on the literature of concave distribution \citep{hansen2002nonparametric,hanson2008modelling}. 
%(Brief introduction). The research questions of interest are XXX.

%\begin{figure}
%     \centering\includegraphics[scale=0.5]{transformed_Pancrease_BN_demo.png}
%     \caption{A \enquote{naïvely} estimated ROC curve in Pancreas data. Red line %indicates the chance line.}
%        \label{fig:Pancreas_BN}
%\end{figure}

The rest of the article is organized as follows. Section \ref{Part:Method} describes the detailed development of the proposed approach, including model formulations and estimations by the Bayesian paradigm. Both parametric and semiparametric frameworks will be considered. Section \ref{Part:Simulation} demonstrates the performance of the developed methodology through detailed simulation studies, and  Section \ref{Part:Data_analysis} presents a real data application that evaluates the diagnostic accuracy of carbohydrate antigen CA199 for pancreatic cancer. We conclude with a brief discussion in Section \ref{Part:Discussion}.

%{\color{red} The receiver operating characteristic (ROC) curve is a popular two-dimensional measure of classification to differentiate two population. The area under the curve (AUC) is the scalar measure to quantify ROC. By nature, ROC curve can be both concave and convex. But concavity is a property of a \textit{proper} ROC curve \citep{dorfman1996proper}, which ensures the resultant ROC will never go below the chance line, keeping the range of AUC in (0.5, 1). Often while discriminating between two very similar distributions, we expect the AUC to be close to 0.5. An AUC value less than 0.5 becomes hard to interpret since a random guess by a predictor to distinguish two populations is 0.5. However if the ROC is not concave, it can not be guaranteed that AUC will always more than 0.5. The widely popular Bi-Normal model doesn't ensure concavity (when $b \neq 1$,\citet{huang2009parametric}); hence the corresponding AUC theoretically could be less than 0.5. Also in ROC-based regression framework AUC could be less than 0.5. But if a ROC is proper (i.e. concave), the AUC will always be more than 0.5. Hence concavity is often desired in ROC.}

\section{Modeling framework} 
\label{Part:Method}

\subsection{Notations and background}
\label{Part:ROC}

A receiver operating characteristic curve is a graphical tool to assess the diagnostic ability of a test to discriminate two populations, say reference and affected \citep{pepe2003statistical}. Let $0$ and $1$ respectively be the status of the reference and affected populations, and let $Y_{i}^{0}$ and $Y_{i}^{1}$ be test scores of reference and affected subject $i \text{ } (i=1,\ldots,N$, with the understanding that $N$ can be different for the reference and affected populations). Let $F_{0}$ and $F_{1}$ be the corresponding cumulative distribution functions (CDF). Then the ROC curve is defined as 
\begin{equation}
ROC(t) = 1-F_{1}\left(F_{0}^{-1}(1-t)\right),\text{ } t \in \left(0,1\right). \label{Eqn:ROCgeneral}
\end{equation}
The area under the ROC curve (AUC) is a measure that can be interpreted as the probability that a test score of an affected subject is higher than that of a reference subject (i.e., $P[Y_i^1 > Y_i^0]$) and is given as $AUC=\int_0^1 ROC(t) dt.$
%\subsection{Bi-Normal models} 
%\label{Part:BinormalModel}

Different parametric ROC models can be considered by varying $F_0$ and $F_1$. The Bi-Normal (BN) model is a parametric approach with a normality assumption for affected and reference test scores, 
\begin{equation*}
Y_{i}^{1} \sim N\left(\mu_{1}, \sigma_{1}^2 \right) \text{  and  } Y_{i}^{0} \sim N\left(\mu_{0}, \sigma_{0}^2 \right).
\end{equation*}
The resultant ROC curve and AUC have closed-form expressions given by
\begin{align}
ROC(t) &= \Phi\left(a+b\Phi^{-1}(t) \right), \text{ } t \in \left(0,1\right) \label{Eqn:BiNorm_ROC}\\
AUC &= \Phi\left( \frac{a}{\sqrt{1+b^2}}\right), \label{Eqn:BiNorm_AUC}
\end{align}
where,
\begin{equation*}
a=\frac{\mu_{1}-\mu_{0}}{\sigma_{1}},\text{ } b=\frac{\sigma_{0}}{\sigma_{1}}, 
%\label{Eqn:AUCest}
\end{equation*}
\noindent
and $\Phi(\cdot)$ is the CDF of the standard normal distribution.

\subsection{Placement value-based approaches}
\label{Part:PV}

%The use of placement values (PV) to make inference about ROC curve is not novel \citep{hanley1997sampling}. PV has been extensively used in several parametric, semi-parametric and non-parametric methods to estimate ROC curves \citep{cai2004semi,sullivan2004analysis,alonzo2002distribution,qin2006empirical}.
To seek a direct approach to estimating the ROC curve, Pepe and others \citep{pepe2003statistical, sullivan2004analysis, cai2004semi, lin2012direct} proposed to use placement values. A placement value (PV) is defined as the standardization of the test score from the affected population relative to that of the reference  population:%. That is, the PV $z_{i}$ of the affected score $Y_{i}^{1}$ can be defined as
$$z_{i} = 1- F_{0}(Y_{i}^{1}).$$
In other words, PV $z_{i}$ can be interpreted as the proportion of the reference population with scores more than that of affected score $Y_{i}^{1}$. It can be easily shown that the CDF of $z_{i}$ $(\text{say, } F_{z})$ is the ROC curve to discriminate reference and affected populations. As such, whereas we model $Y_{i}^{1}$ and $Y_{i}^{0}$ in the conventional approach of ROC analysis, we model $z_{i}$ and $Y_{i}^{0}$ in the PV-based approach. 

%%In this manuscript, the later subsections will focus primarily on the modeling framework of the PV while following traditional parametric and semiparametric techniques of estimating the distribution function of the reference score to obtain the PV.

\subsection{Concave ROC curves}
\label{Part:Concave}

The aforementioned connection between ROC curve and CDF of the PV makes it possible to impose a concave curvature on the ROC curve. To that end, we borrow the idea of a concave CDF from \citet{hansen2002nonparametric} as stated below for completeness.
\begin{lemma} \label{intlemma}
A CDF $F$ is concave if and only if there exists a distribution function $G$ on the same support $\Omega$ of $F$ such that $F$ admits the representation
\begin{eqnarray*}
F(z)=\int F_w (z) dG(w),~z\in \Omega,
\end{eqnarray*}
where $F_w$ for $w \ne 0$ is the distribution function corresponding to the uniform distribution on $(0,~w)$, i.e.,
\begin{eqnarray*}
F_w (z) =\frac{1}{w} min\{z,~w\}.
\end{eqnarray*}
\end{lemma}

Equivalently, this result can be stated as follows, which can aid the development of an efficient computational algorithm:
\begin{lemma}\label{prodlemma}
Let $F$ be the CDF of $Z$. $F$ is concave if only if $Z$ can be written as $Z=WV$ where $W$ and $V$ are independent, $V\sim U(0,1)$ and $W$ has some distribution $G$ on the same support of $Z$.
\end{lemma}

It is immediate from Lemma \ref{prodlemma} that $AUC=1-E(W)/2$. The essence of these results is simple: in order to model the distribution function $F_{z}$ of $z$, we can model the distribution function $G$ of $w$. Our strategy to model a concave ROC curve using $\z$ therefore involves three steps: 
\begin{enumerate}
\item Construct a hierarchical model
\begin{eqnarray}
& z_i | w_i \sim U(0,w_i), \nonumber \\
& w_i | \bdelta \sim G(w_i|\bdelta), \nonumber 
\end{eqnarray}
\noindent
where $\bdelta$ is the parameter vector indexing $G$.
\item Develop an MCMC algorithm to obtain a sample of $\w=(\w^{(1)},\dots,\w^{(L)})$ from the posterior distribution $p(G|\z)$, where $\w^{(l)}=(w_{1}^{(l)},\dots,w_{S}^{(l)})$ is a sample of size $S$ at the $l^{\text{th}}$ iteration.
\item Obtain an estimate $\hat{F}_{z}$ of $F_{z}$ (hence an estimate of the ROC curve) as,
\begin{align}
\hat{F}_{z}^{(l)} (t)=\frac{1}{S}\sum_{s=1}^S F_{w_{s}^{(l)}} (t) =\frac{1}{S}\sum_{s=1}^S \frac{1}{w_{s}^{(l)}} min\{z,~w_{s}^{(l)}\},\text{ }t \in (0,1). 
\label{Eqn:Fest}
\end{align}
\item Given the sample $\w^{(l)}$, we can also obtain $\widehat{AUC}^{(l)}=\frac{1}{2S} \sum_{s=1}^S (1-w_{s}^{(l)})$. 
\end{enumerate}

There are several ways to specify $G$. Parametrically, we can specify $G\sim N(\mu,\sigma^2)$ to propose a parametric representation of the concave ROC curve model (called pCN thereafter)
\begin{eqnarray}
& z_i | w_i \sim U(0,w_i), \nonumber \\
& \eta^{-1}(w_i) \sim N(\mu,\sigma^2). \label{Eqn:pCN} 
\end{eqnarray}
Alternatively, we can propose a semiparametric way of modeling $G$ to allow more flexibility. By following a DPM modeling approach \citep{sethuraman1994constructive}, we specify
 \begin{align*}
G(w_i|\mu_i,\sigma^2) &= \int K(w_i; \mu_i,\sigma^2)dH(\mu_i), \\
H(\mu_i) &\sim \mathcal{D}(\alpha H_0(\mu_i)),
\end{align*}
\noindent
where $\mathcal{D}$ denotes a Dirichlet process \citep{ferguson1973bayesian} with base measure $H_0$ and precision parameter $\alpha$ and $K(\cdot; \mu, \sigma^2)$ is a kernel with parameters $\mu$ and $\sigma$. Following the parametric approach of concavity modeling, we can write our semi-parametric concave model (called spCN thereafter) using DPM prior for $G$ as:
\begin{eqnarray}
& z_i | w_i \sim U(0, w_i), \nonumber \\
& \eta^{-1}(w_i | \mu_i, \sigma^2) \sim G(w_i|\mu_i, \sigma^2) = \int K(w_i;\mu_i, \sigma^2) dH(\mu_i), \nonumber \\
&K(w; \mu_i, \sigma^2) = N(w_i |\mu_i,\sigma^2),  \nonumber \\
&H(\cdot) \sim \mathcal{D}(\alpha_0 H_{0} (\cdot)),  \nonumber\\
&H_{0}(\mu_i) = N(\mu_i|\mu_0,\sigma_0^2),  \nonumber
\end{eqnarray}
where $\alpha_0,~\mu_0,~\sigma_\mu^2$ are all constants. Equivalently, this spCN model can be written follows:
\begin{eqnarray}
& z_i | w_i \sim U(0, w_i), \nonumber \\
& \eta^{-1}(w_i |\mu_i,\sigma^2) \sim N(\mu_i,\sigma^2), \nonumber \\
& \mu_i | H \sim H=\sum_{k} p_k \delta_{\mu_k^{*}}, \label{Eqn:Semipar_con} \\
& p_1=V_1,~p_k=V_k (1-V_{k-1})\cdots(1-V_{1}), \nonumber \\
& V_i \sim \text{Beta}(1,\alpha), \nonumber \\
& \mu_i^{*} \sim H_0= N(\mu_0,\sigma_0^2).  \nonumber
\end{eqnarray}

%Following the structure of equation (\ref{Eqn:Semipar_con}), we can also propose a parametric counterpart of the model (say, pCN) which can be written as:
%\begin{eqnarray}
%& z_i | w_i \sim U(0,\eta^{-1}(w_i)), \nonumber \\
%& w_i |\mu_i,\sigma^2 \sim N(\mu,\sigma^2), \label{Eqn:Par_con} \\
%& \mu \sim N(\mu_0, \sigma_0^2), \nonumber 
%\end{eqnarray}
%\noindent
%where $\mu_0$ and $\sigma_0$ are hyperparameters.

%Another subsidiary model that arises from spCN in equation (\ref{Eqn:Semipar_con}), that uses a quadratic ($\text{spCN}_q$) transformation of the raw test scores. Following \citet{hillis2016equivalence}, we empirically estimate $c_1$ as $$c_1 = -\frac{ab}{(1-b^2)},$$ where $a$ and $b$ are empirically estimated from the raw test scores. Then the quadratic transformed test scores ($Y^{0*}$ and $Y^{1*}$ respectively) have the form
 %\begin{eqnarray}
%&Y^{0*} = (Y^0-c_1)^2 \text{ and } Y^{1*} = (Y^1-c_1)^2. \nonumber 
%\end{eqnarray}
%Further, these transformed scores are used to estimate the placement values and further in equation (\ref{Eqn:Semipar_con}).

\subsection{Estimation, inference, and computation} 
\label{Part:Likelihood}

We take a Bayesian approach for the inference and use proper objective prior. Specifically, each $\mu_i$ follows $N(0,100)$ priors and variance parameter $\sigma^2$ follows $IG(0.01, 0.01)$.

We use \texttt{RJAGS} to implement the Monte Carlo Markov chain (MCMC) algorithms to generate samples from the posterior distribution of the model parameters given the data. Both visual inspection of the trace plots and diagnostic tools \citep{gelman1992inference} are used to ensure convergence of the MCMC chains. After convergence, we thin the iterations to produce a sample of 5000 to produce posterior means, standard deviations and 95\% credible intervals. R code of implementing simulation and real data analysis will be made available online.

\section{Simulation studies}
%\label{Part:posterior_inference}
\label{Part:Simulation}

As PV-based models have been studied elsewhere \citep{ghosal2019discriminatory, chen2019, stanley2018beta, ghosal2022estimation, sullivan2004analysis}, we focus on the concave ROC curves in our simulations. We generate data from Binormal (PBN), Bigamma (BG), and pCN and evaluate the performance of existing and proposed approaches. We examine bias and efficiency in AUC estimates and empirical mean square error (EMSE) in ROC curve estimates, where $$EMSE = \int_0^1 \left[\widehat{ROC(t)}-ROC(t)\right]^2 dt.$$ EMSE is the preferred metric of performance since ROC curves with different curvatures can have the same AUC value. 

\subsection{Simulation scenarios}
%\label{Part:posterior_inference}
\label{Part:Sim_scenarios}

\begin{enumerate}
%\begin{description}
\item \textbf{Proper Bi-Normal (PBN)}. We generate the reference $(Y^0)$ and affected $(Y^1)$ test scores respectively from $Y^0 \sim N(0,1)$ and $Y^1 \sim N\left(\frac{\alpha_0}{\alpha_1}, \frac{1}{\alpha_1^2}\right)$. Following the equivalence of PBN and the bi-chi-square distribution \citep{hillis2016equivalence}, we calculate $\lambda$ and $\theta$ as 
\begin{align*}
\lambda  = \frac{1}{\alpha_1^2} \text{ and } \theta = \frac{\alpha_0^2 \alpha_1^2}{(1-\alpha_1^2)^2},
\end{align*}
so that the true ROC curve and corresponding AUC can be obtained as
\begin{eqnarray}
& ROC(t) = 
\begin{dcases}
	1-F_{\lambda\theta}\left(\frac{1}{\lambda}F_{\theta}^{-1}(1-t)\right), & \lambda > 1 \\
	F_{\lambda\theta}\left(\frac{1}{\lambda}F_{\theta}^{-1}(1t)\right), & \lambda < 1
\end{dcases} \\
& AUC = \Phi\left( \frac{\sqrt{\theta}\sqrt{\lambda-1}}{\sqrt{\lambda+1}}\right)+2F_{BVN} \left( -\frac{\sqrt{\theta}\sqrt{\lambda-1}}{\sqrt{\lambda+1}},0; -\frac{2\sqrt{\lambda}}{\lambda+1}\right),
\end{eqnarray}
\noindent
%{\color{red}{Check latex code of formulas above to make sure sqrt signs are appropriate. If a single parameter inside sqrt sign, there is no need to have parentheses.}}
where $F_{\nu}$ is then CDF of a chi-square distribution with noncentrality parameter $\nu$ and $F_{BVN}(\cdot,\cdot; \rho)$ denotes the CDF of a standardized bivariate normal distribution with correlation $\rho$. The case of $\lambda=1$, the true AUC and ROC have the same forms as that from the Bi-Normal (BN) model: $AUC = \Phi\left( \frac{\alpha_0}{\sqrt{1+\alpha_1^2}}\right)$, $ROC(t) = \Phi\left(\alpha_0+\alpha_1 \Phi^{-1}(t)\right)$.

\item \textbf{Bi-Gamma (BG)}: The BG model \citep{dorfman1996proper} postulates that $Y^0 \sim Gam(k,\phi_0)$ and $Y^1 \sim Gam(k,\phi_1)$, where $Gam(k,\phi)$ denotes a gamma distribution with mean $k\phi$. True ROC and AUC are given by
\begin{eqnarray}
%& ROC(t) = 1-G_1 \left(\frac{1}{\phi_1} G_1^{-1}(1-t)\right) \\
& ROC(t) = 1-\mathbb{G}_1 \left(\mathbb{G}_0^{-1}(1-t)\right) \\
& AUC = 1-H_{(2k, 2k)} \left( \frac{\phi_0}{\phi_1}\right), \text{ }t\in (0,1).
\end{eqnarray}
\noindent 
where $\mathbb{G}_l(\cdot)$ is the CDF of $Gam(k,\phi_l)$, $l=0,1$, and $H_{\nu_1,\nu_2}$ is CDF of the F-distribution with degrees of freedom $\nu_1$ and $\nu_2$.

\item \textbf{Parametric Concave (pCN)}: The third data-generating scenario follows from the proposed pCN model. We first generate the healthy score $Y^0 \sim N(0,1)$ then the PV $z \sim U(0,\Phi^{-1}(w^{*}))$, where $w^{*}$ is from $N\left(\frac{\alpha_0}{\alpha_1}, \frac{1}{\alpha_1^2}\right)$. Assuming normality on the reference population scores, the affected scores can be calculated as $Y^1 = \Phi^{-1}(1-z)$. Since the AUC and ROC curve do not have closed forms under pCN, we take an empirical approach to obtain their truth. In particular, we first obtain $z$ and then calculate its empirical CDF and corresponding $\text{AUC}=1-\bar{z}$. We replicate this process 10000 times and treat the average of empirical CDFs and AUCs as the underlying truth.
%\end{description}
\end{enumerate}

We vary the underlying parameters of each of the three data-generating mechanisms to obtain three levels of AUC: low, medium, and high. Under PBN, the AUC level is low with AUC of 0.673 $(\alpha_0=0.5,\text{ }\alpha_1=0.7)$, medium with AUC of 0.762 $(\alpha_0=0.5,\text{ }\alpha_1=0.45)$ and high with AUC of 0.842 $(\alpha_0=0.5,\text{ }\alpha_1=0.28)$. Under BG, the corresponding true AUCs are 0.665 $(k=1,\text{ }\phi_0=1,\text{ }\phi_1=2)$, 0.774 $(k=1,\text{ }\phi_0=1,\text{ }\phi_1=3.5)$ and 0.869 $(k=1,\text{ }\phi_0=1,\text{ }\phi_1=7)$. Similarly, when the data are generated from the pCN model, the corresponding AUCs are 0.619 $(\alpha_0=1,\text{ }\alpha_1=1)$, 0.741 $(\alpha_0=0.1,\text{ }\alpha_1=3)$ and 0.916 $(\alpha_0=-15,\text{ }\alpha_1=15)$. For each of these nine combinations, we generate test scores for reference ($N=1000$) and affected ($N=1000$) subjects and fit five fitting models to the resultant data: BN, BG, PBN, pCN, and spCN.
 
We create 1000 data replicates and report average posterior mean and bias of AUC and EMSE (times 1000) of ROC in Table \ref{tab:Sim_noCov}. %For each model in each scenario, we run 100000 iterations, thinning every 20 iterations after burning 100000 iterations  for 3 chains. 
  We also plot 200 randomly selected estimated ROC curves from the 1000 replicated datasets in Figures \ref{fig:Sim_ROC_noCov_PBN}-\ref{fig:Sim_ROC_noCov_pCN}.

\subsection{Simulation results}
%\label{Part:posterior_inference}
\label{Part:Sim_results}

\begin{table}[htbp]
\caption{Simulation results}
\label{tab:Sim_noCov}
\begin{center}
\adjustbox{max width=0.75\textwidth}{%
\begin{tabular}{lllrrrr}
\hline
\multicolumn{2}{l}{\textbf{Generating Model}}   & \multirow{2}{*}{\textbf{Fitting model}} & \multirow{2}{*}{\textbf{True AUC}} & \multirow{2}{*}{\textbf{Mean}} & \multirow{2}{*}{\textbf{Bias}} & \multirow{2}{*}{\textbf{$\text{EMSE } \times 1000$}} \\ \cline{1-2}
\textbf{Model}        & \textbf{AUC level} &                                         &                                    &                                &                                &                                                      \\ \hline
\multirow{15}{*}{PBN} & \multirow{5}{*}{Low}    & BN                                      & 0.673                              & 0.656                          & -0.017                         & 0.532                                                \\
                      &                         & BG                                      & 0.673                              & 0.646                          & -0.027                         & 2.733                                                \\
                      &                         & PBN                                     & 0.673                              & 0.675                          & 0.002                         & 0.120 \\
                      &                         & pCN                                    & 0.673                              & 0.661                          & -0.012                         & 0.814                                                \\
                      &                         & spCN                                   & 0.673                              & 0.666                          & -0.007                         & 0.328                                                \\\cline{2-7}       & \multirow{5}{*}{Medium} & BN                                      & 0.762                              & 0.672                          & -0.089                         & 8.789                                                \\
                      &                         & BG                                      & 0.762                              & 0.665                          & -0.097                         & 18.293                                               \\
                      &                         & PBN                                     & 0.762                              & 0.766                          & 0.004                          & 0.068 \\
                      &                         & pCN                                    & 0.762                              & 0.699                          & -0.062                         & 5.976                                                \\
                      &                         & spCN                                   & 0.762                              & 0.728                          & -0.033                         & 1.564                                                \\\cline{2-7}			& \multirow{5}{*}{High}   & BN                                      & 0.842                              & 0.682                          & -0.160                         & 26.604                                               \\
                      &                         & BG                                      & 0.842                              & 0.674                          & -0.167                         & 49.171                                               \\
                      &                         & PBN                                     & 0.842                              & 0.847                          & 0.006                          & 0.033 \\
                      &                         & pCN                                    & 0.842                              & 0.730                          & -0.112                         & 17.542                                               \\
                      &                         & spCN                                   & 0.842                              & 0.768                          & -0.074                         & 7.256                                                \\\hline
\multirow{15}{*}{BG}  & \multirow{5}{*}{Low}    & BN                                      & 0.665                              & 0.674                          & 0.009                          & 4.656                                                \\
                      &                         & BG                                      & 0.665                              & 0.665                          & 0.000                          & 0.118                                                \\
                      &                         & PBN                                     & 0.665                              & 0.744                          & 0.079                          & 9.192                                               \\
                      &                         & pCN                                    & 0.665                              & 0.651                          & -0.014                         & 0.749                                                \\
                      &                         & spCN                                   & 0.665                              & 0.664                          & -0.001                         & 0.297                                                \\\cline{2-7} 
                      & \multirow{5}{*}{Medium} & BN                                      & 0.774                              & 0.776                          & 0.002                          & 5.290                                                \\
                      &                         & BG                                      & 0.774                              & 0.774                          & 0.000                          & 0.091                                                \\
                      &                         & PBN                                     & 0.774                              & 0.864                          & 0.090                          & 12.629 \\
                      &                         & pCN                                    & 0.774                              & 0.767                          & -0.007                         & 0.689                                                \\
                      &                         & spCN                                   & 0.774                              & 0.773                          & -0.001                         & 0.277                                                \\\cline{2-7}  & \multirow{5}{*}{High}   & BN                                      & 0.869                              & 0.879                          & 0.010                          & 2.573                                                \\
                      &                         & BG                                      & 0.869                              & 0.869                          & 0.000                          & 0.054                                                \\
                      &                         & PBN                                     & 0.869                              & 0.938                          & 0.068                          & 7.272 \\
                      &                         & pCN                                    & 0.869                              & 0.884                          & 0.014                          & 0.692                                                \\
                      &                         & spCN                                   & 0.869                              & 0.863                          & -0.006                         & 0.322                                                \\\hline
\multirow{15}{*}{pCN} & \multirow{5}{*}{Low}    & BN                                      & 0.619                              & 0.620                          & 0.002                          & 0.225                                                \\
                      &                         & BG                                      & 0.619                              & 0.610                          & -0.009                         & 0.924                                                \\
                      &                         & PBN                                     & 0.619                              & 0.622                          & 0.003                         & 0.216                                                \\
                      &                         & pCN                                    & 0.619                              & 0.618                          & -0.001                         & 0.215                                                \\
                      &                         & spCN                                   & 0.619                              & 0.618                          & -0.001                         & 0.216                                                \\\cline{2-7} 
                      & \multirow{5}{*}{Medium} & BN                                      & 0.741                              & 0.743                          & 0.002                          & 0.799                                                \\
                      &                         & BG                                      & 0.741                              & 0.715                          & -0.026                         & 8.116                                                \\
                      &                         & PBN                                     & 0.741                              & 0.754                          & 0.013                         & 1.225 \\
                      &                         & pCN                                    & 0.741                              & 0.741                          & 0.000                          & 0.188                                                \\
                      &                         & spCN                                   & 0.741                              & 0.741                          & 0.000                          & 0.189                                                \\\cline{2-7} 
                      & \multirow{5}{*}{High}   & BN                                      & 0.916                              & 0.913                          & -0.003                         & 1.206                                                \\
                      &                         & BG                                      & 0.916                              & 0.857                          & -0.060                         & 14.106                                               \\
                      &                         & PBN                                     & 0.916                              & 0.922                          & 0.006                         & 1.648 \\
                      &                         & pCN                                    & 0.916                              & 0.916                          & 0.000                          & 0.182                                                \\
                      &                         & spCN                                   & 0.916                              & 0.916                          & 0.000                          & 0.184                                                \\\hline
\end{tabular}%
}
\end{center}
\end{table}

\begin{figure}[htbp]
\begin{adjustbox}{addcode={\begin{minipage}{\width}}{\caption{%
      ROC estimates for case when data is generated from PBN model
      }\label{fig:Sim_ROC_noCov_PBN}
      \end{minipage}},rotate=90,center}
\settoheight{\tempheight}{\includegraphics[width=.20\linewidth]{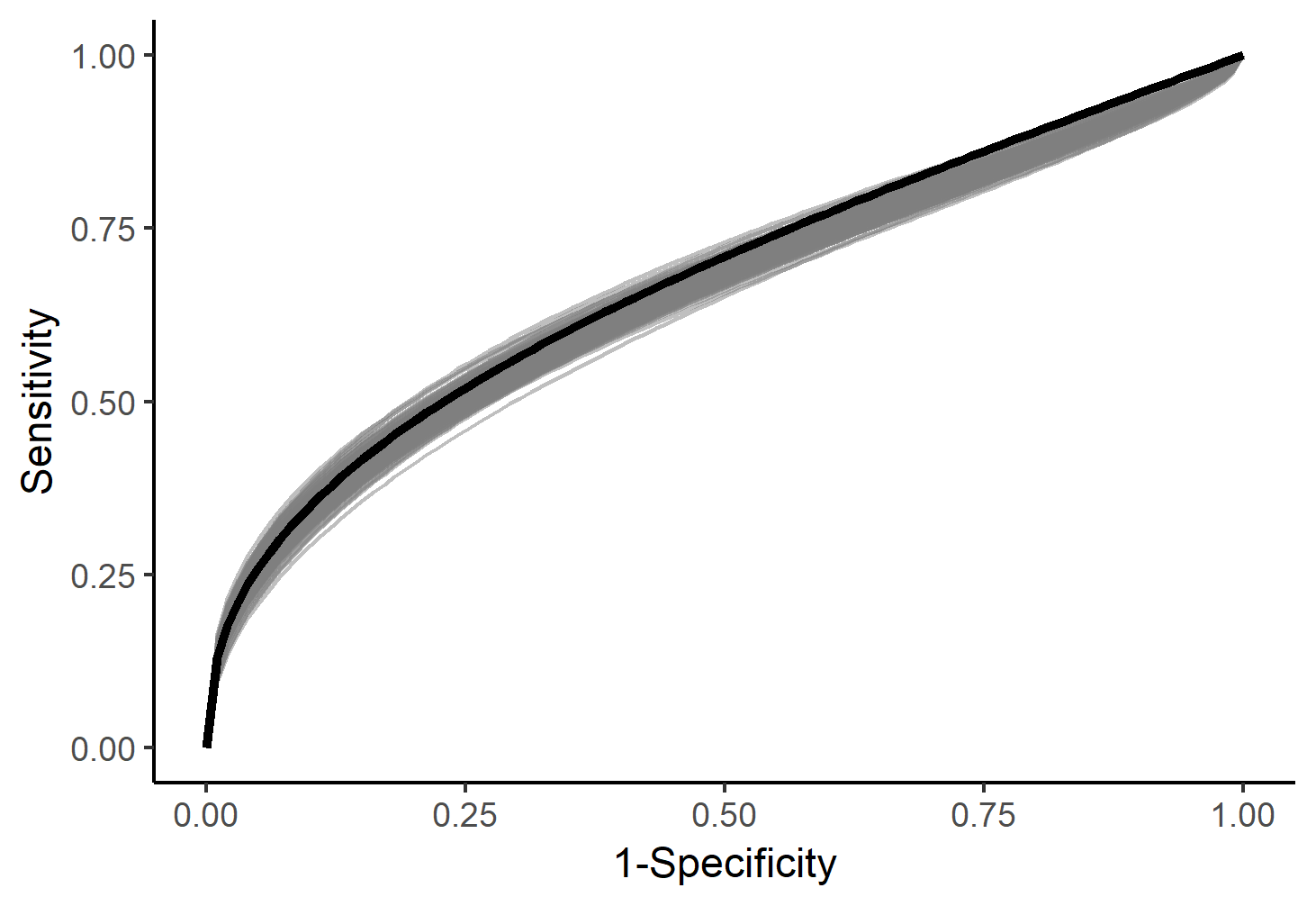}}%
\centering\begin{tabular}{@{}c@{ }c@{ }c@{ }c@{ } c@{ } c@{ }}
&\textbf{BN} & \textbf{BG} & \textbf{PBN} & \textbf{pCN} & \textbf{spCN} \\
\rowname{Low}&
\includegraphics[height=.25\columnwidth,width=.20\linewidth]{NoCov_BN_ROC_PBN_Low.png}&
\includegraphics[height=.25\columnwidth,width=.20\linewidth]{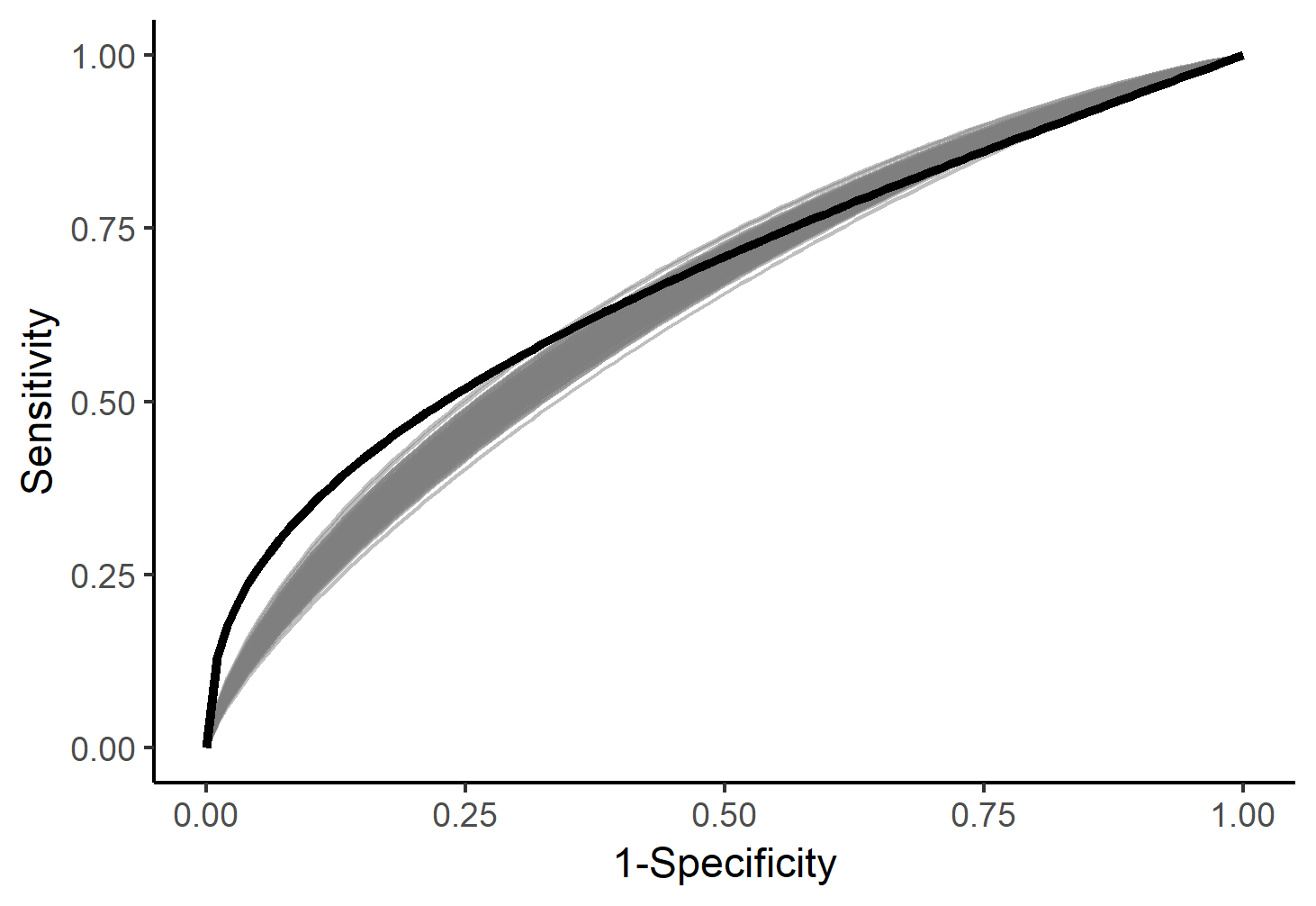}&
\includegraphics[height=.25\columnwidth,width=.20\linewidth]{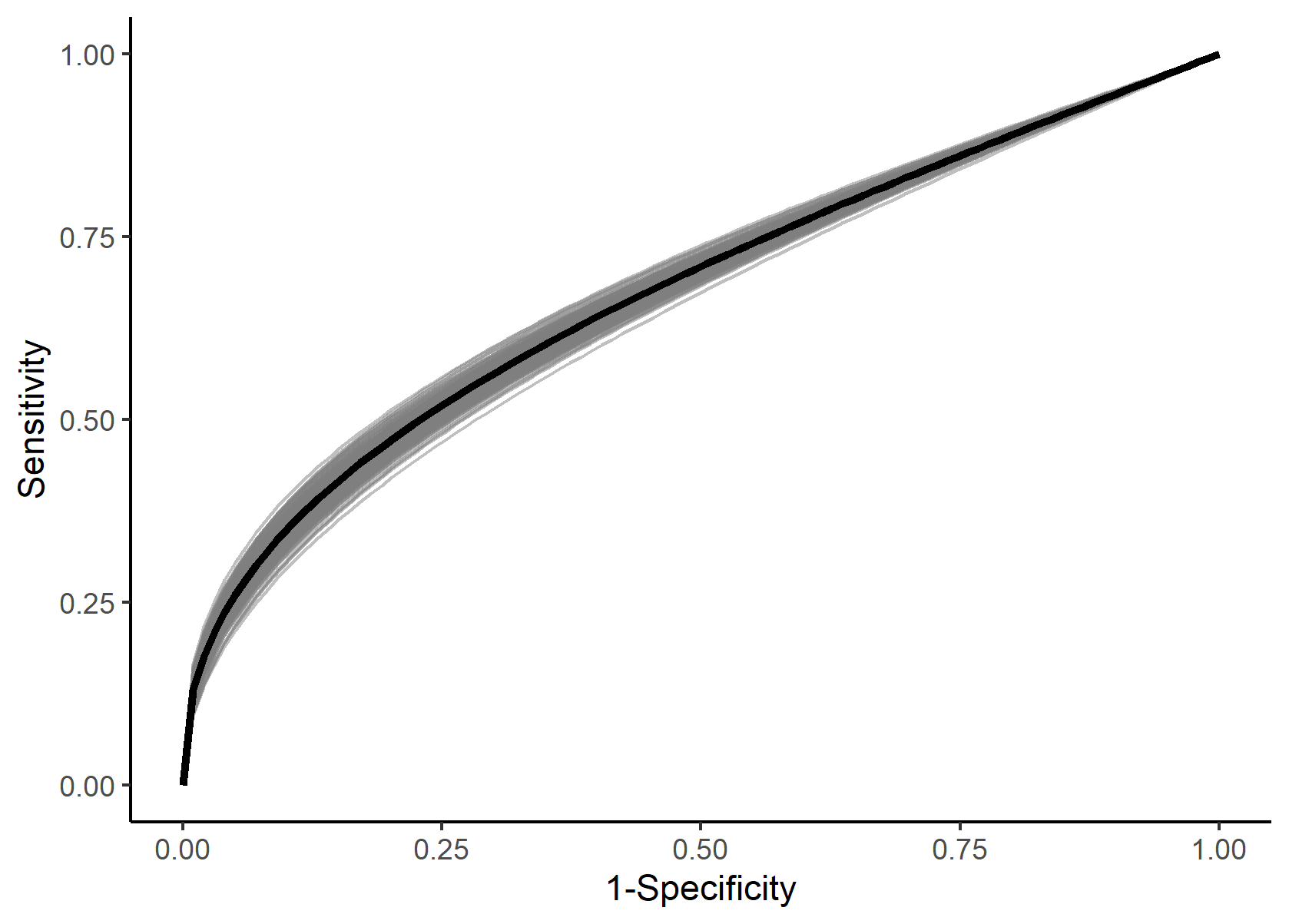}&
\includegraphics[height=.25\columnwidth,width=.20\linewidth]{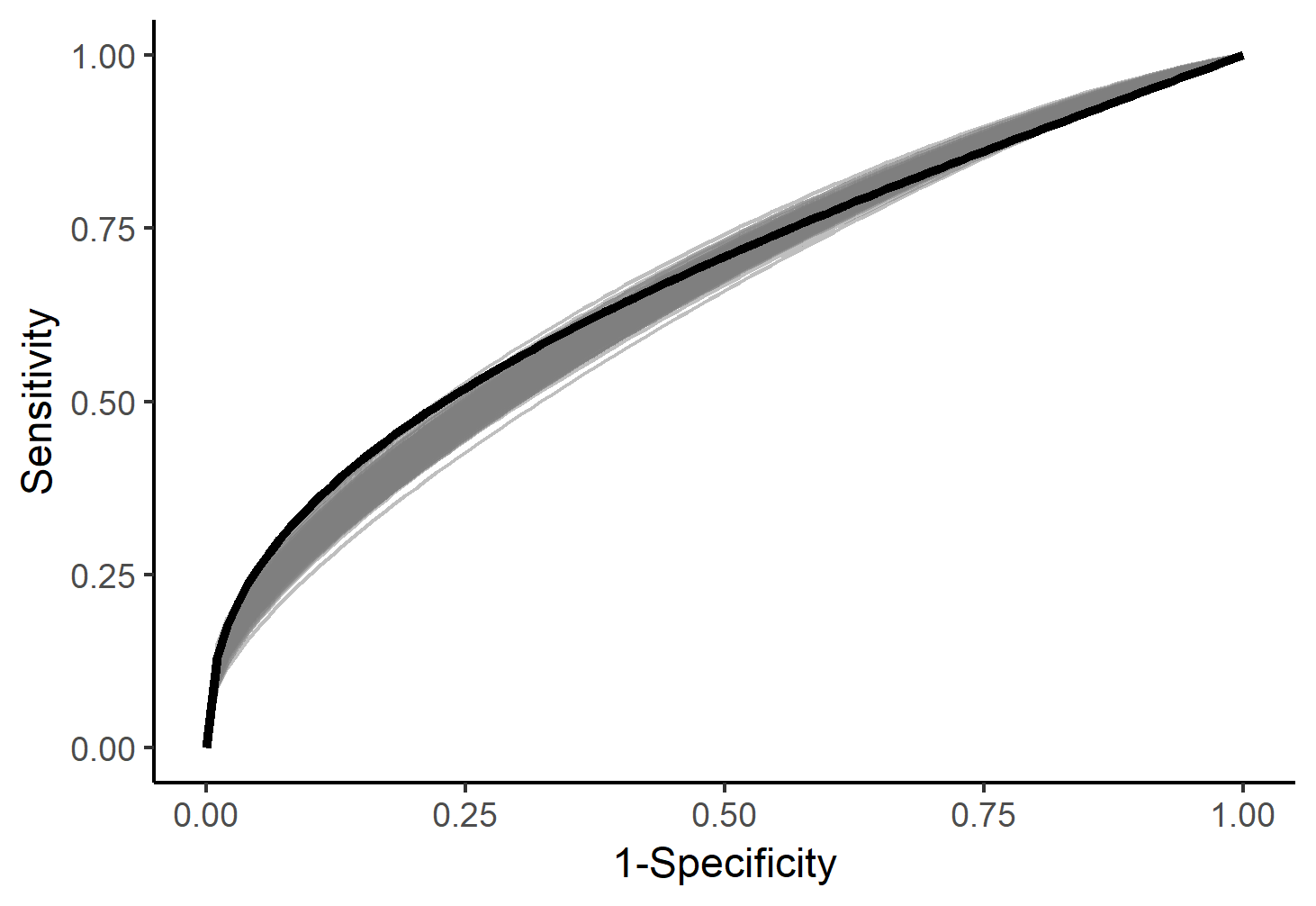}&
\includegraphics[height=.25\columnwidth,width=.20\linewidth]{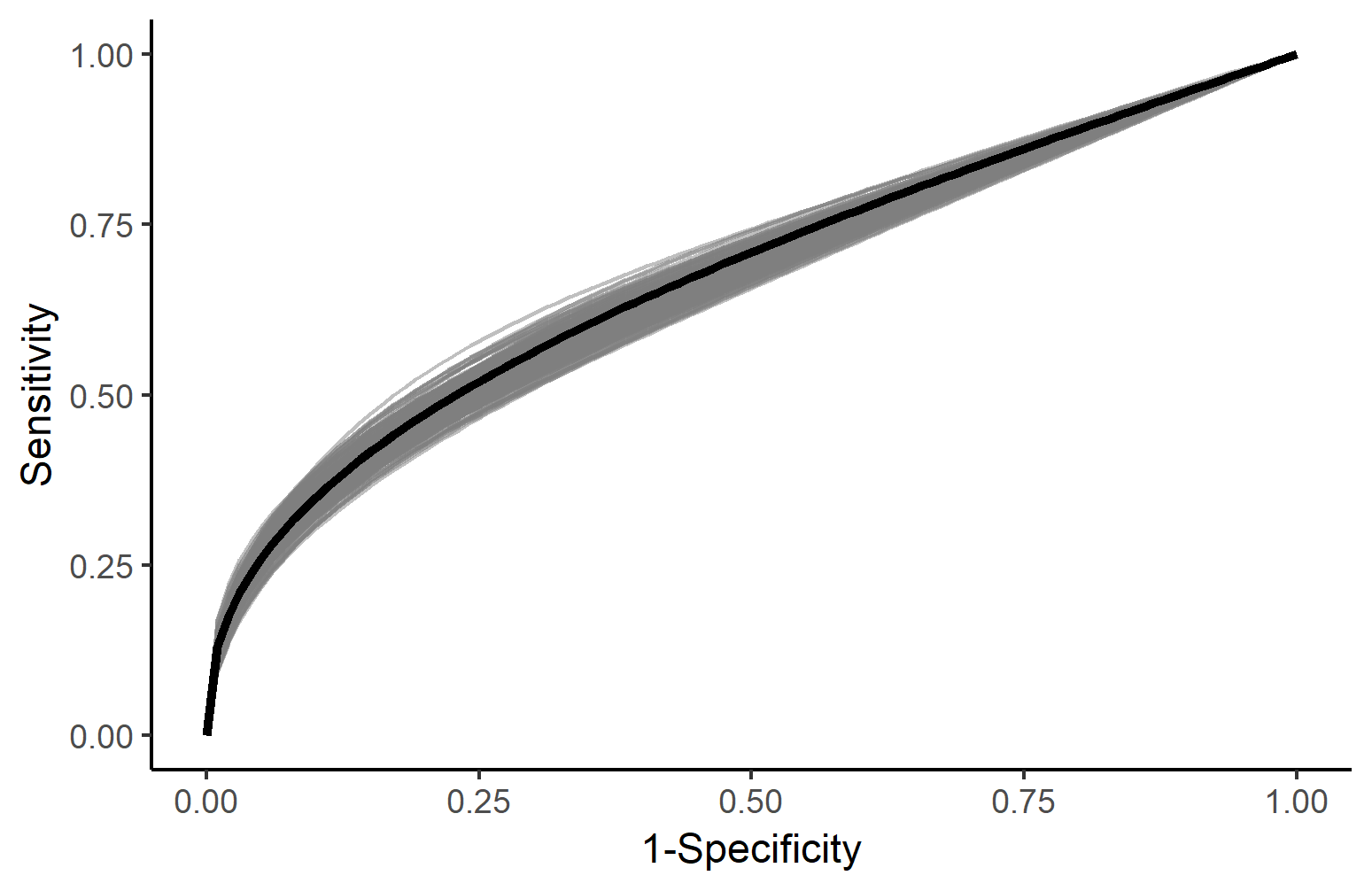}\\
%\includegraphics[height=.25\columnwidth,width=.20\linewidth]{NoCov_PBN_spCBN2_ROC_Low.png} \\
%[-1ex] &\mycaption{0.2} & \mycaption{0.2} & \mycaption{0.3}\\
\rowname{Medium}&
\includegraphics[height=.25\columnwidth,width=.20\linewidth]{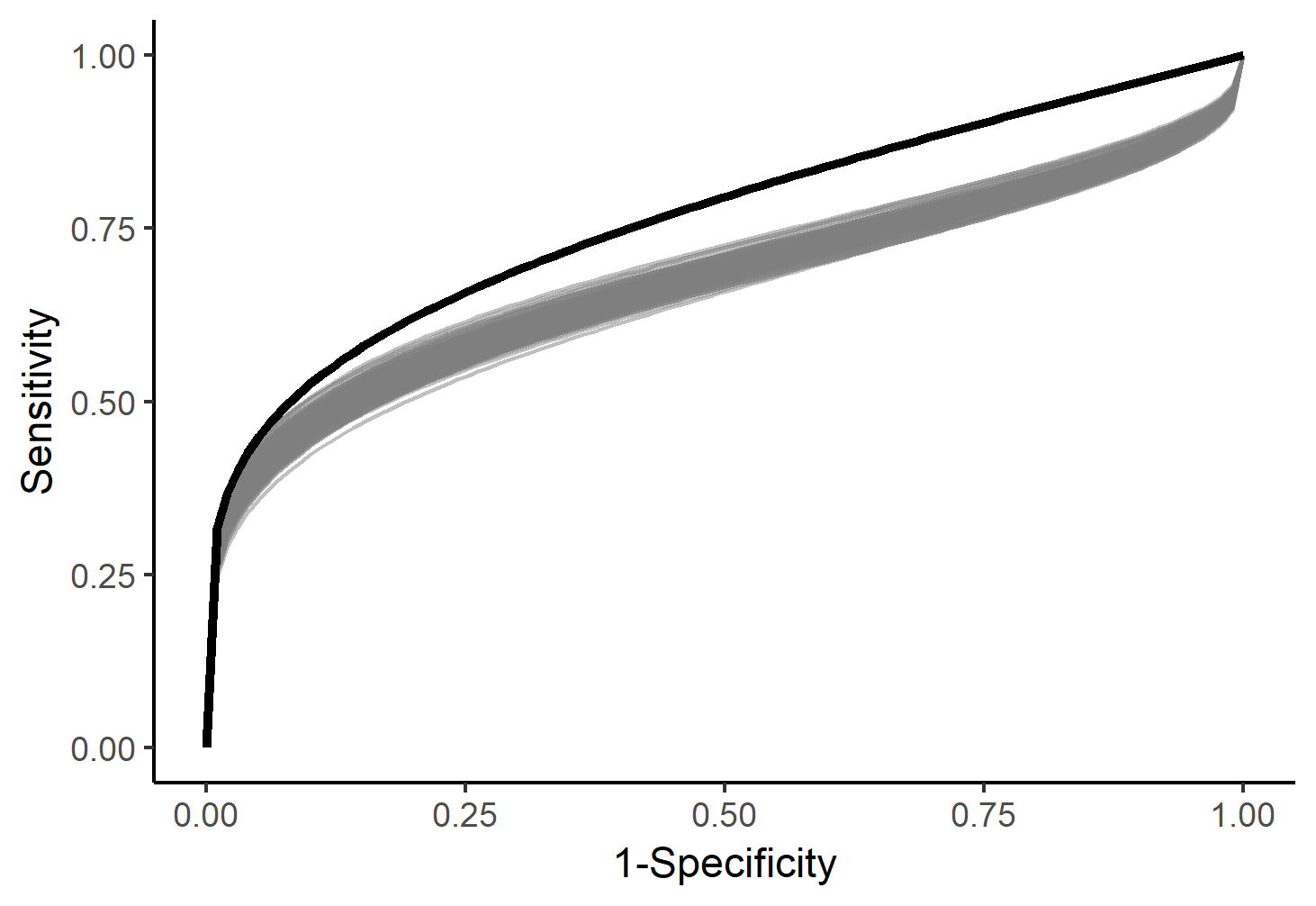}&
\includegraphics[height=.25\columnwidth,width=.20\linewidth]{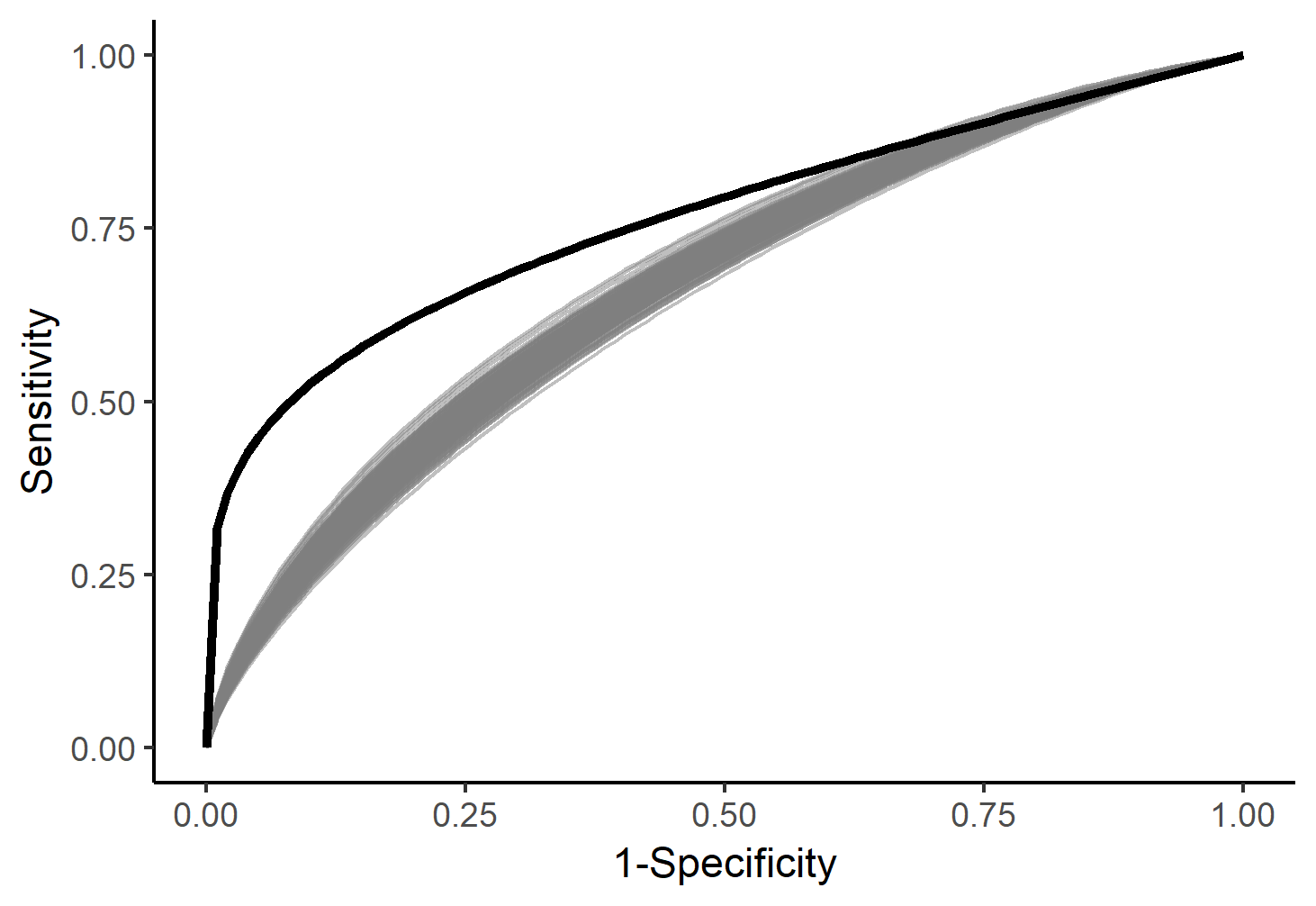}&
\includegraphics[height=.25\columnwidth,width=.20\linewidth]{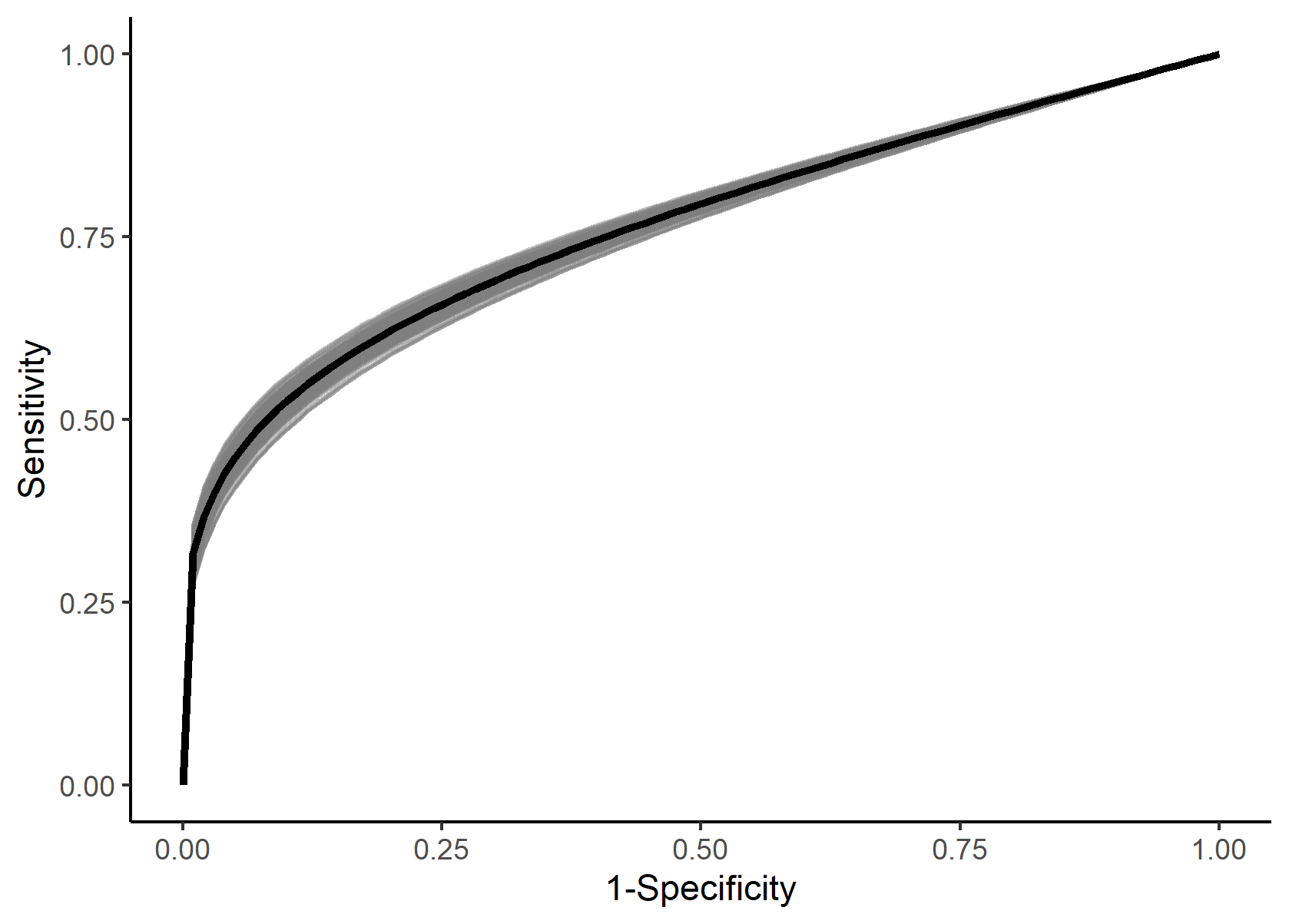}&
\includegraphics[height=.25\columnwidth,width=.20\linewidth]{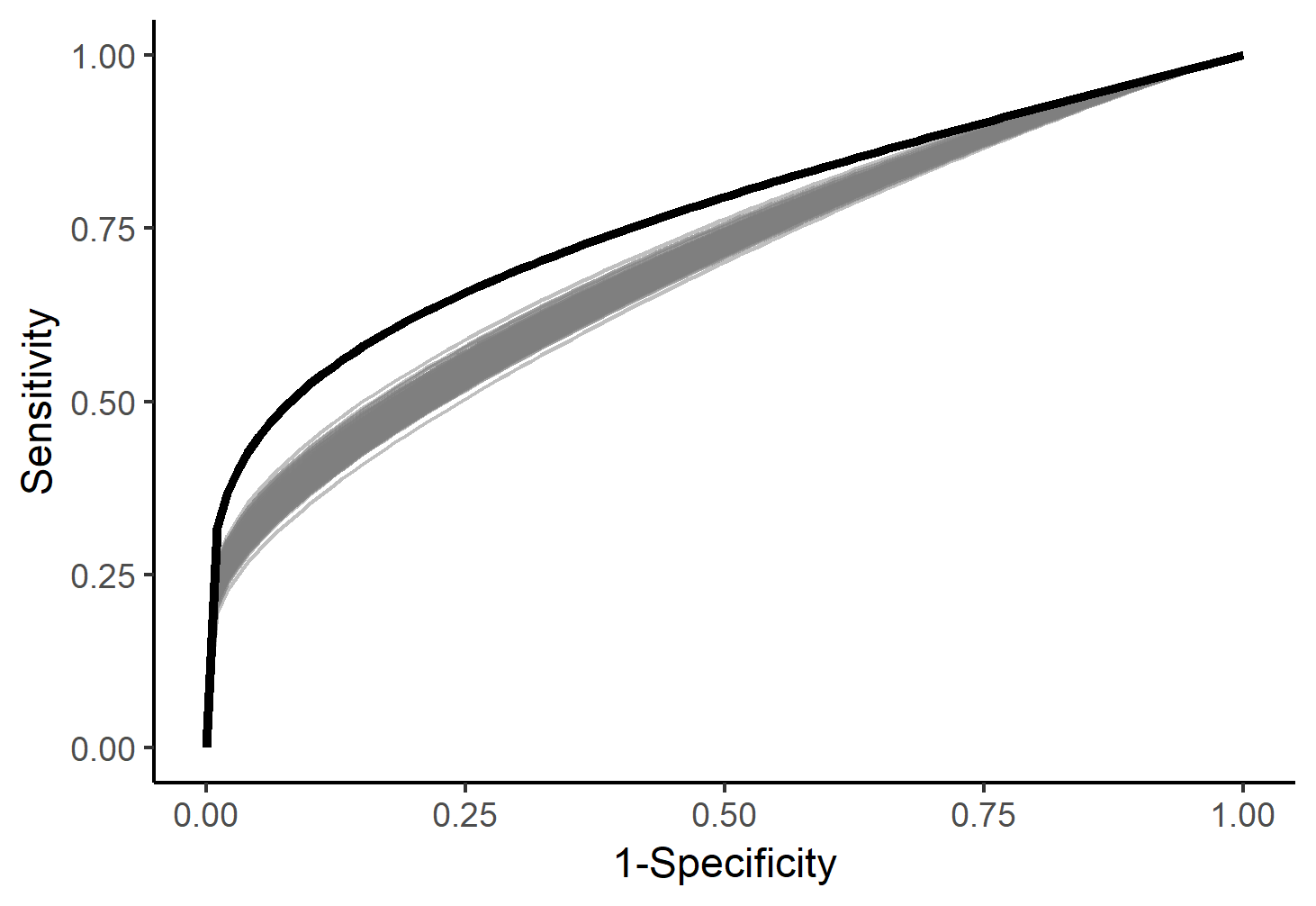}&
\includegraphics[height=.25\columnwidth,width=.20\linewidth]{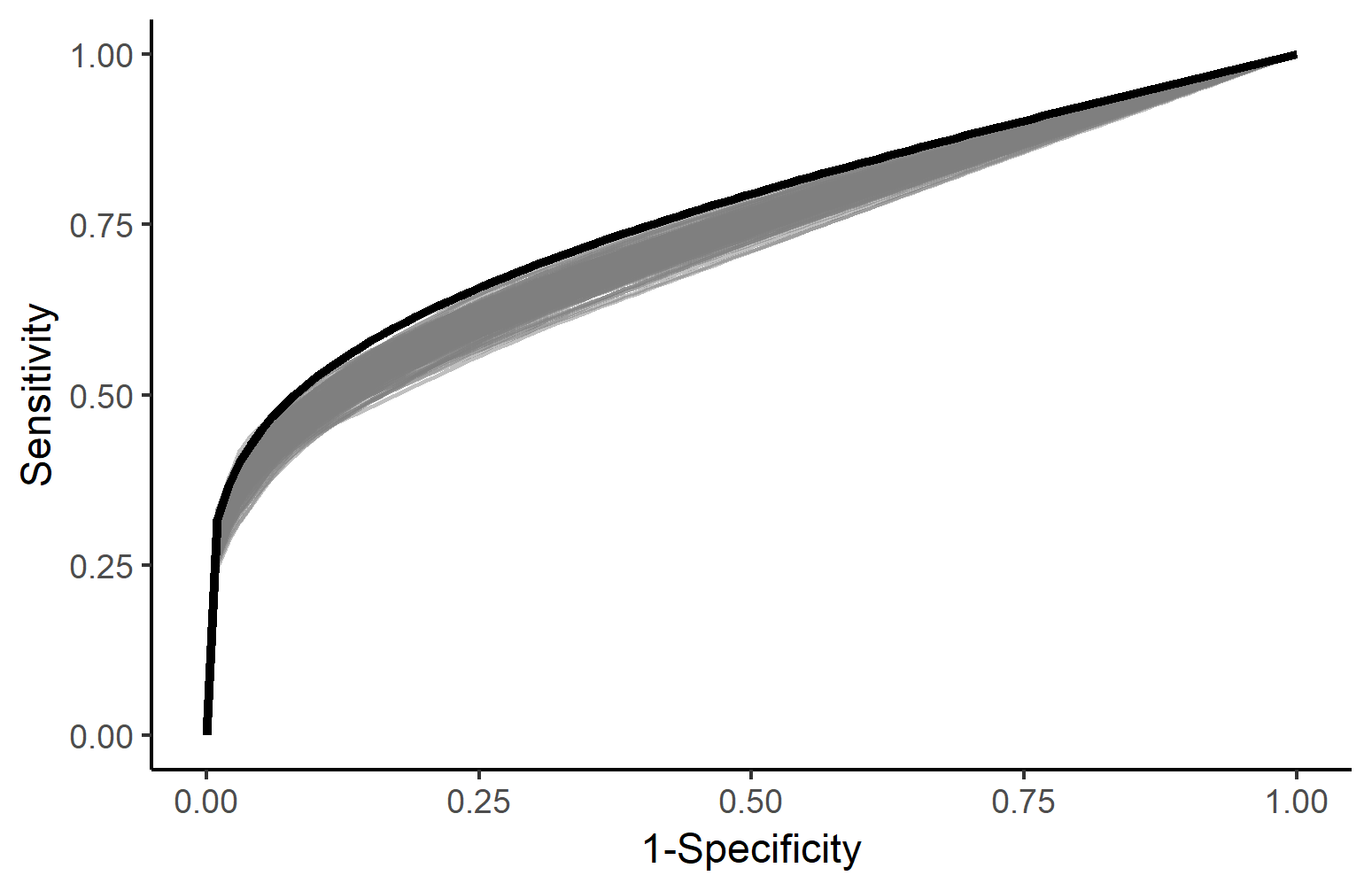}\\
%\includegraphics[height=.25\columnwidth,width=.20\linewidth]{NoCov_PBN_spCBN2_ROC_Medium.png}\\
%[-1ex]&\mycaption{0.5} & \mycaption{0.4} & \mycaption{0.6}\\
\rowname{High}&
\includegraphics[height=.25\columnwidth,width=.20\linewidth]{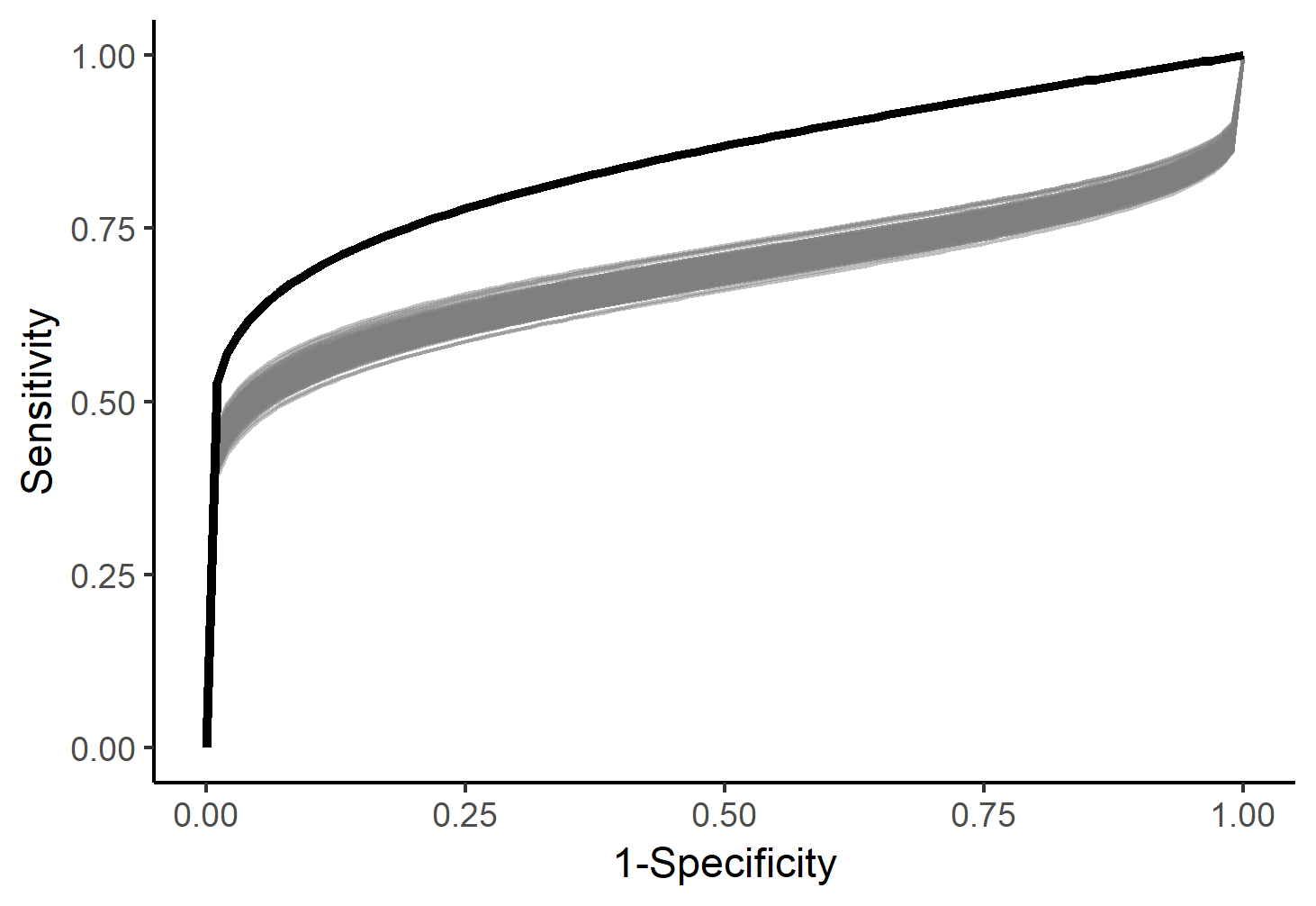}&
\includegraphics[height=.25\columnwidth,width=.20\linewidth]{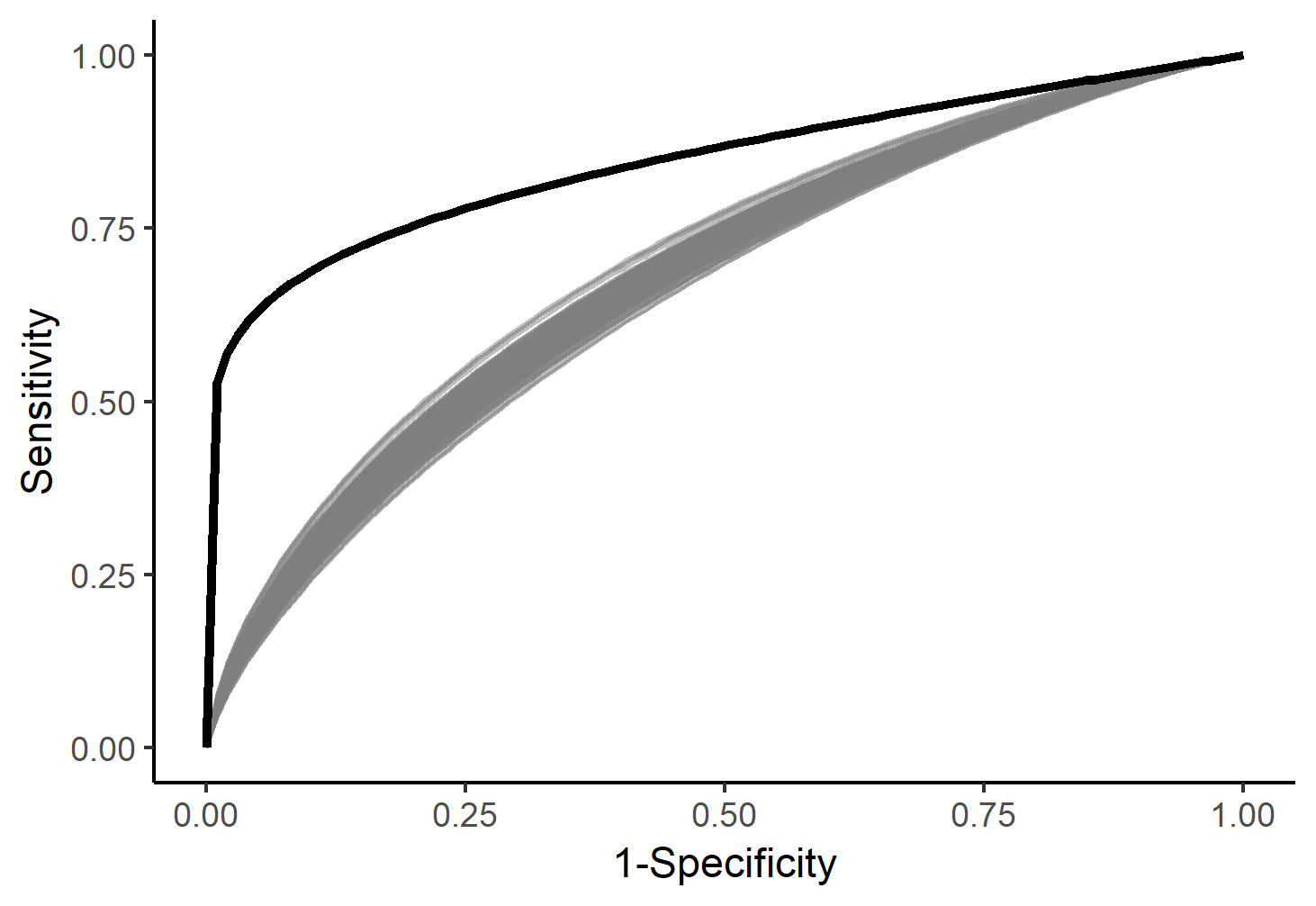}&
\includegraphics[height=.25\columnwidth,width=.20\linewidth]{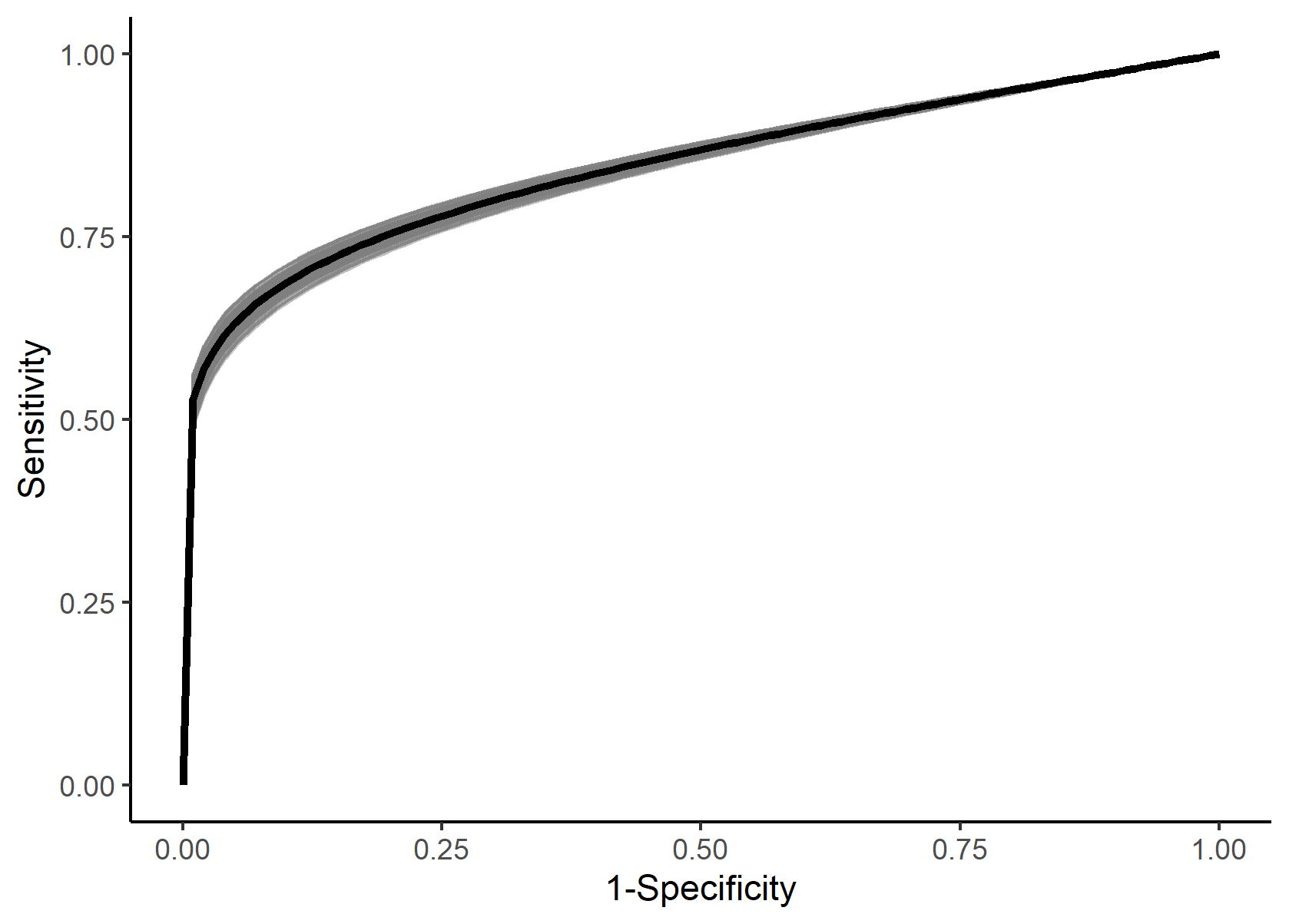}&
\includegraphics[height=.25\columnwidth,width=.20\linewidth]{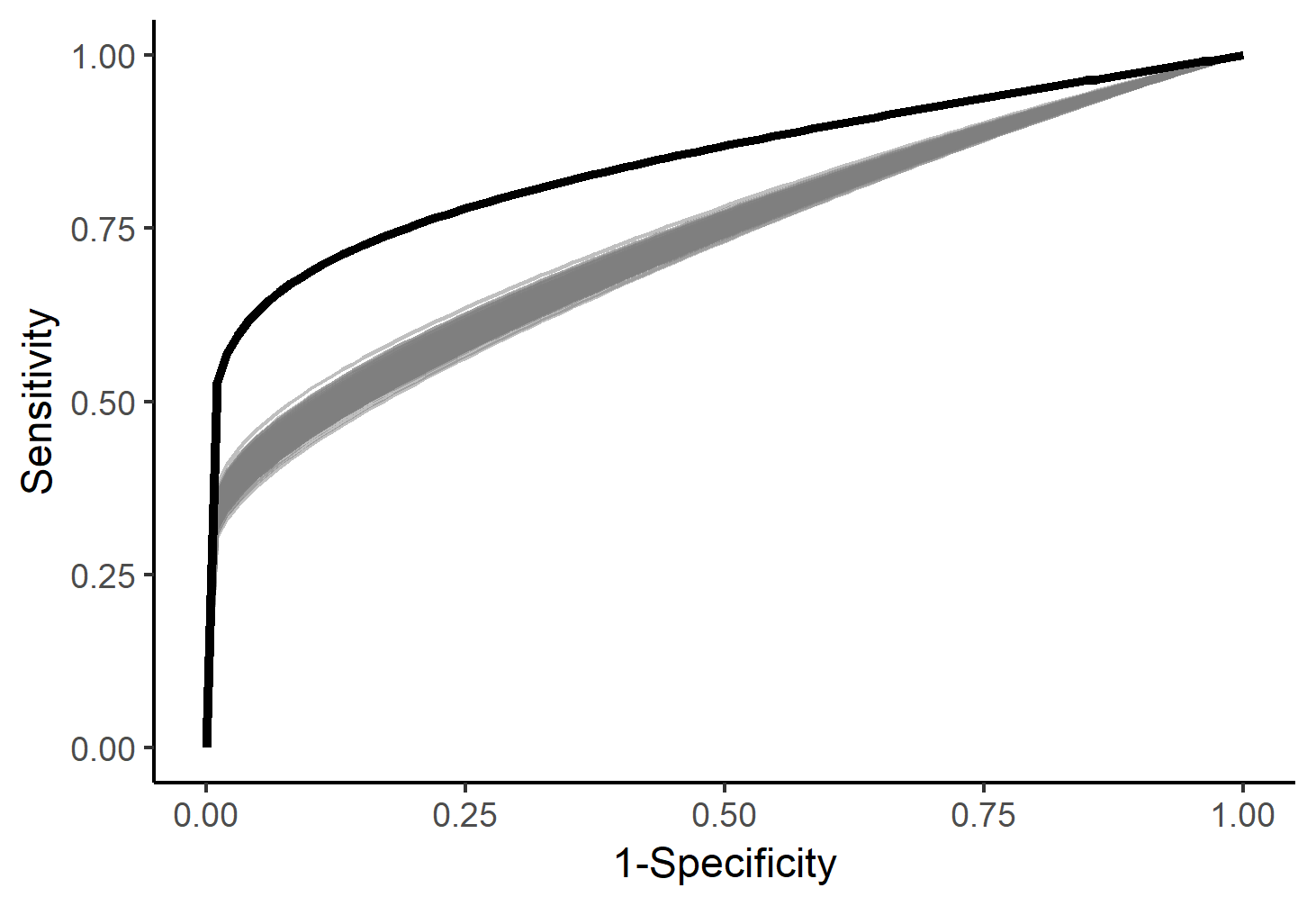}&
\includegraphics[height=.25\columnwidth,width=.20\linewidth]{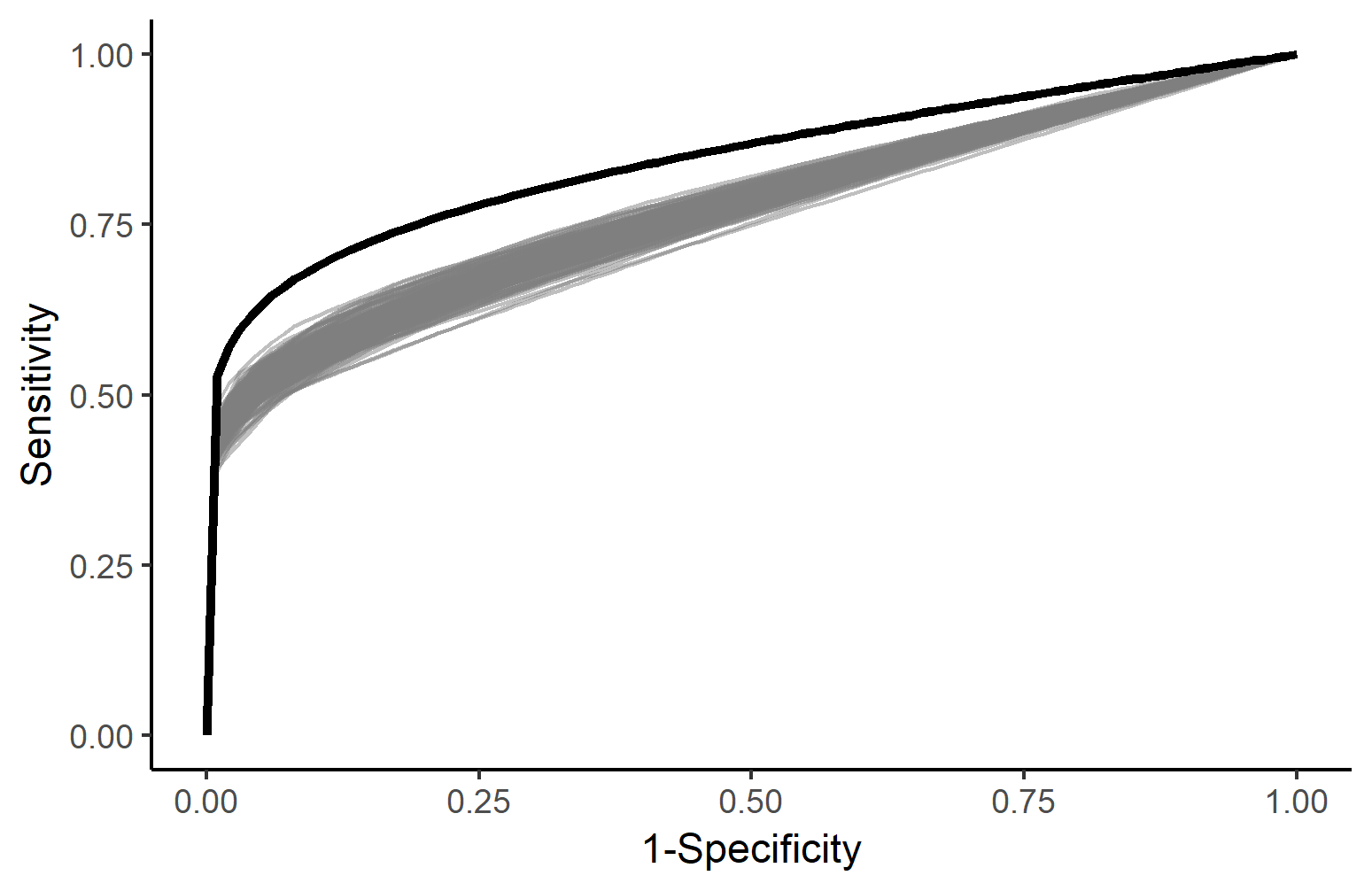}\\
%\includegraphics[height=.25\columnwidth,width=.20\linewidth]{NoCov_PBN_spCBN2_ROC_High.png}\\
%[-1ex]&\mycaption{0.5} & \mycaption{0.5} & \mycaption{0.7} \\
\end{tabular}
        \end{adjustbox}
\end{figure}

%%From Table \ref{tab:Sim_noCov}, it is evident that the performance of the proposed semiparametric method is consistent. More specifically, 
When the data generating mechanism is PBN, the EMSE is lowest for the PBN fitting model across the three levels of AUCs, an expected outcome since the model fitting model is correctly specified. The proposed semiparametric model spCN achieves the second lowest EMSE, followed by the parametric concave model pCN. BG produces highest EMSE in estimating ROCs and highest bias in estimating AUCs. BN model also produces high bias in AUC and non-concave ROCs as illustrated in the first column of Figure \ref{fig:Sim_ROC_noCov_PBN}. 

When data are generated from the BG model, the BG model itself is unbiased in estimating AUC and produces the lowest EMSEs in estimating ROC curves. Similar to the previous scenario, the semiparametric model spCN consistently has superior EMSEs compared to other models across varying levels of AUC. Similarly, the performance of the pCN model closely trailed that of the spCN model. However, in this scenario, the PBN model performs poorly both in terms of bias in estimating AUCs and EMSE in estimating ROCs. 

\begin{figure}[htbp]
\begin{adjustbox}{addcode={\begin{minipage}{\width}}{\caption{%
      ROC estimates for case when data is generated from BG model
      }\label{fig:Sim_ROC_noCov_BG}
      \end{minipage}},rotate=90,center}
\settoheight{\tempheight}{\includegraphics[width=.20\linewidth]{NoCov_BN_ROC_PBN_Low.png}}%
\centering\begin{tabular}{@{}c@{ }c@{ }c@{ }c@{ } c@{ } c@{ } }
&\textbf{BN} & \textbf{BG} & \textbf{PBN} & \textbf{pCN} & \textbf{spCN}\\
\rowname{Low}&
\includegraphics[height=.25\columnwidth,width=.20\linewidth]{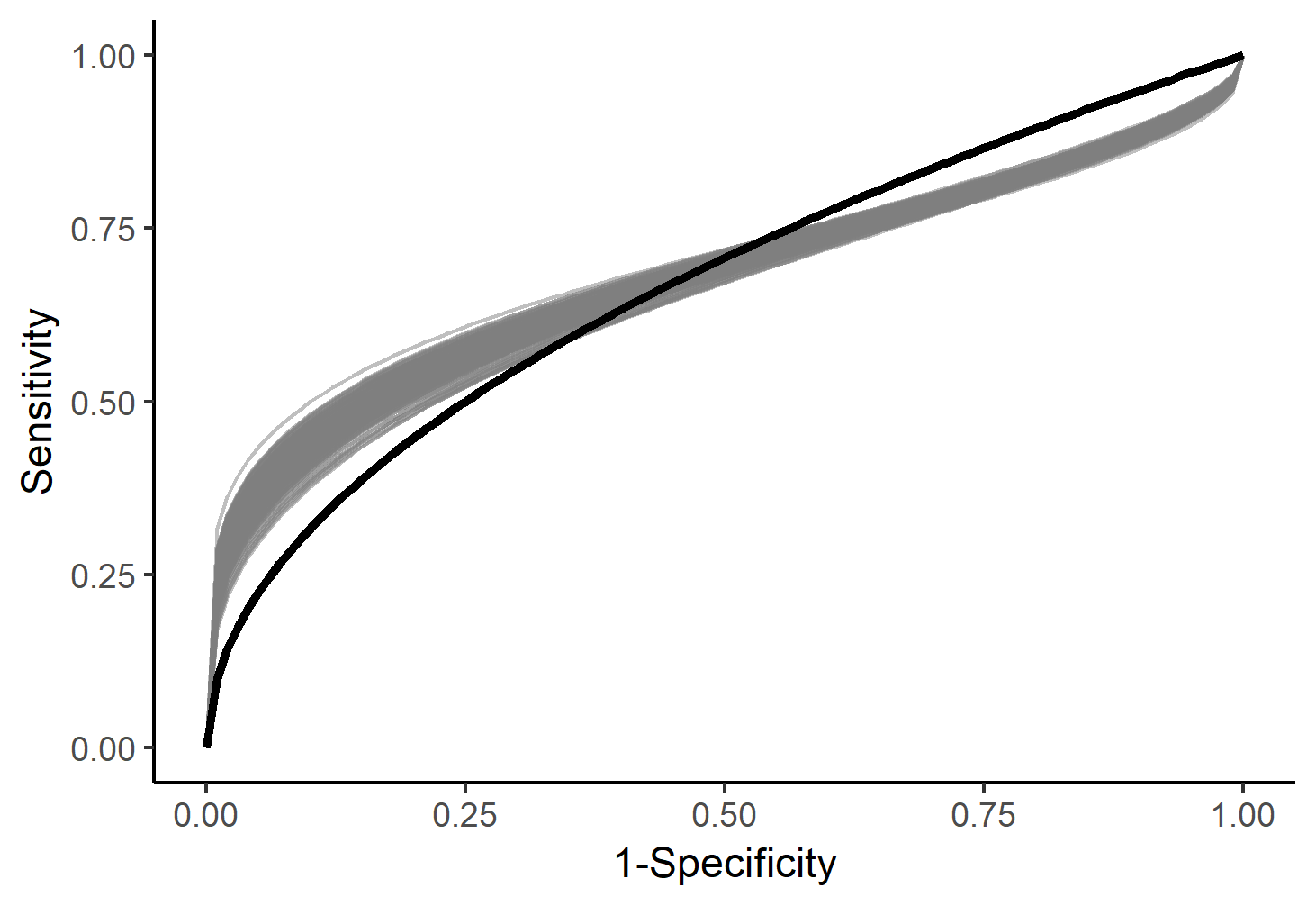}&
\includegraphics[height=.25\columnwidth,width=.20\linewidth]{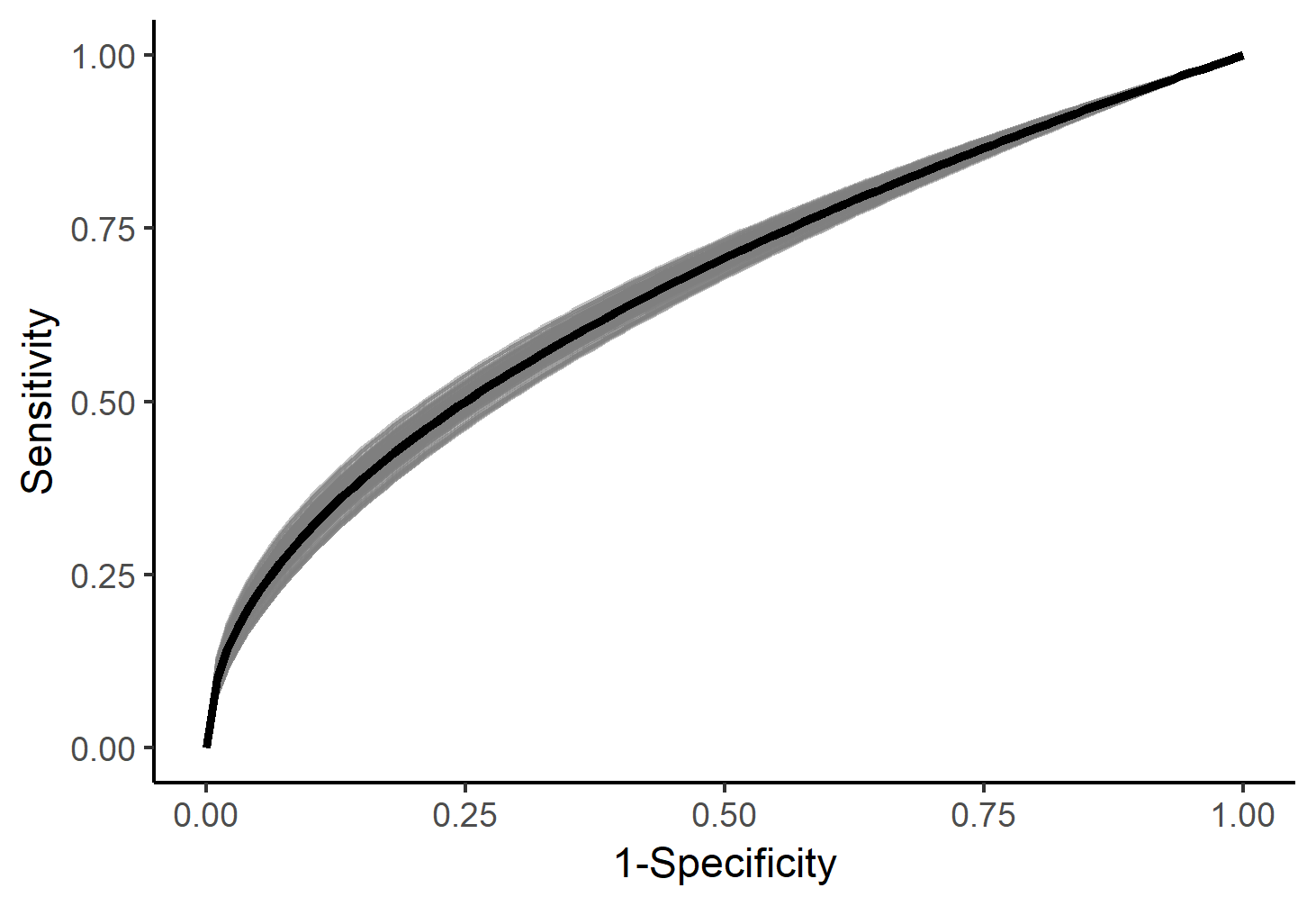}&
\includegraphics[height=.25\columnwidth,width=.20\linewidth]{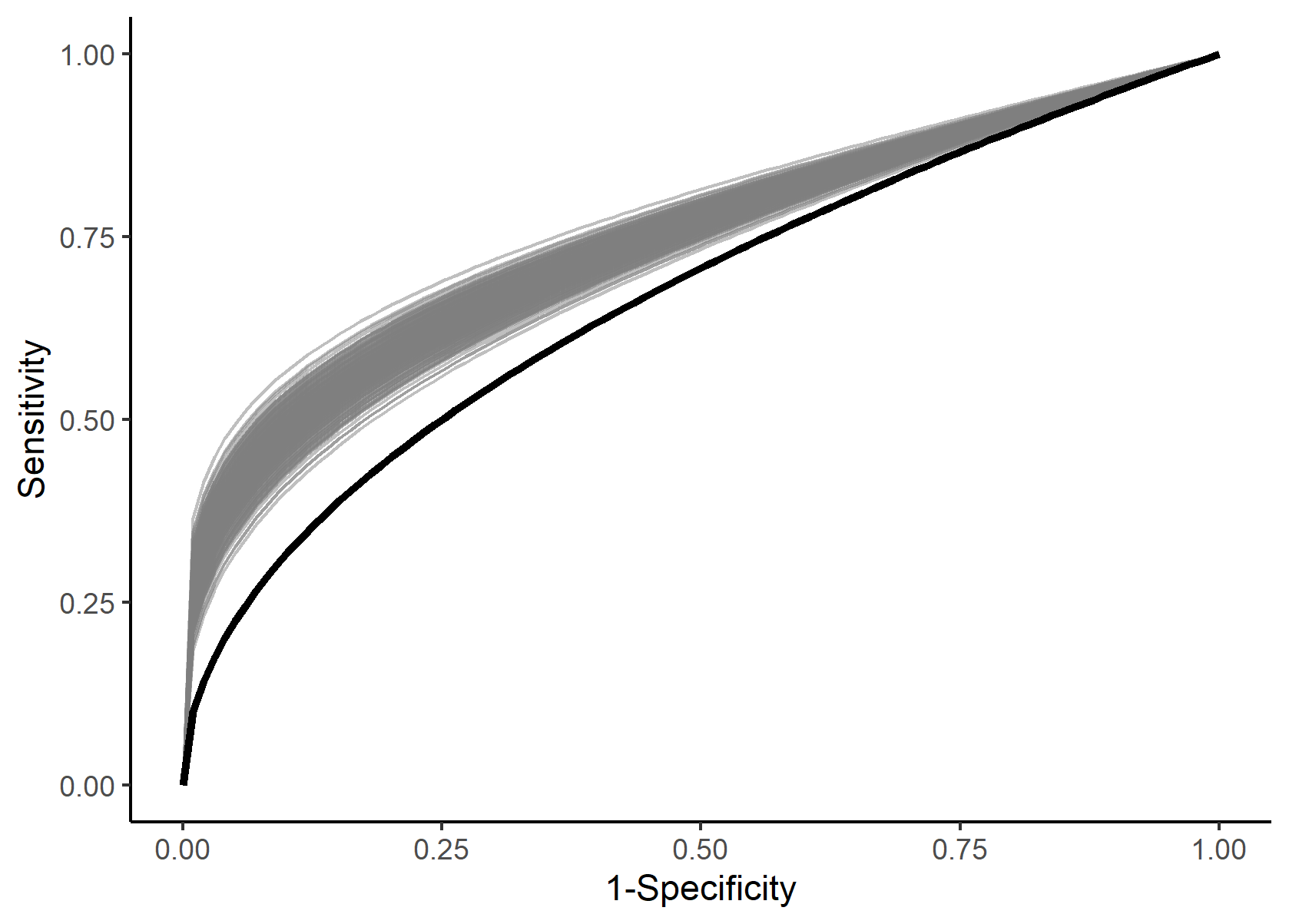}&
\includegraphics[height=.25\columnwidth,width=.20\linewidth]{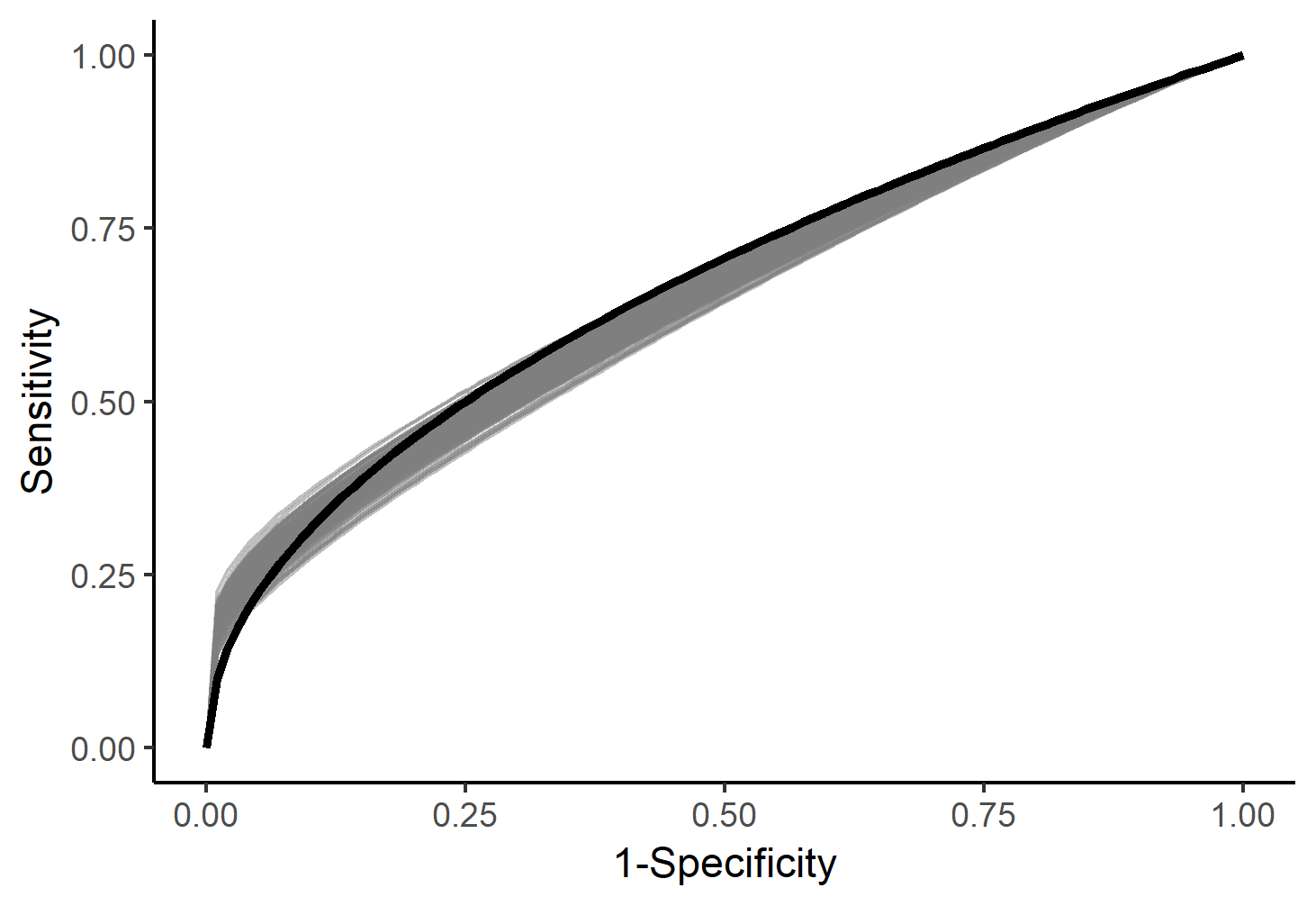}&
\includegraphics[height=.25\columnwidth,width=.20\linewidth]{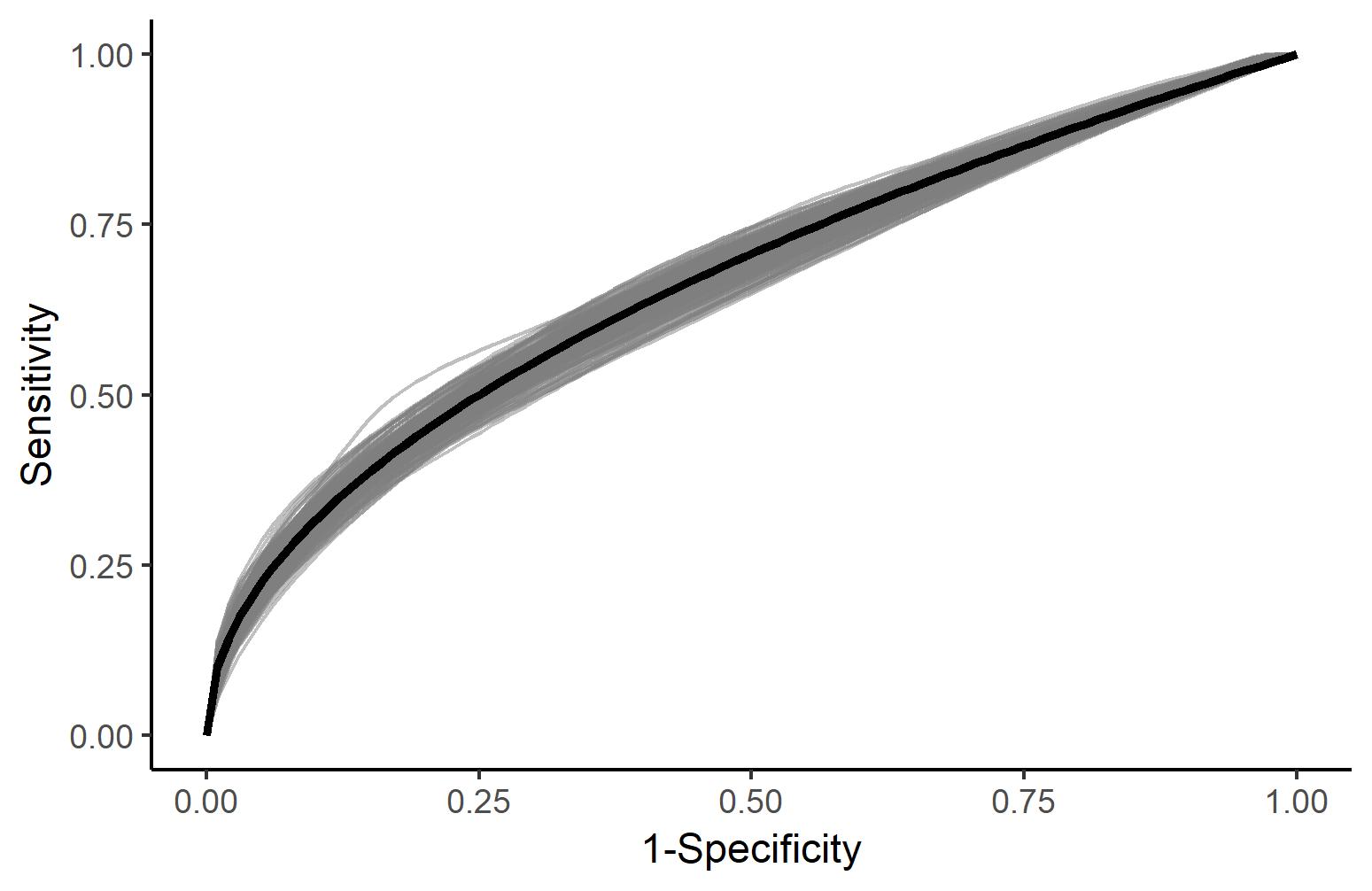}\\
%\includegraphics[height=.25\columnwidth,width=.20\linewidth]{NoCov_BG_spCBN2_ROC_Low.png} \\
%[-1ex] &\mycaption{0.2} & \mycaption{0.2} & \mycaption{0.3}\\
\rowname{Medium}&
\includegraphics[height=.25\columnwidth,width=.20\linewidth]{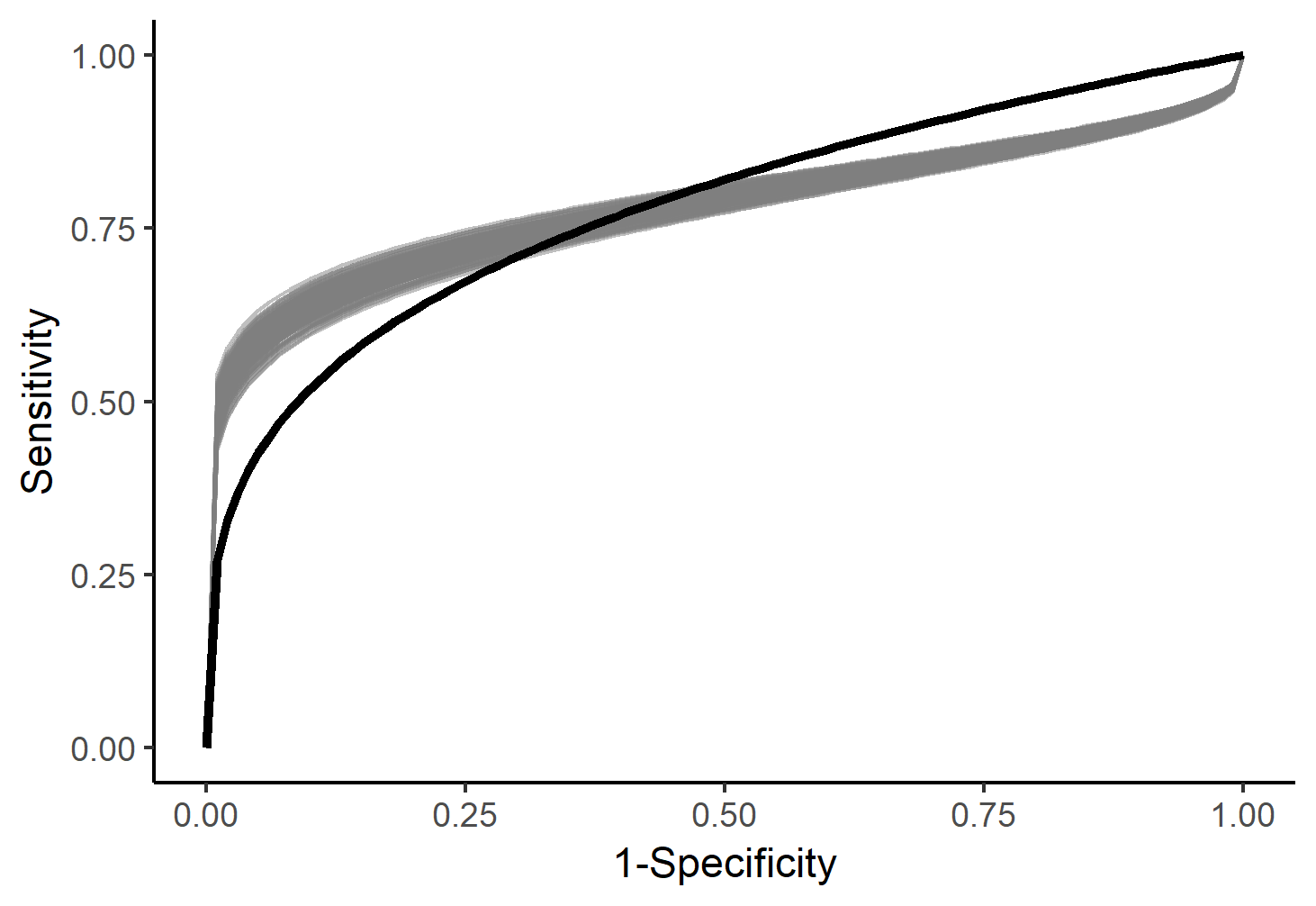}&
\includegraphics[height=.25\columnwidth,width=.20\linewidth]{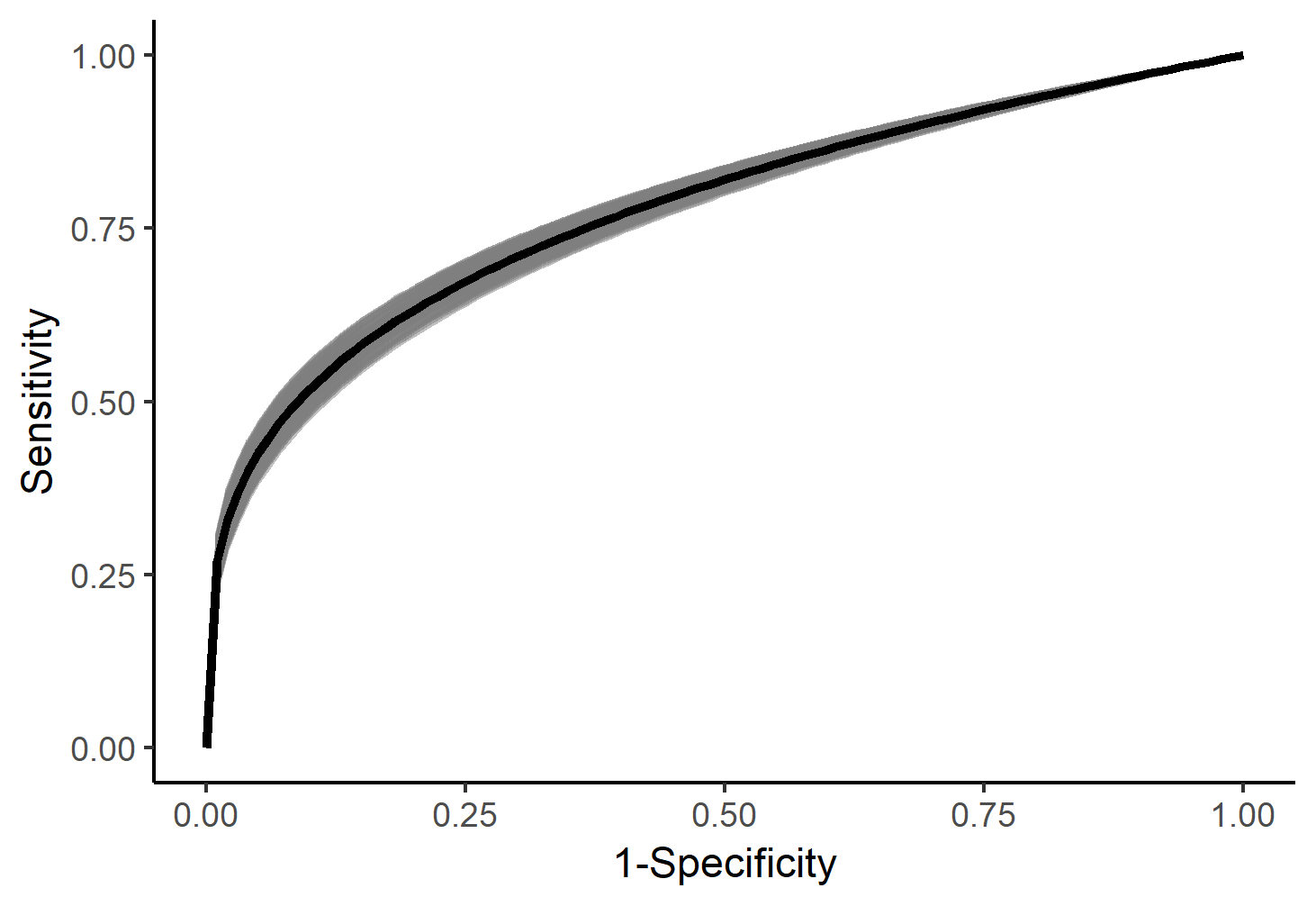}&
\includegraphics[height=.25\columnwidth,width=.20\linewidth]{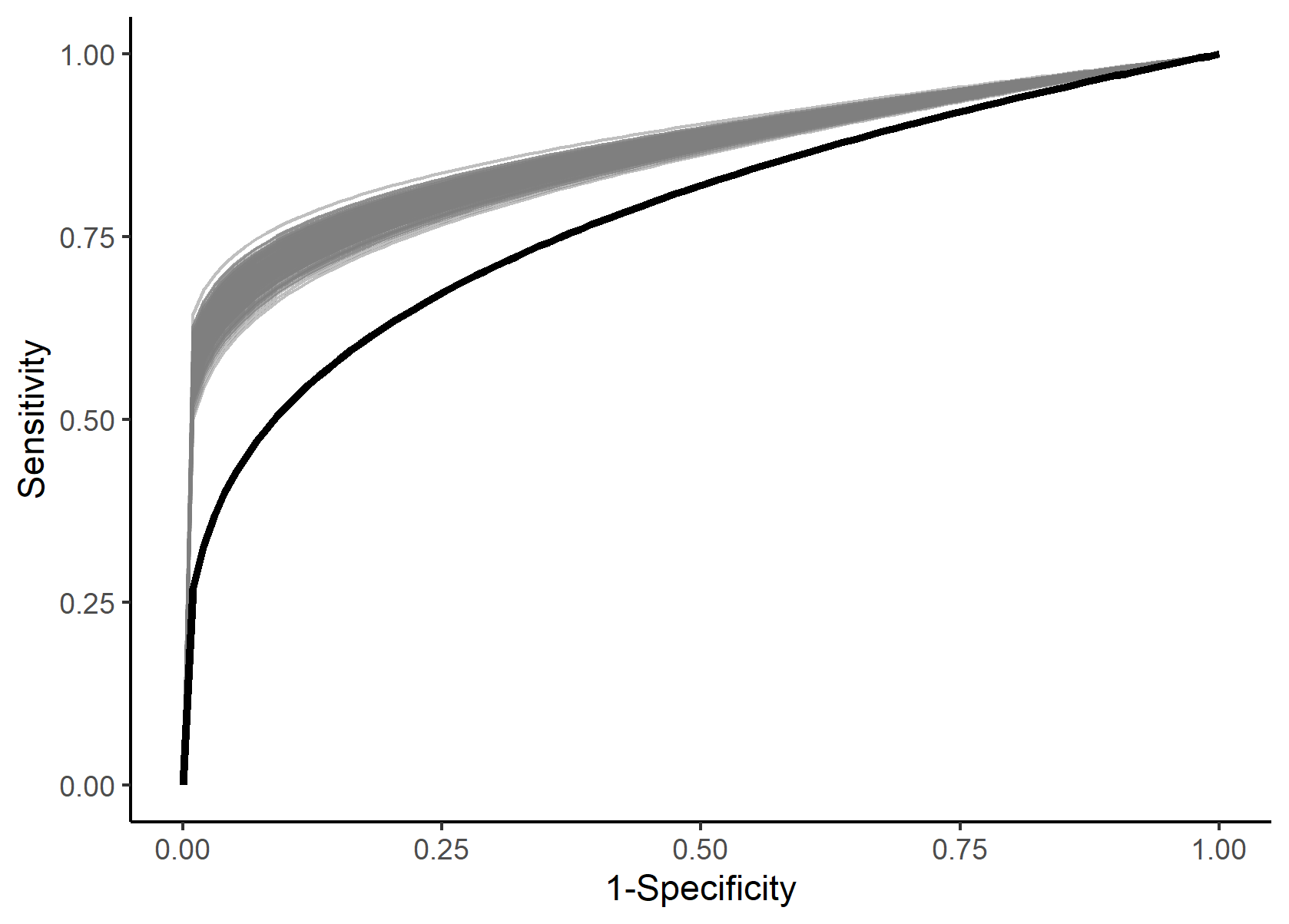}&
\includegraphics[height=.25\columnwidth,width=.20\linewidth]{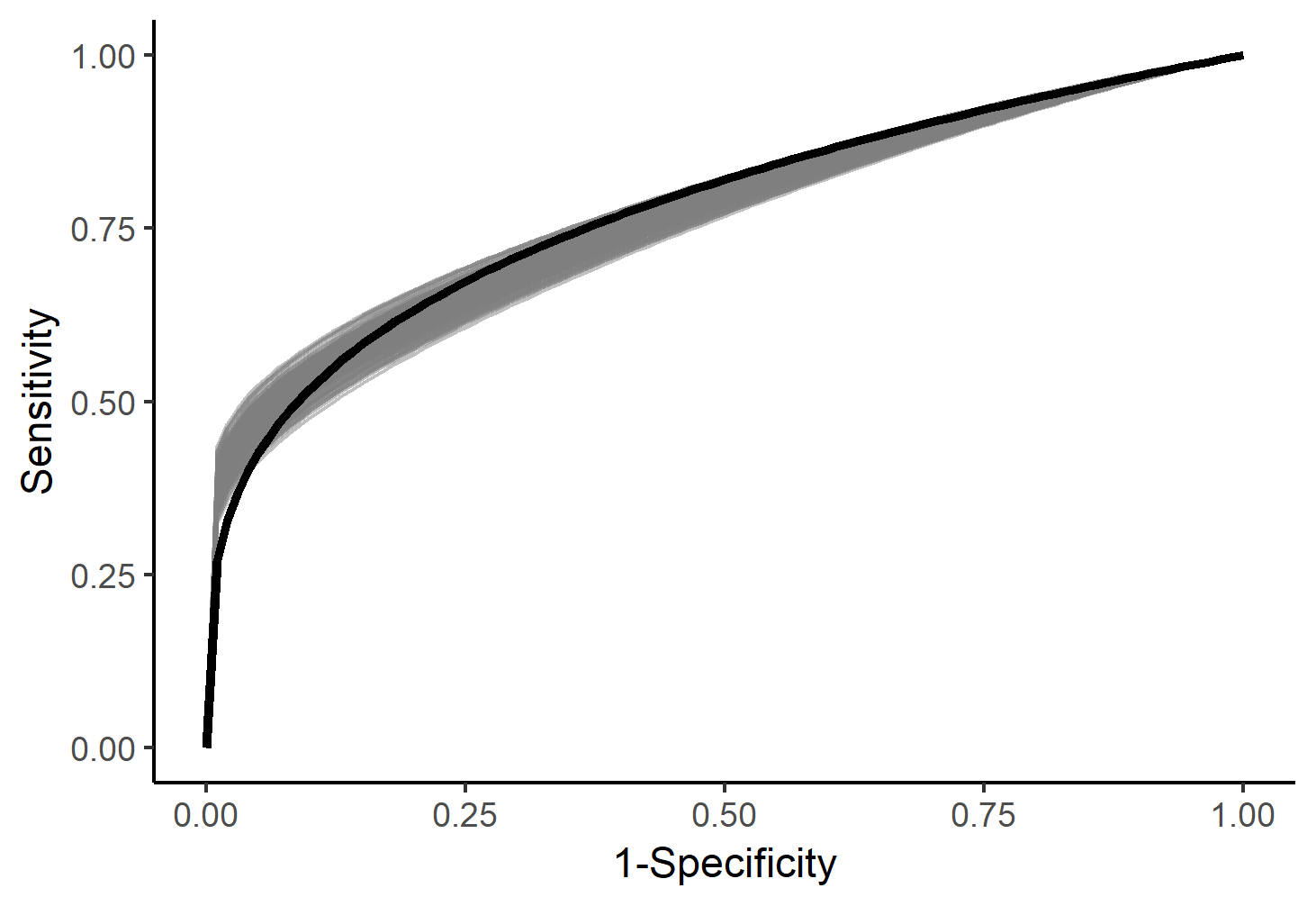}&
\includegraphics[height=.25\columnwidth,width=.20\linewidth]{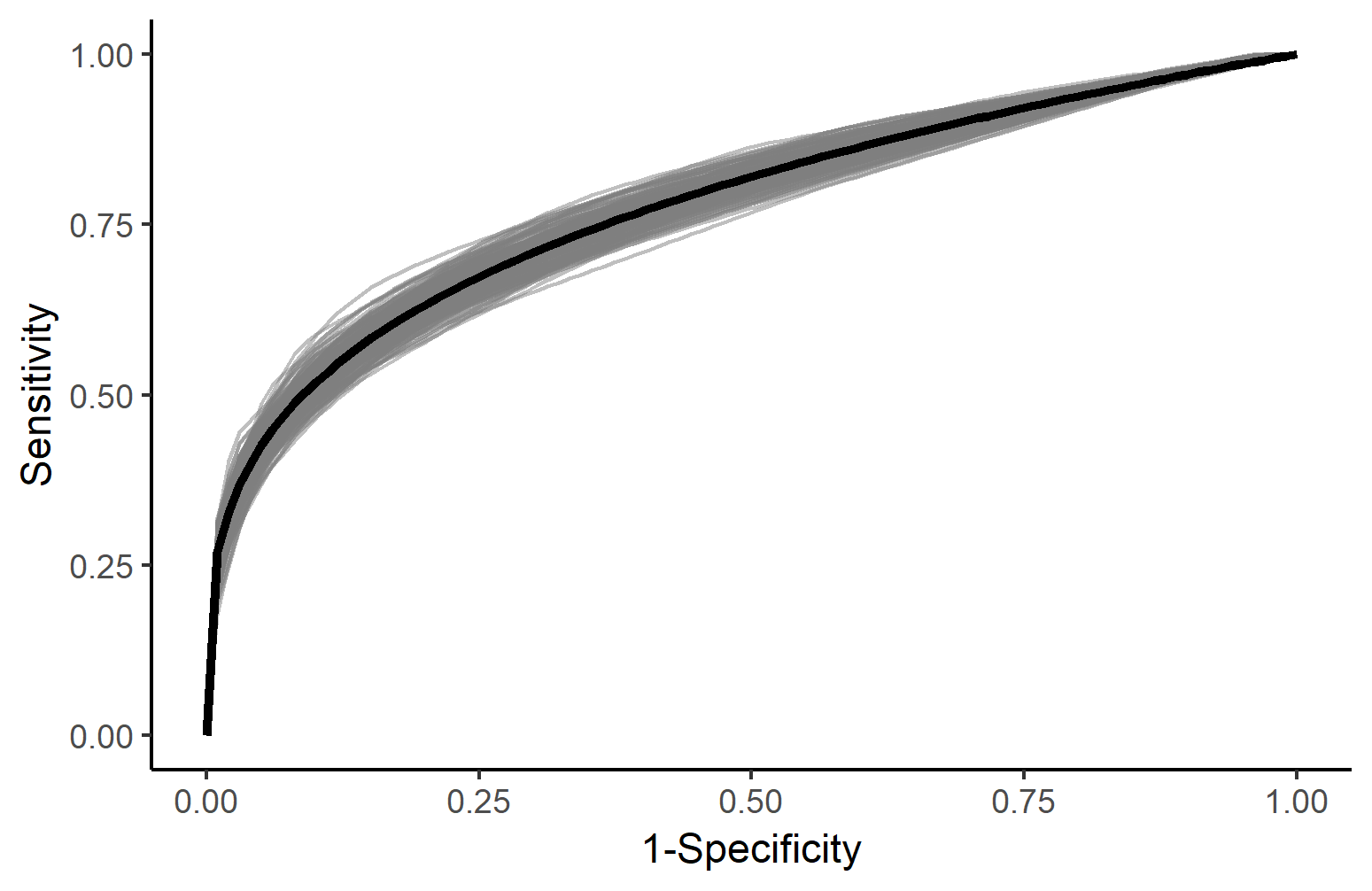}\\
%\includegraphics[height=.25\columnwidth,width=.20\linewidth]{NoCov_BG_spCBN2_ROC_Medium.png}\\
%[-1ex]&\mycaption{0.5} & \mycaption{0.4} & \mycaption{0.6}\\
\rowname{High}&
\includegraphics[height=.25\columnwidth,width=.20\linewidth]{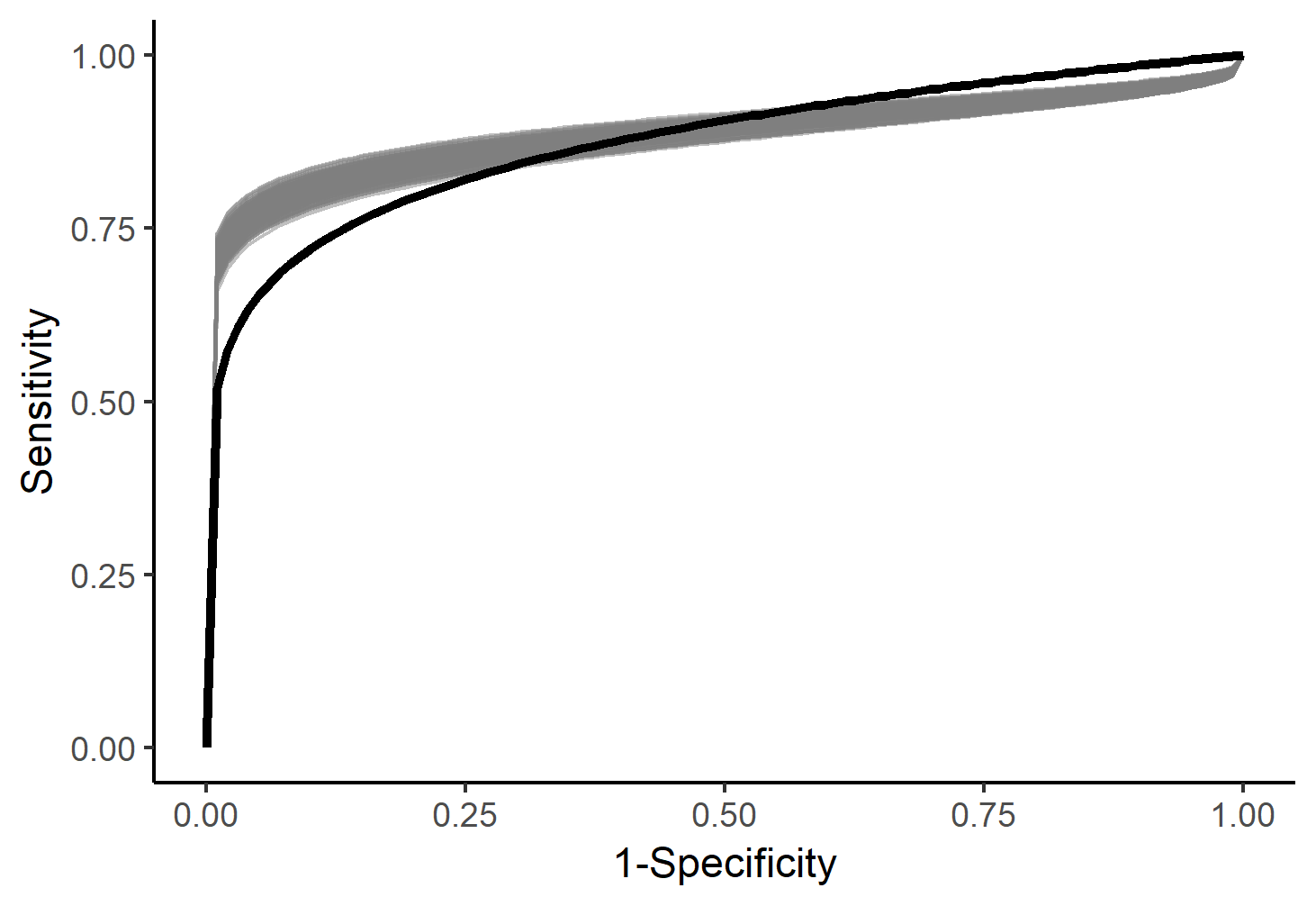}&
\includegraphics[height=.25\columnwidth,width=.20\linewidth]{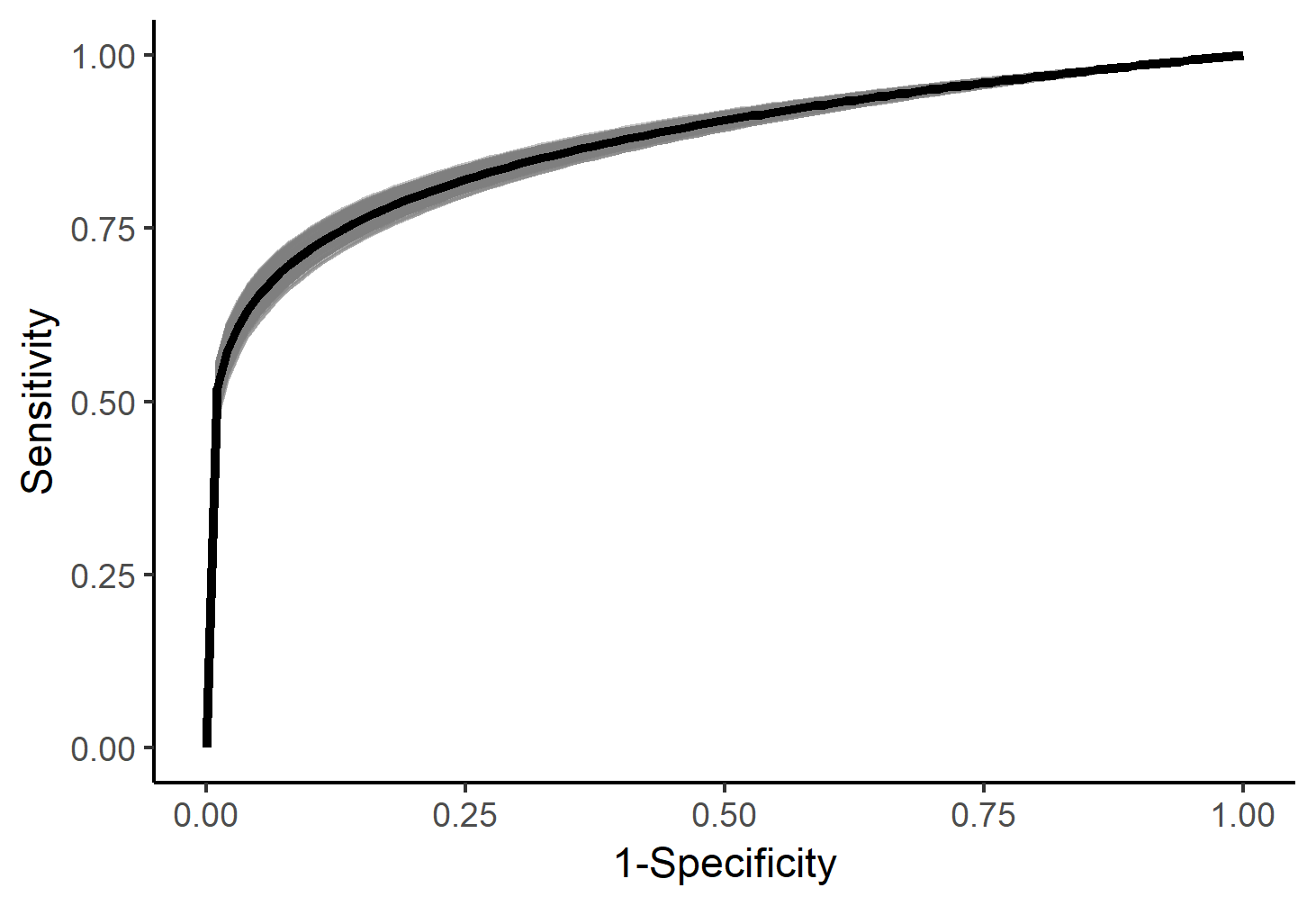}&
\includegraphics[height=.25\columnwidth,width=.20\linewidth]{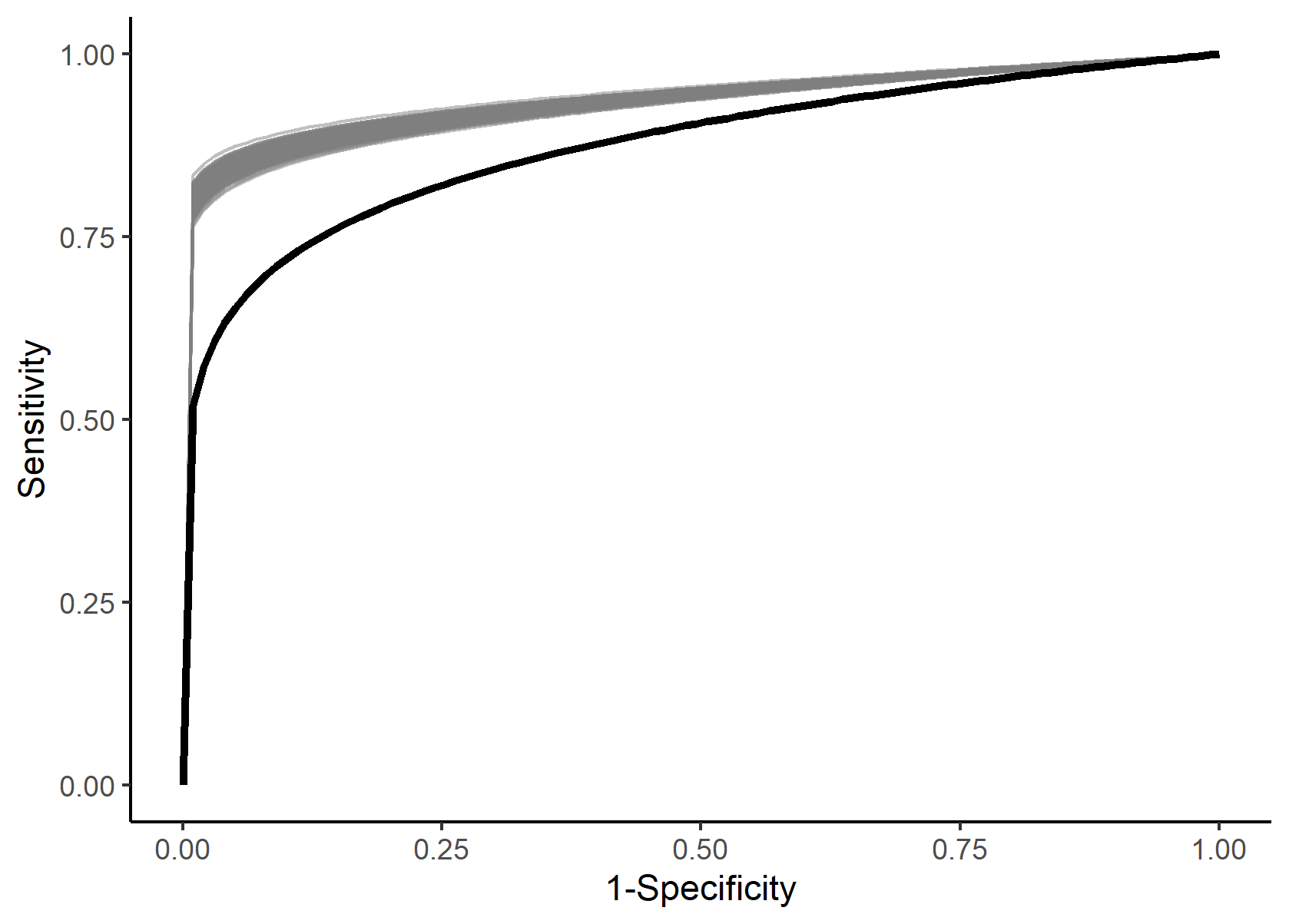}&
\includegraphics[height=.25\columnwidth,width=.20\linewidth]{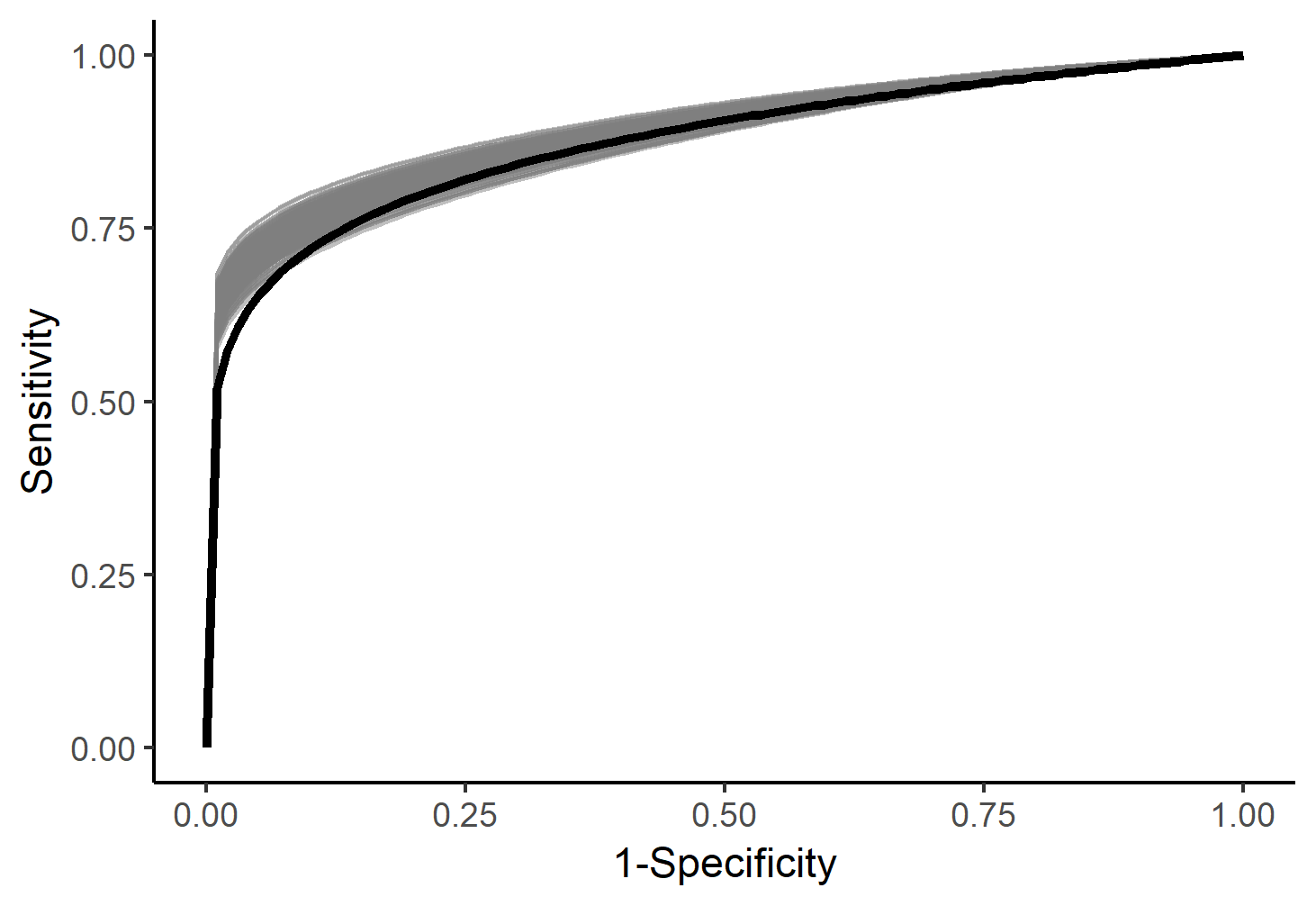}&
\includegraphics[height=.25\columnwidth,width=.20\linewidth]{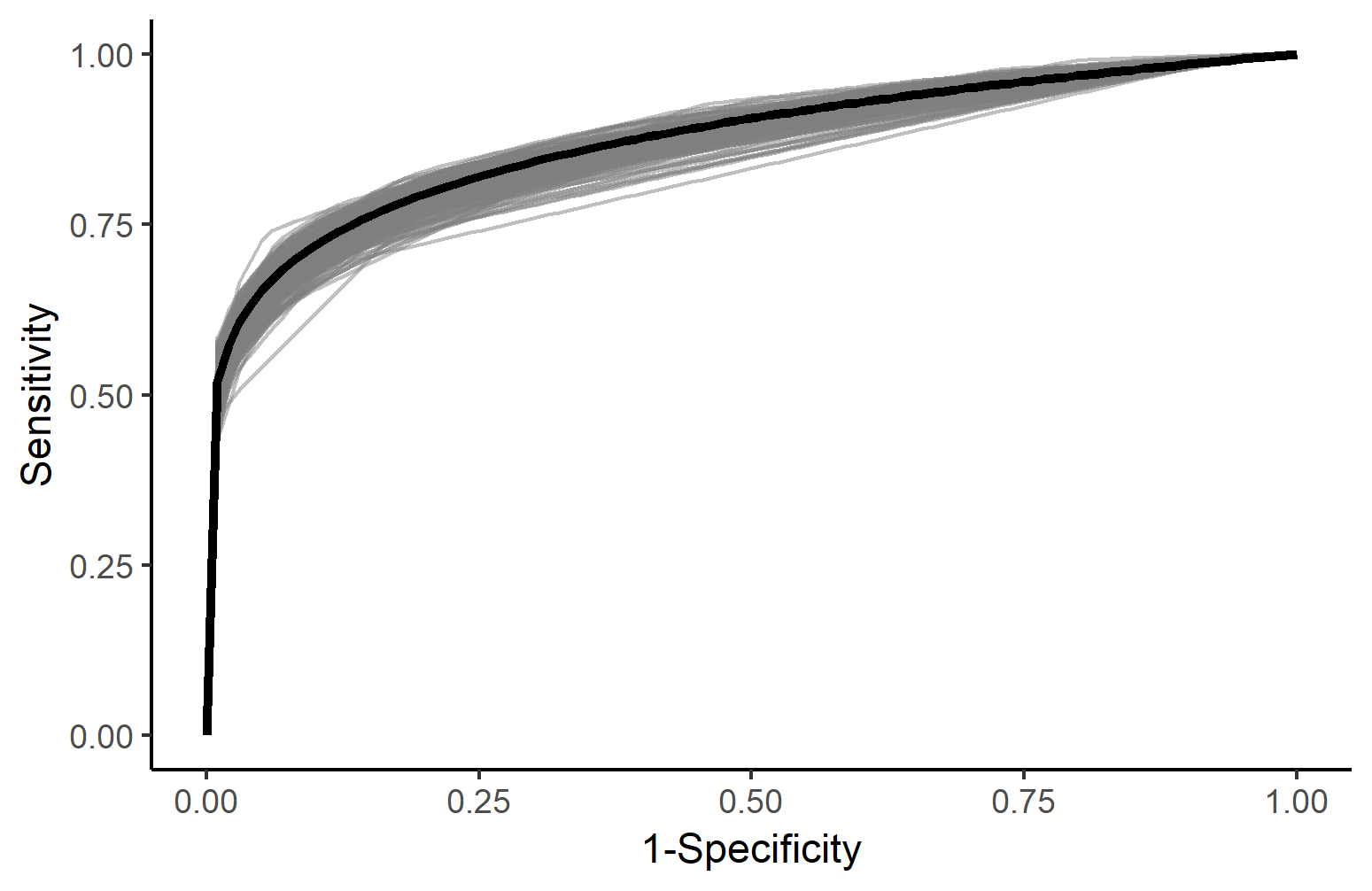}\\
%\includegraphics[height=.25\columnwidth,width=.20\linewidth]{NoCov_BG_spCBN2_ROC_High.png}\\
%[-1ex]&\mycaption{0.5} & \mycaption{0.5} & \mycaption{0.7} \\
\end{tabular}
        \end{adjustbox}
\end{figure}

\begin{figure}[htbp]
\begin{adjustbox}{addcode={\begin{minipage}{\width}}{\caption{%
      ROC estimates for case when data is generated from pCN model
      }\label{fig:Sim_ROC_noCov_pCN}
      \end{minipage}},rotate=90,center}
\settoheight{\tempheight}{\includegraphics[width=.20\linewidth]{NoCov_BN_ROC_PBN_Low.png}}%
\centering\begin{tabular}{@{}c@{ }c@{ }c@{ }c@{ } c@{ } c@{ } }
&\textbf{BN} & \textbf{BG} & \textbf{PBN} & \textbf{pCN} & \textbf{spCN} \\
\rowname{Low}&
\includegraphics[height=.25\columnwidth,width=.20\linewidth]{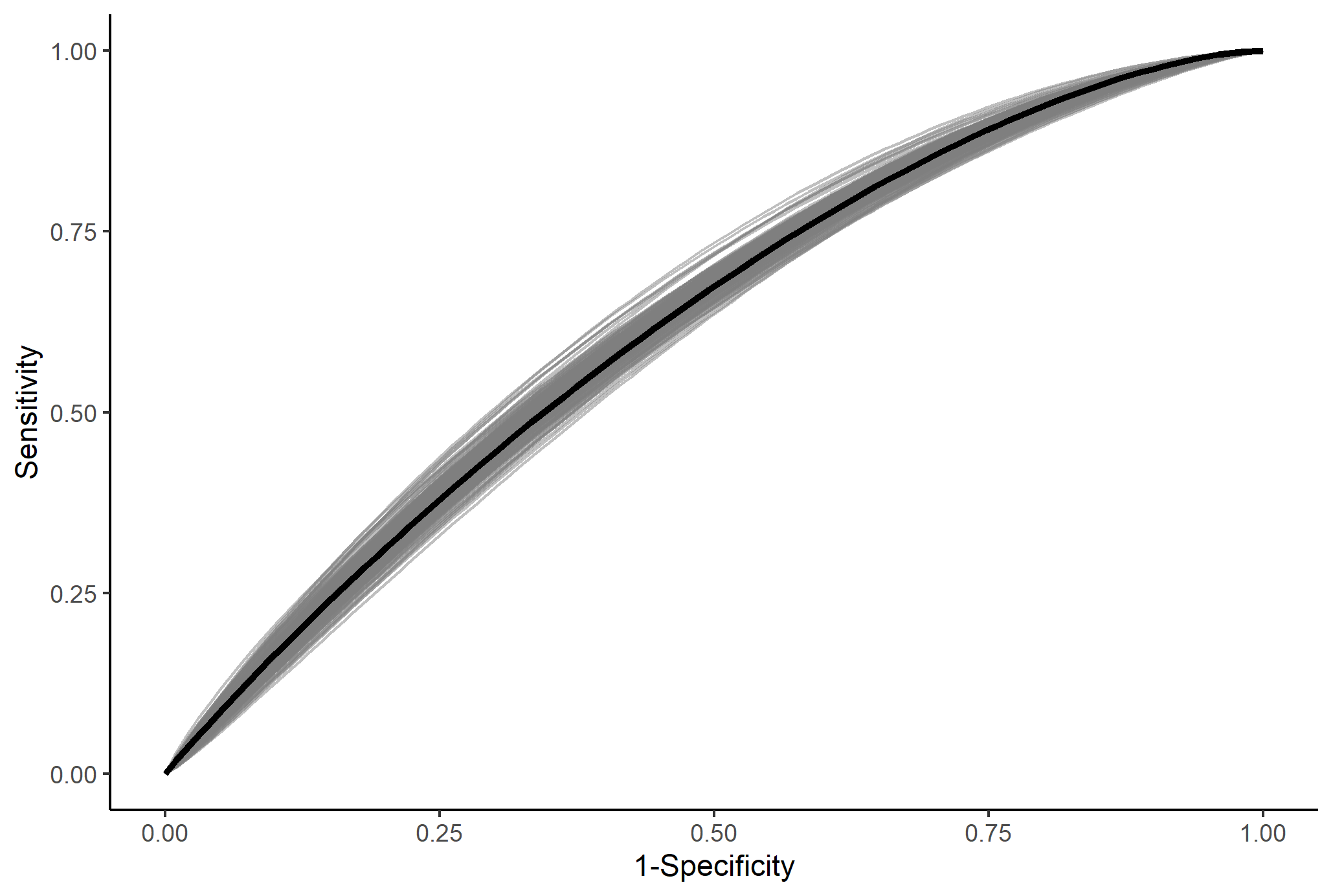}&
\includegraphics[height=.25\columnwidth,width=.20\linewidth]{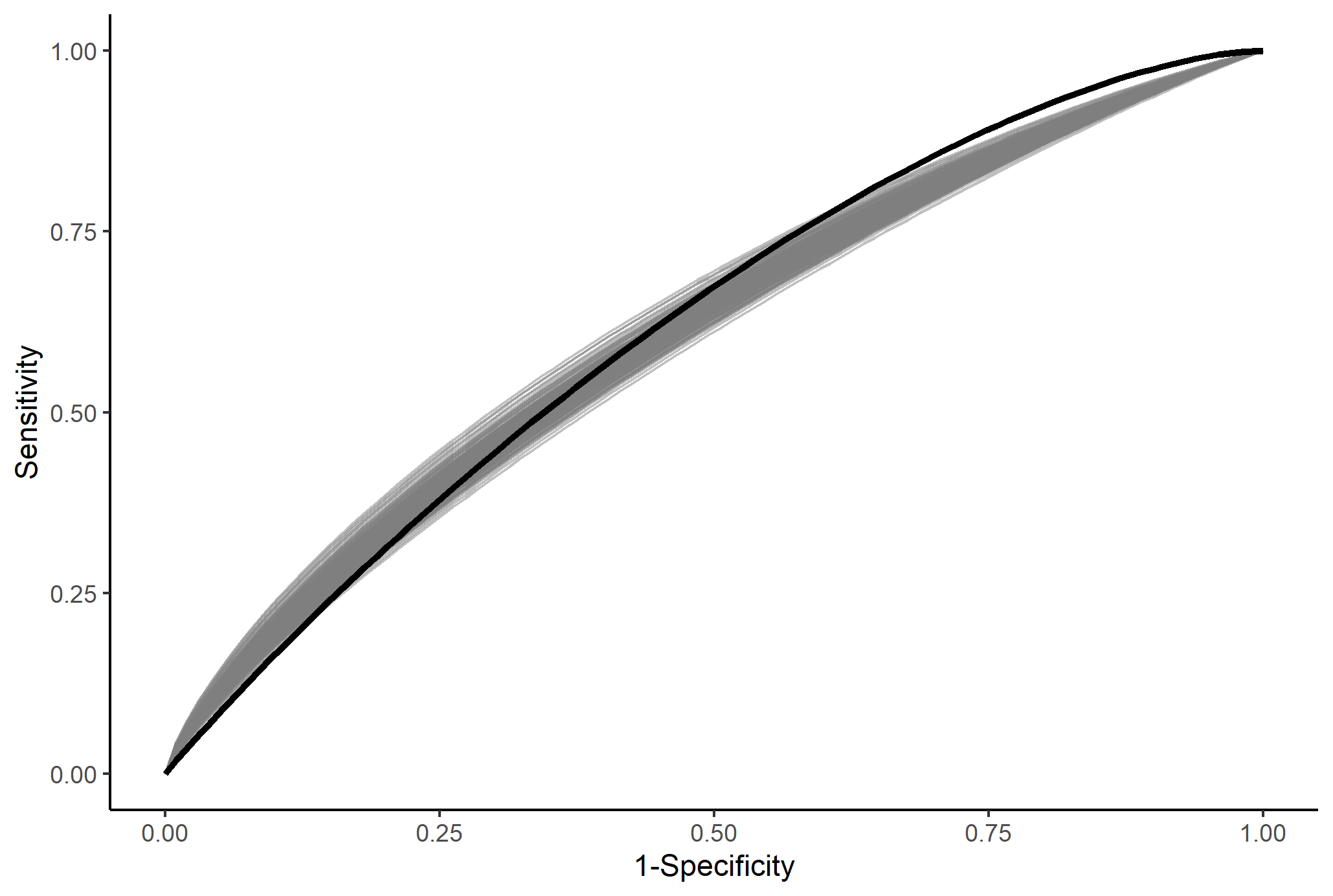}&
\includegraphics[height=.25\columnwidth,width=.20\linewidth]{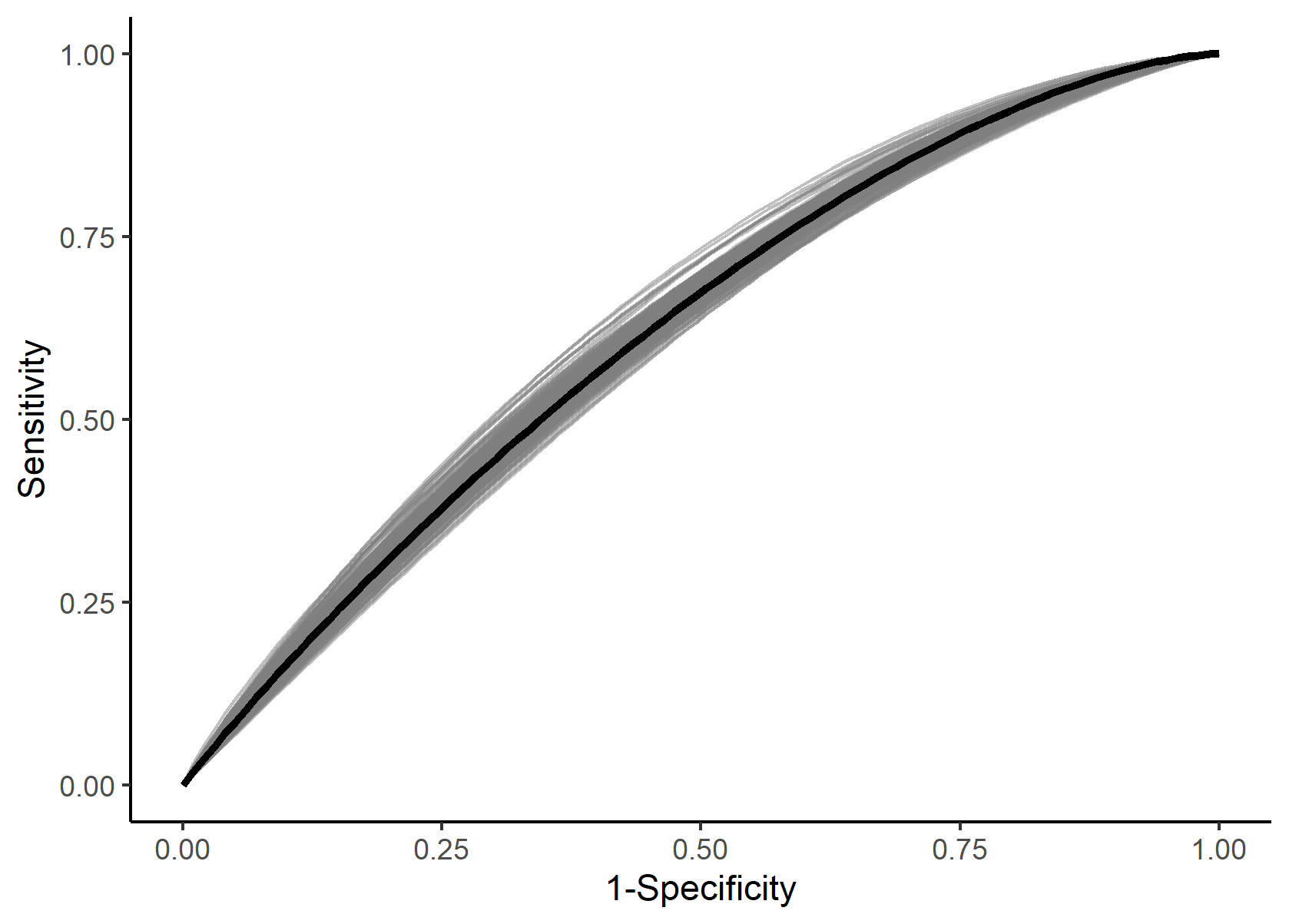}&
\includegraphics[height=.25\columnwidth,width=.20\linewidth]{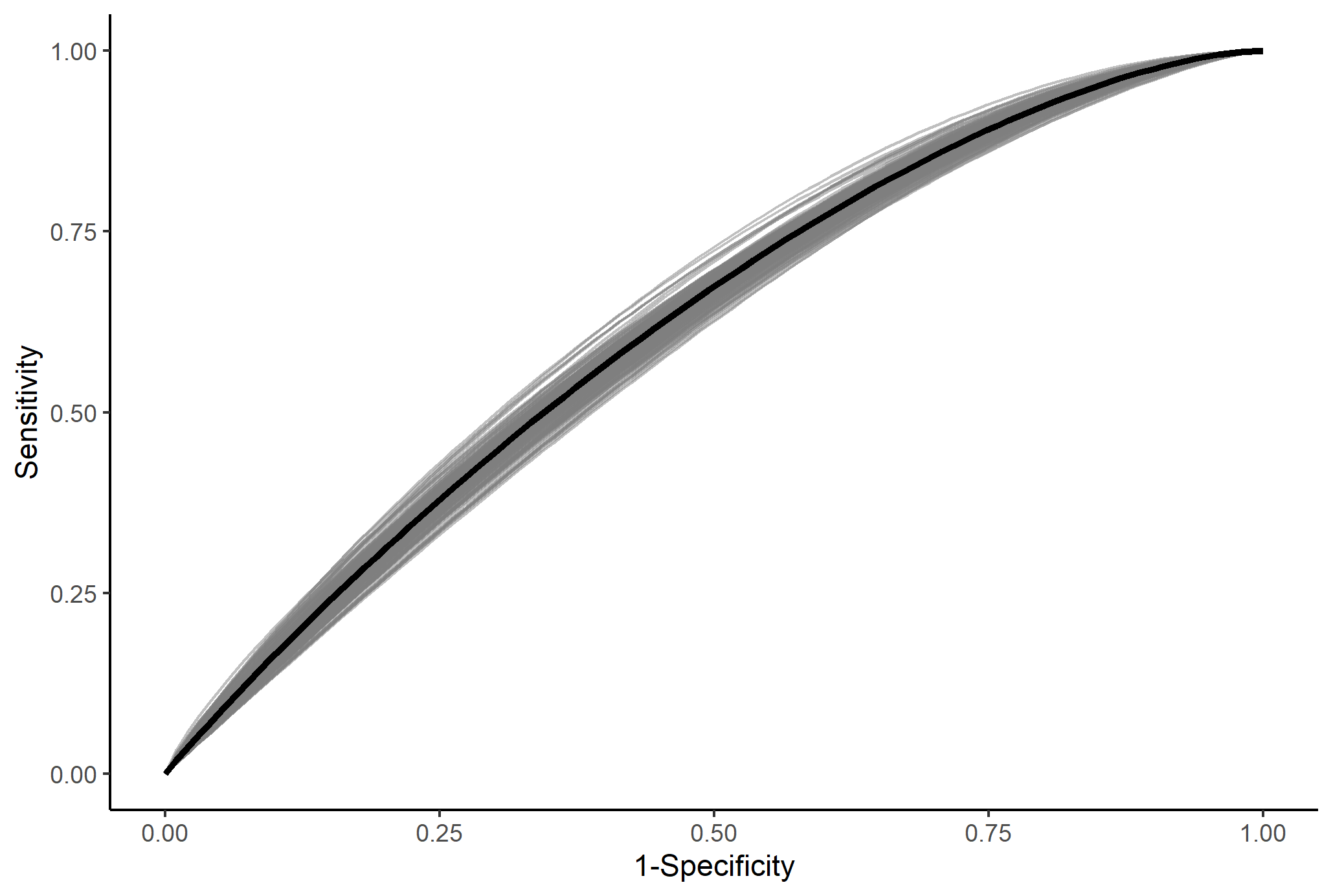}&
\includegraphics[height=.25\columnwidth,width=.20\linewidth]{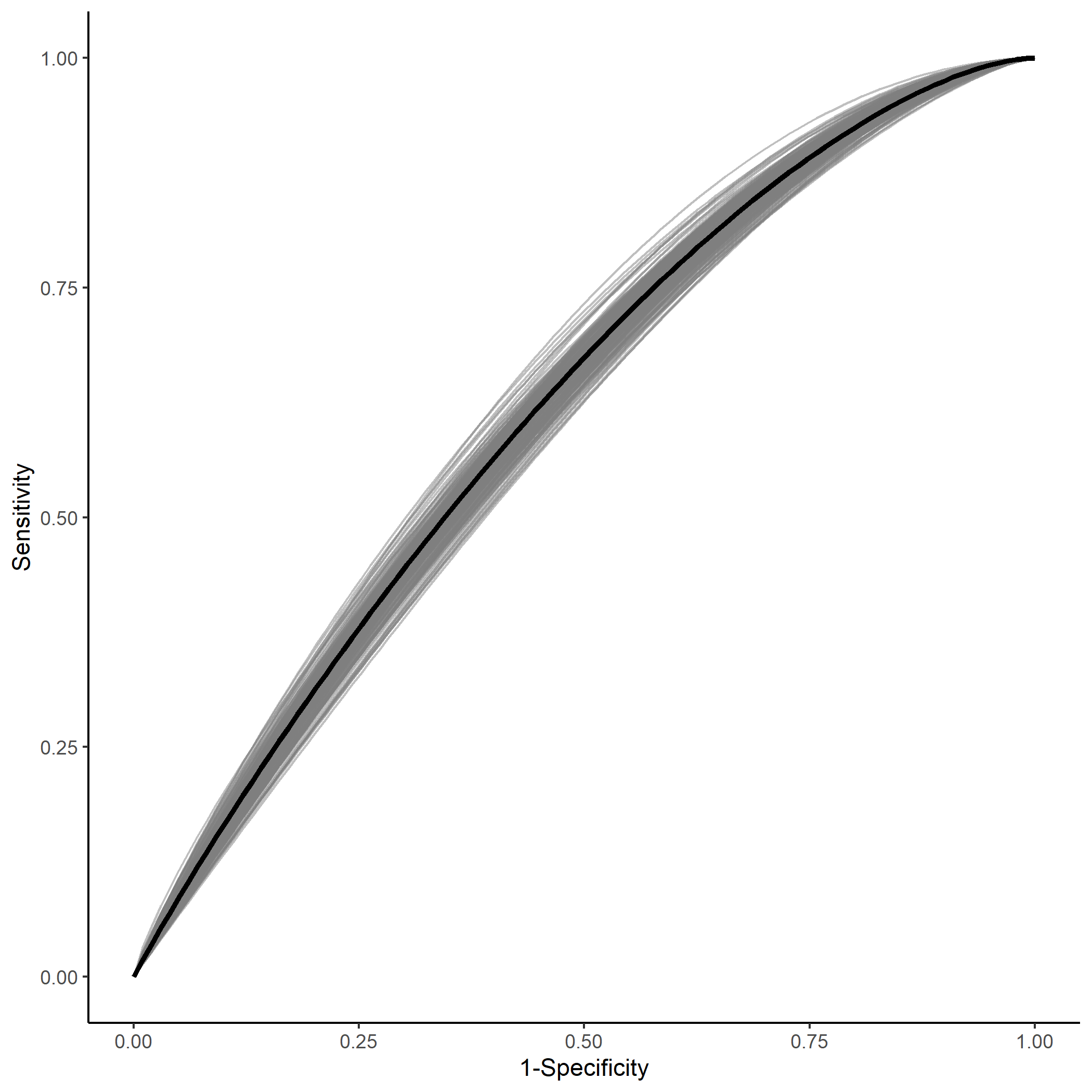}\\
%\includegraphics[height=.25\columnwidth,width=.20\linewidth]{NoCov_ParCon_spCBN2_ROC_Low.png}\\
%[-1ex] &\mycaption{0.2} & \mycaption{0.2} & \mycaption{0.3}\\
\rowname{Medium}&
\includegraphics[height=.25\columnwidth,width=.20\linewidth]{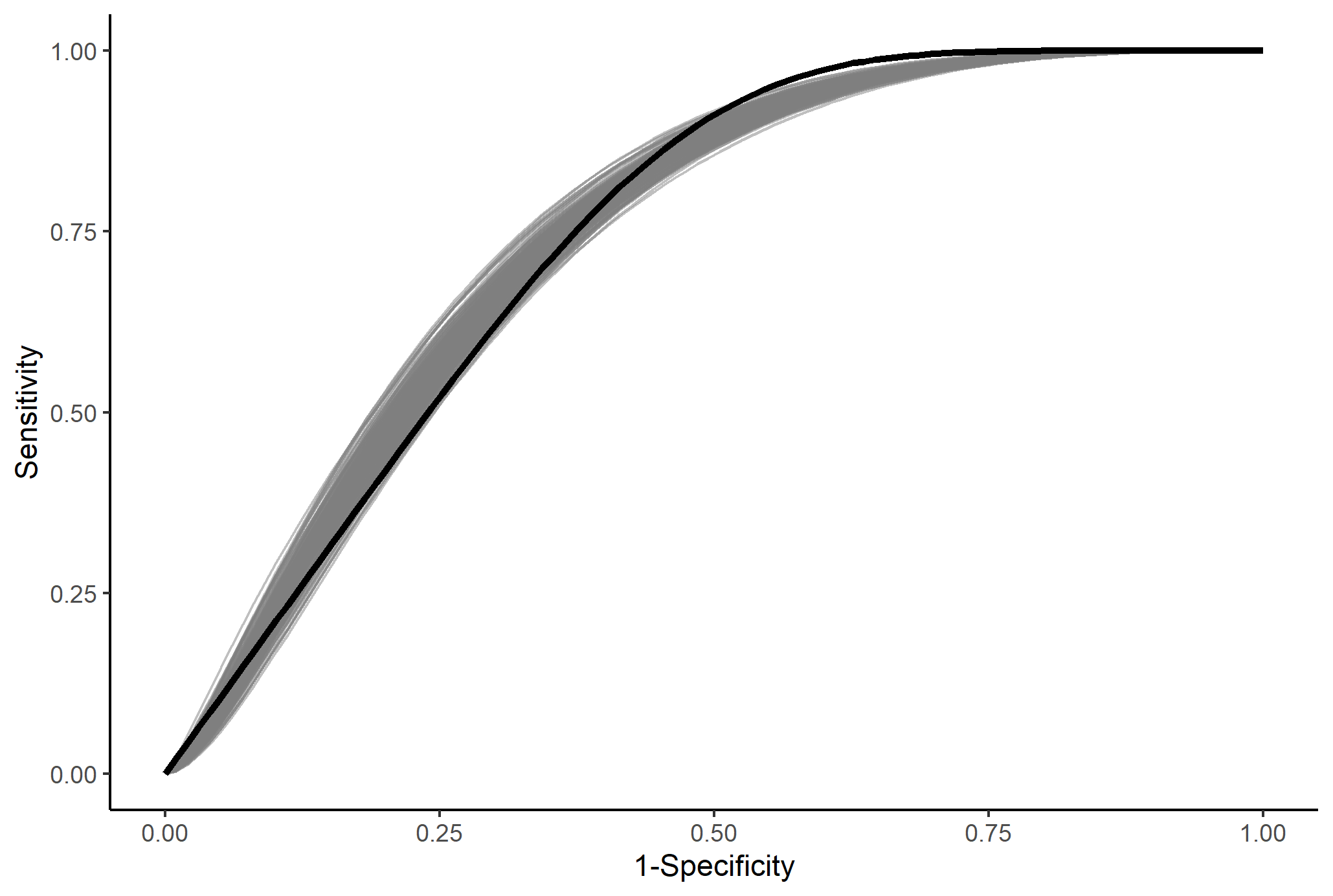}&
\includegraphics[height=.25\columnwidth,width=.20\linewidth]{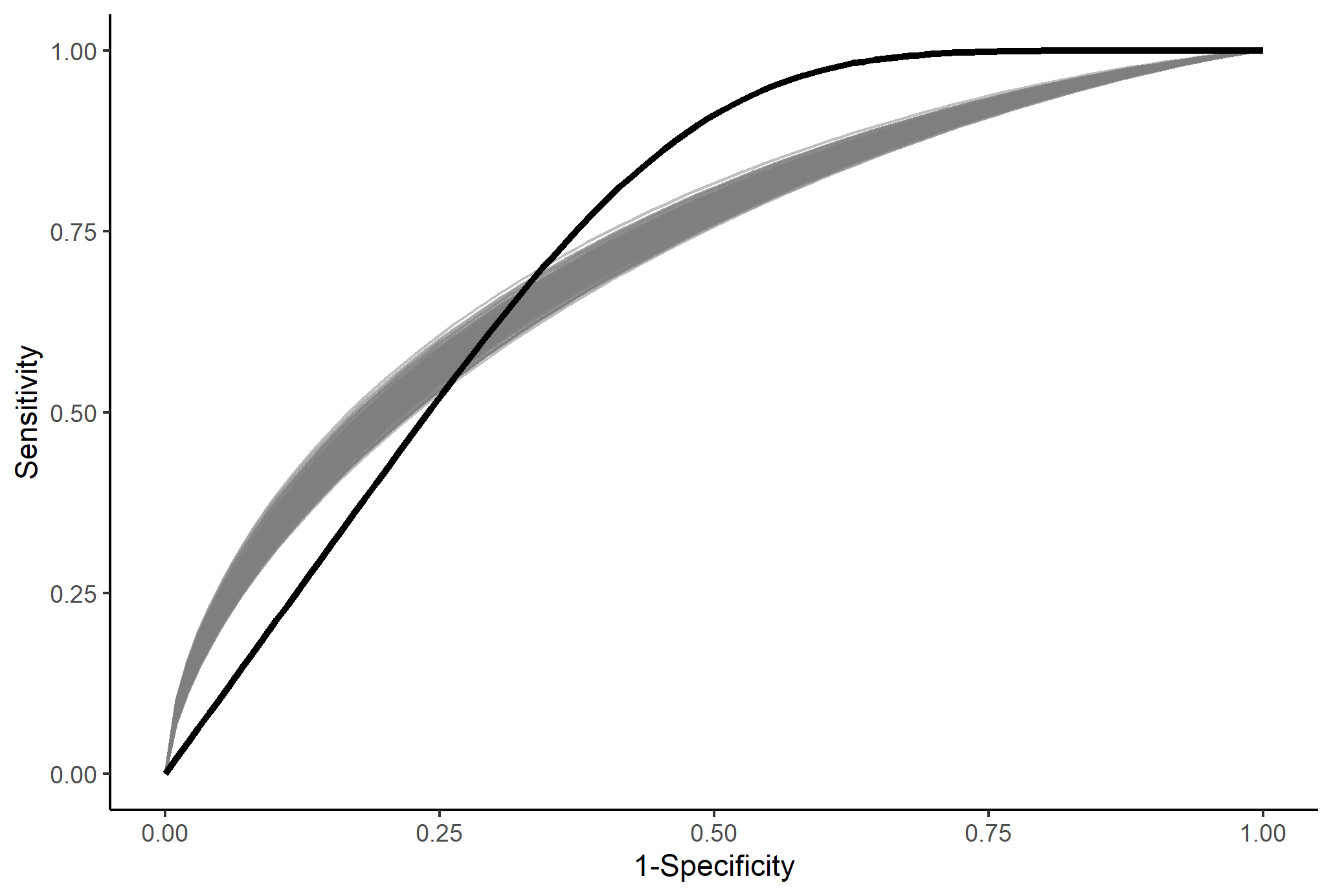}&
\includegraphics[height=.25\columnwidth,width=.20\linewidth]{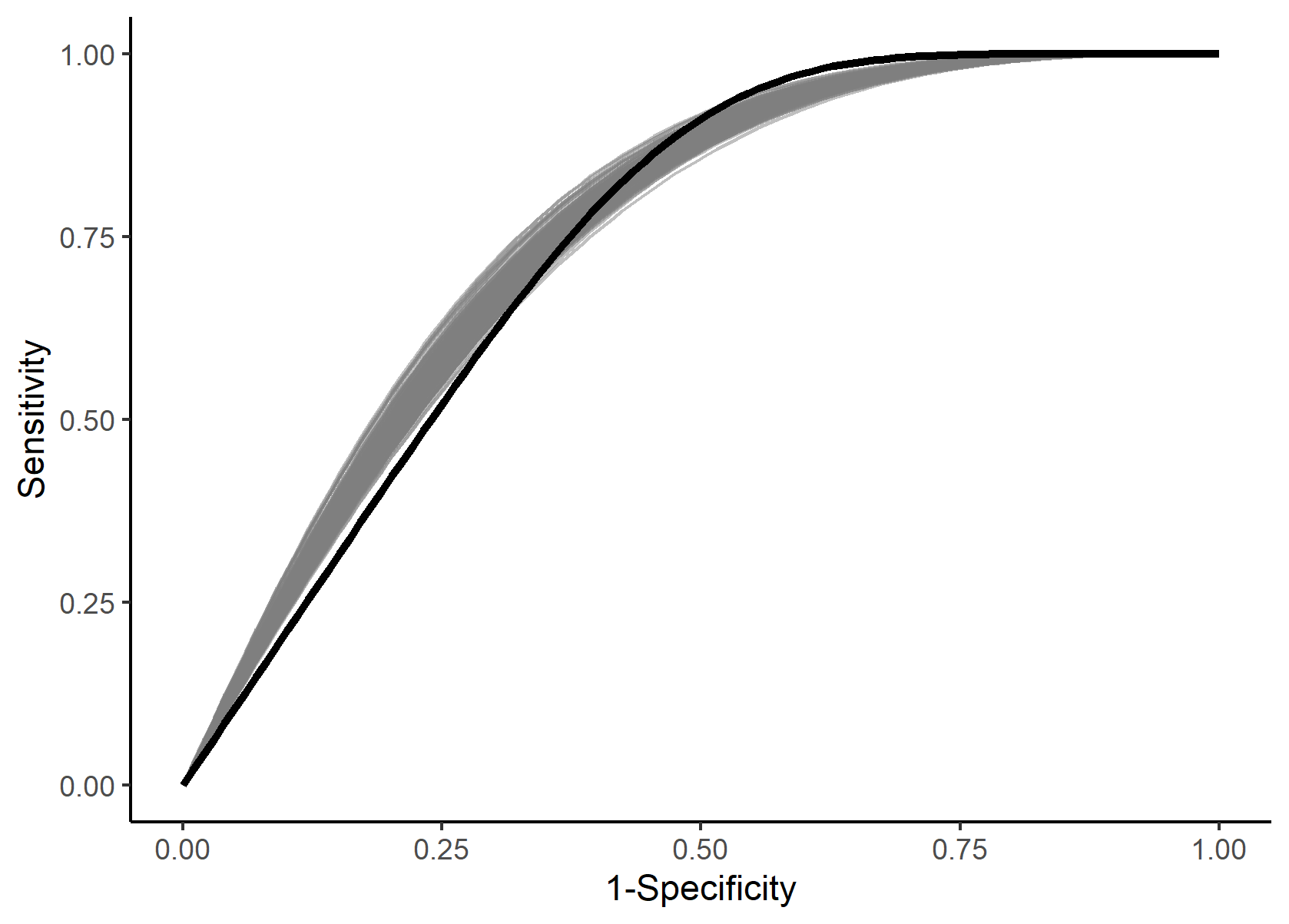}&
\includegraphics[height=.25\columnwidth,width=.20\linewidth]{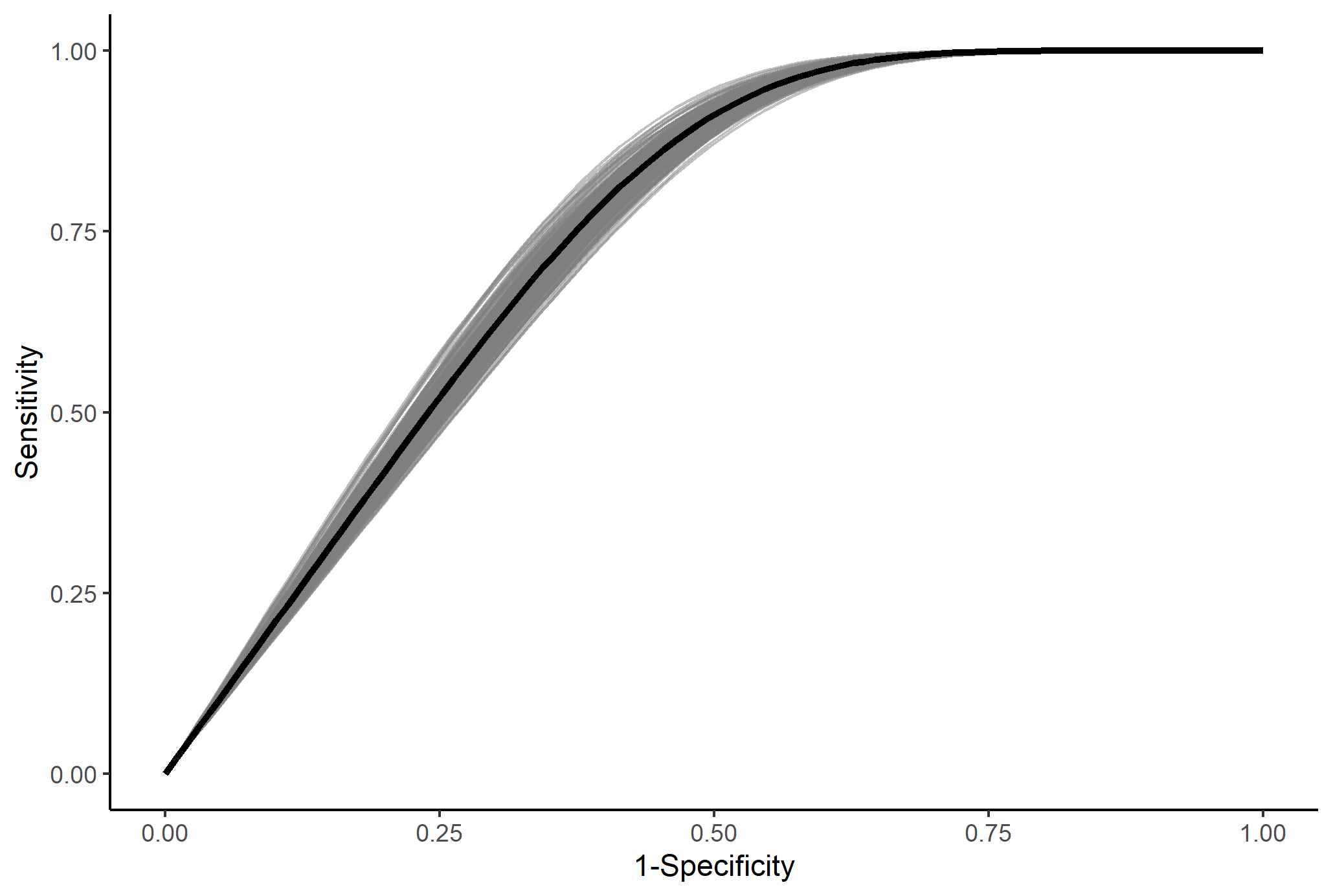}&
\includegraphics[height=.25\columnwidth,width=.20\linewidth]{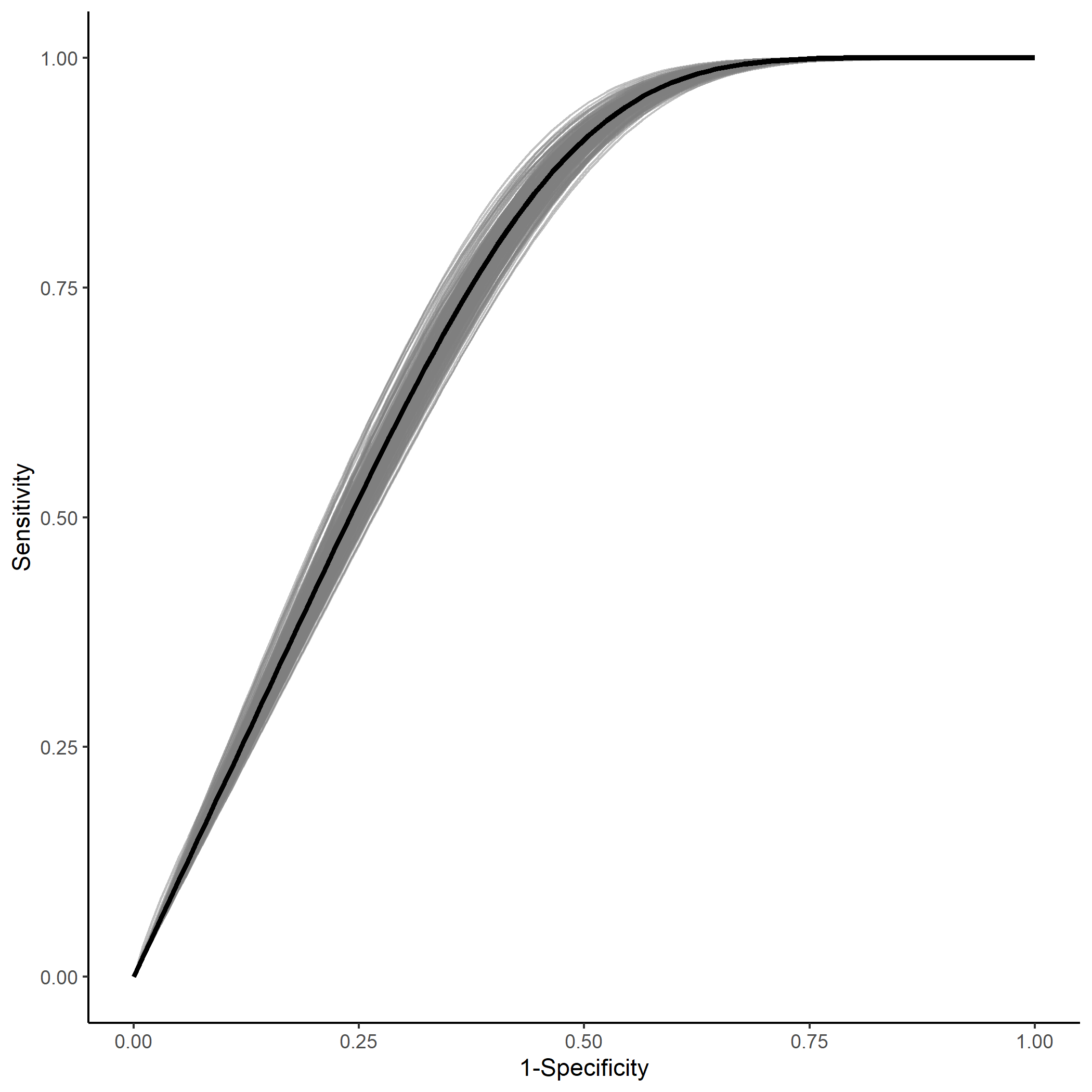}\\
%\includegraphics[height=.25\columnwidth,width=.20\linewidth]{NoCov_ParCon_spCBN2_ROC_Medium.png}\\
%[-1ex]&\mycaption{0.5} & \mycaption{0.4} & \mycaption{0.6}\\
\rowname{High}&
\includegraphics[height=.25\columnwidth,width=.20\linewidth]{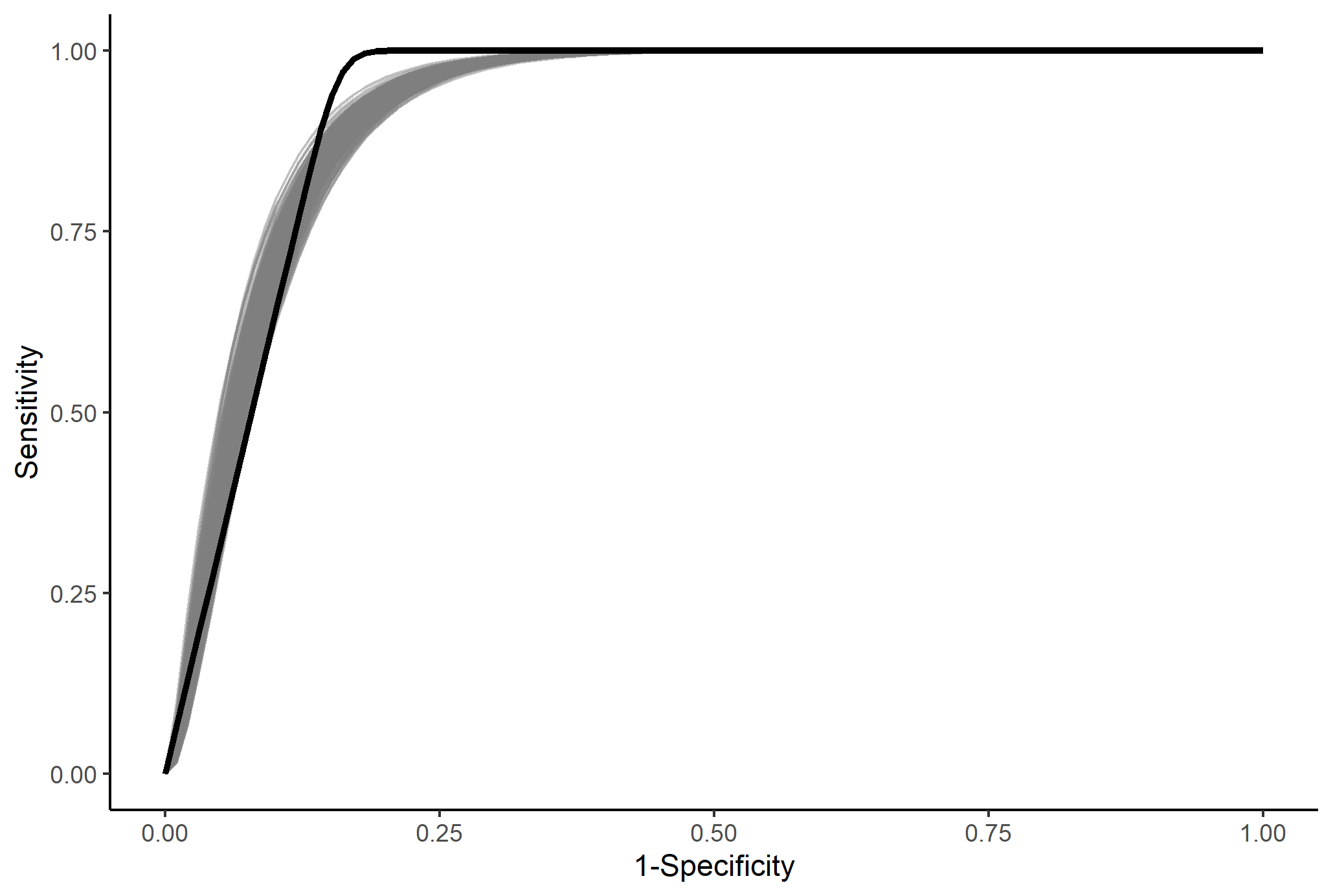}&
\includegraphics[height=.25\columnwidth,width=.20\linewidth]{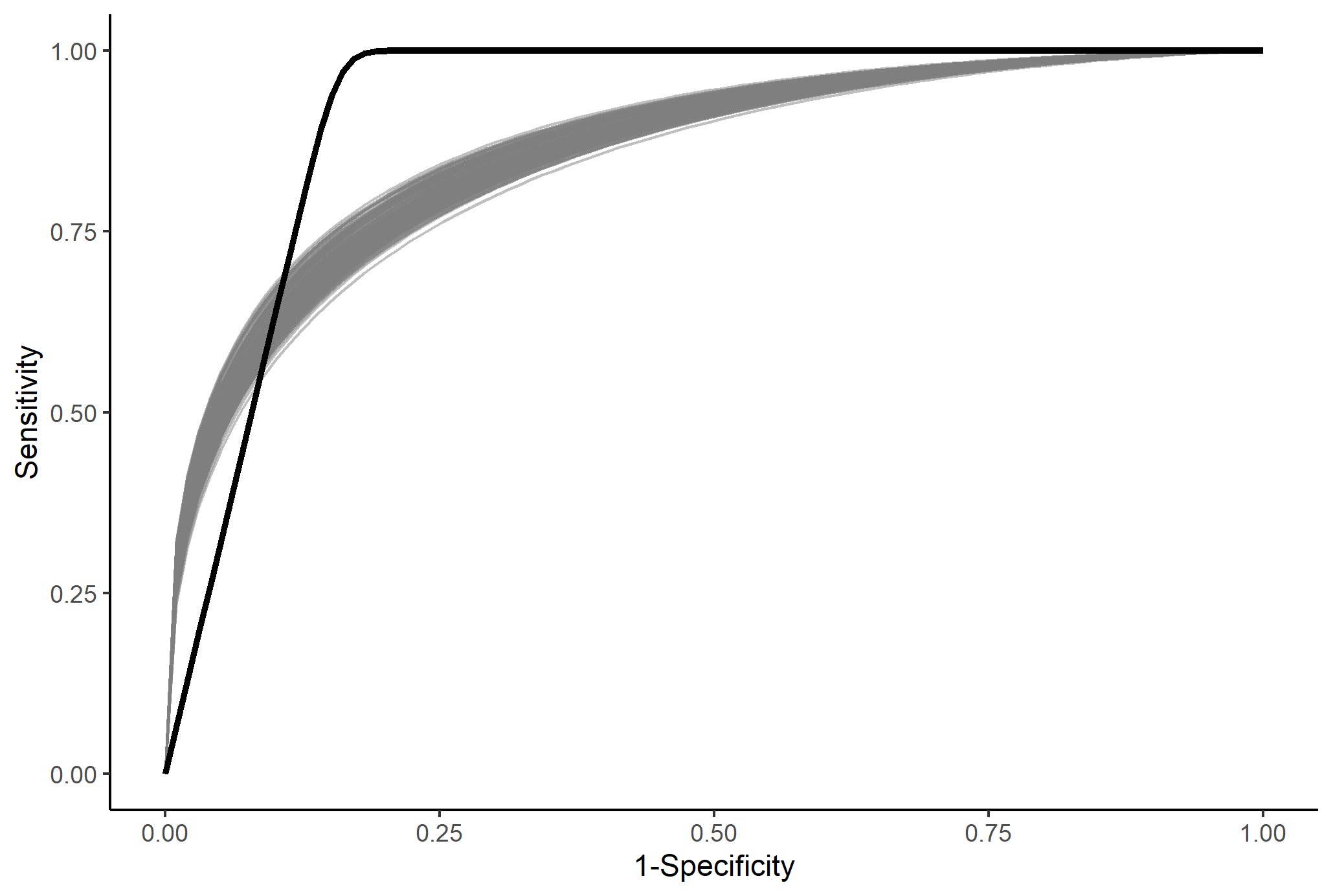}&
\includegraphics[height=.25\columnwidth,width=.20\linewidth]{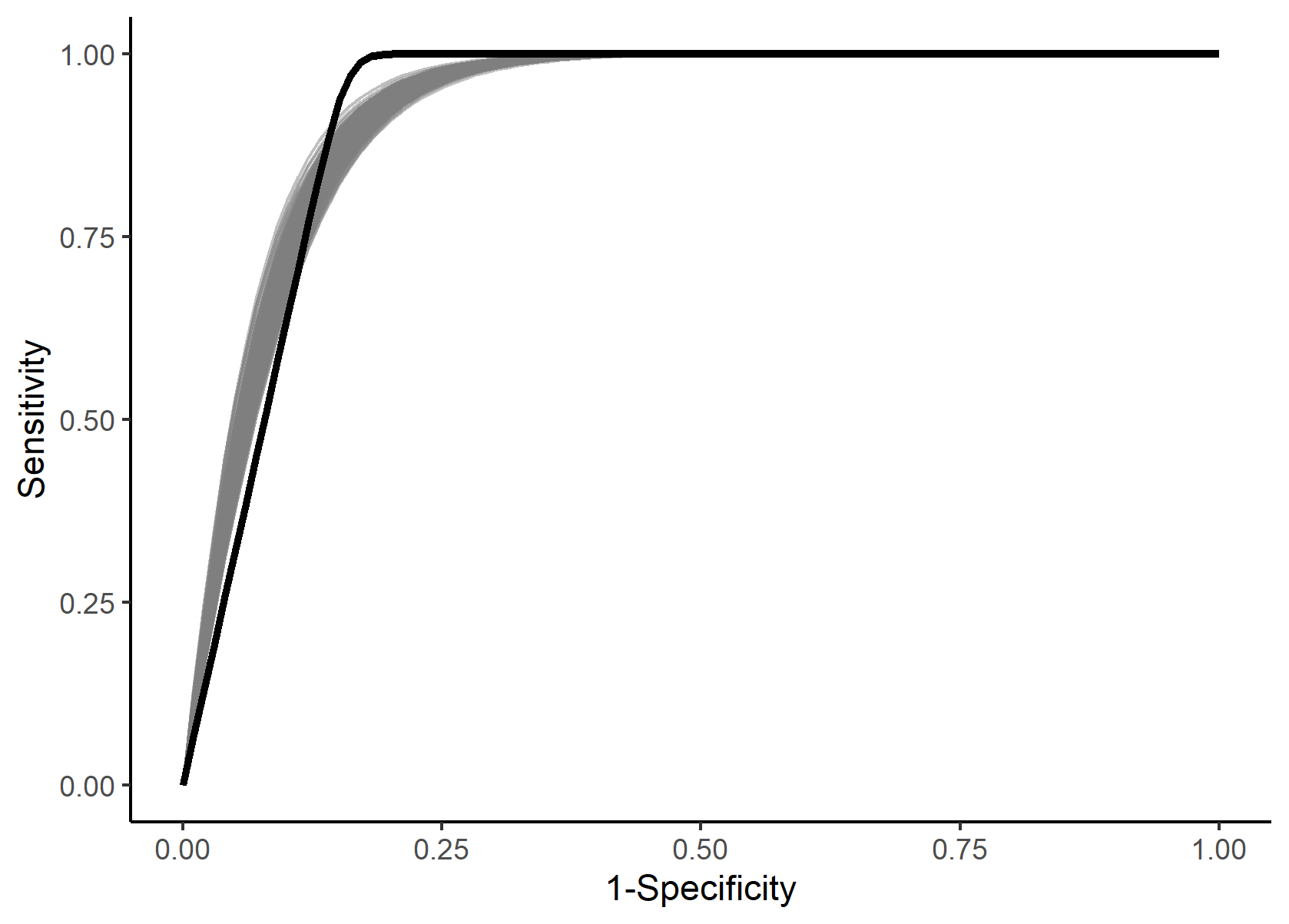}&
\includegraphics[height=.25\columnwidth,width=.20\linewidth]{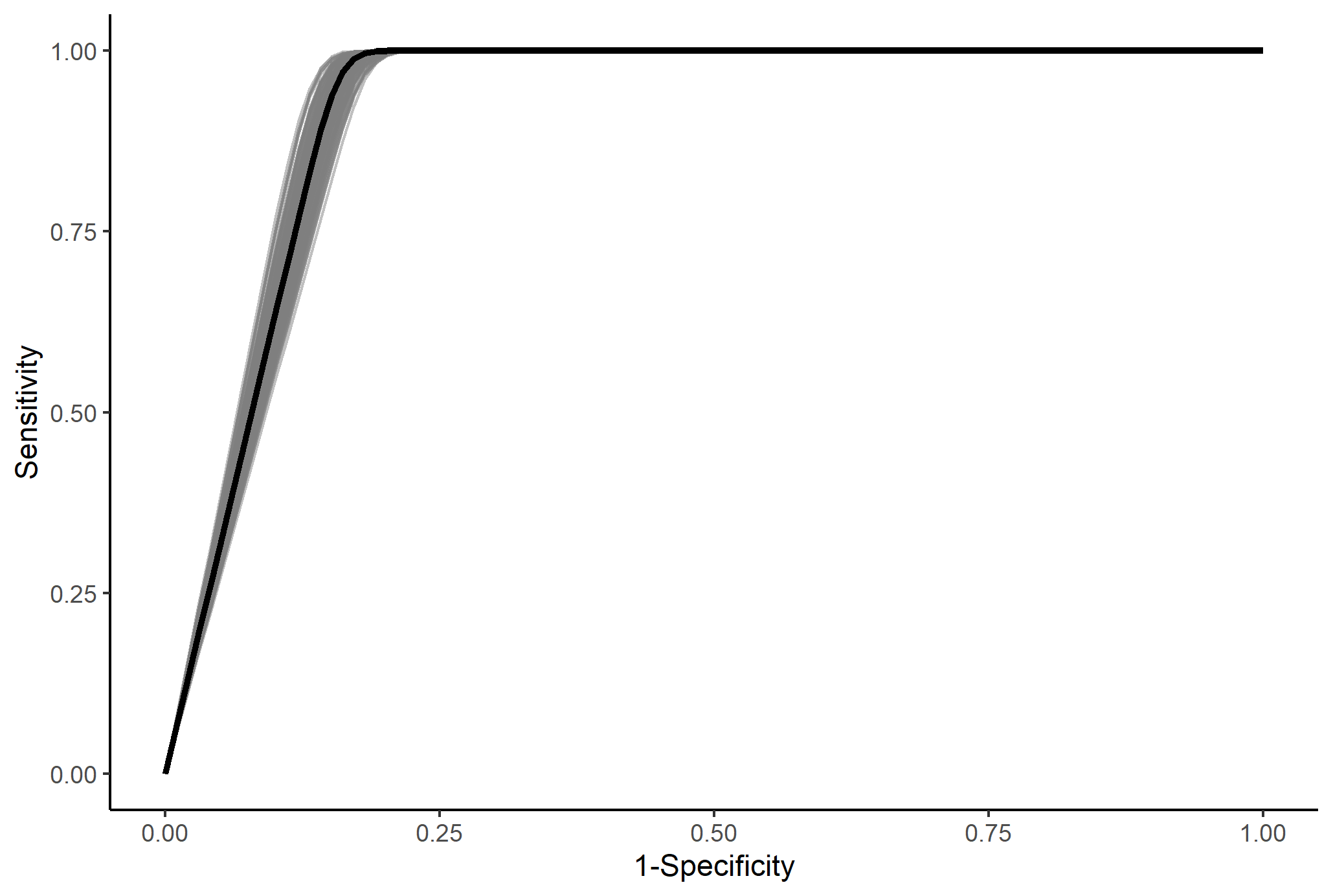}&
\includegraphics[height=.25\columnwidth,width=.20\linewidth]{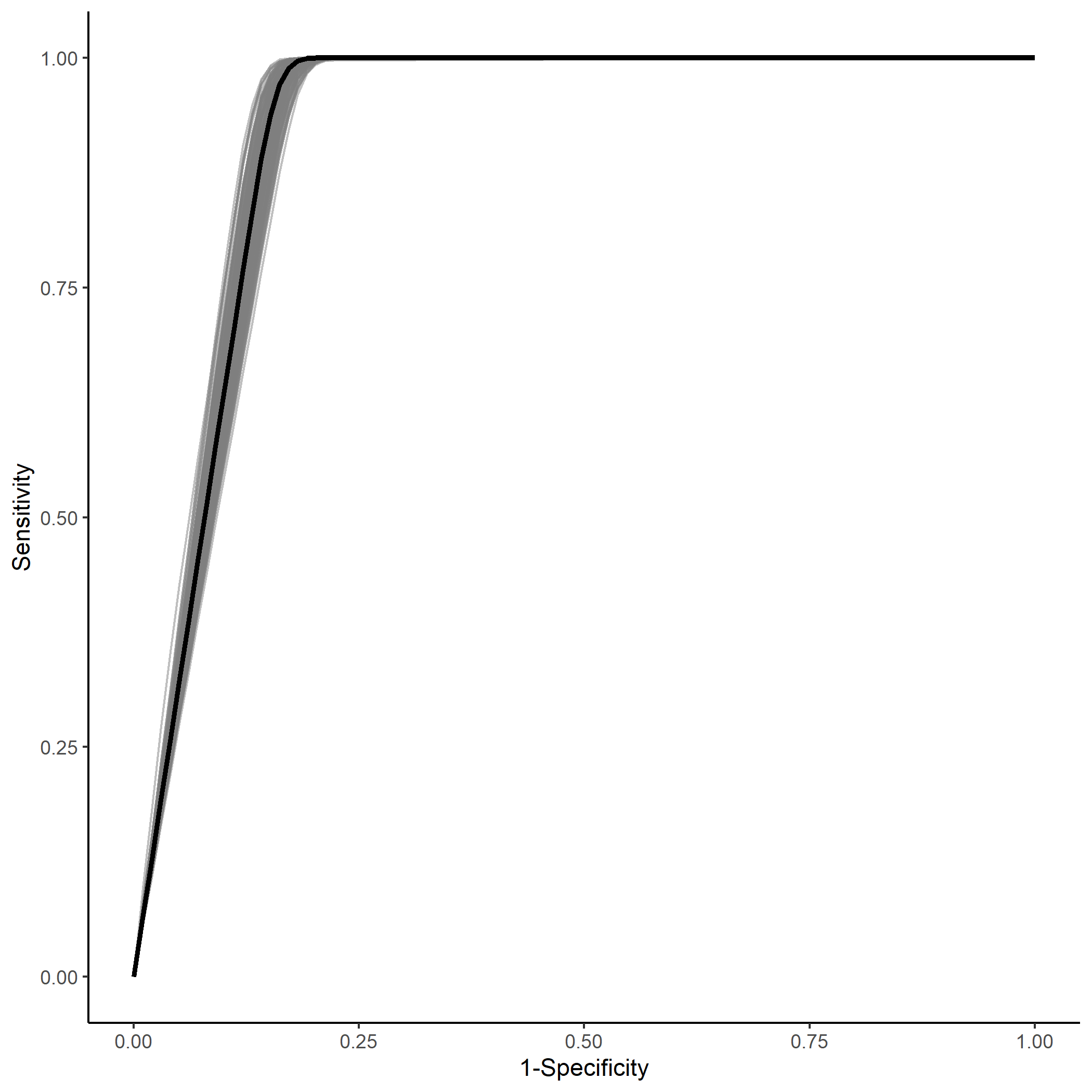}\\
%\includegraphics[height=.25\columnwidth,width=.20\linewidth]{NoCov_ParCon_spCBN2_ROC_High.png}\\
%[-1ex]&\mycaption{0.5} & \mycaption{0.5} & \mycaption{0.7} \\
\end{tabular}
        \end{adjustbox}
\end{figure}

Finally when the data generating mechanism is pCN, the pCN model itself is unbiased and yields the smallest EMSEs. The semiparametric model spCN has almost identical performance to the pCN model. On the other hand, the PBN model performs poorly except when the AUC level is low. Figures \ref{fig:Sim_ROC_noCov_PBN}-\ref{fig:Sim_ROC_noCov_pCN} reflect these observations.

In summary, if the goal is to estimate ROC curves that are proper in the sense of satisfying optimal decision theory, it is preferred to apply the proposed semiparametric or parametric PV-based concave ROC approaches. Existing concave ROC approaches such as PBN and BG may have similar or slightly better performance when we know the true underlying models, the fact that we never possess such knowledge makes them less desirable. As popular as it is, the BN model will not guarantee ROC curves estimates that are proper. 
%\centering\begin{tabular}{@{}c@{ }c@{ }c@{ }c@{}}

%\centering\begin{tabular}{@{}c@{ }c@{ }c@{ }c@{}}

%\centering\begin{tabular}{@{}c@{ }c@{ }c@{ }c@{}}

\section{Application} 
\label{Part:Data_analysis}

\subsection{Pancreatic cancer data} \label{Part:Data_description}

For application and illustration, we use the pancreatic cancer data that was studied by the Mayo Clinic. The study aimed to compare the diagnostic assessments of carbohydrate antigen (CA199) and cancer antigen (CA125) in screening for pancreatic cancer. More on this can be found in \citet{wieand1989family}.

\begin{figure}[htbp]
    \centering
    \includegraphics[width = 0.45\linewidth]{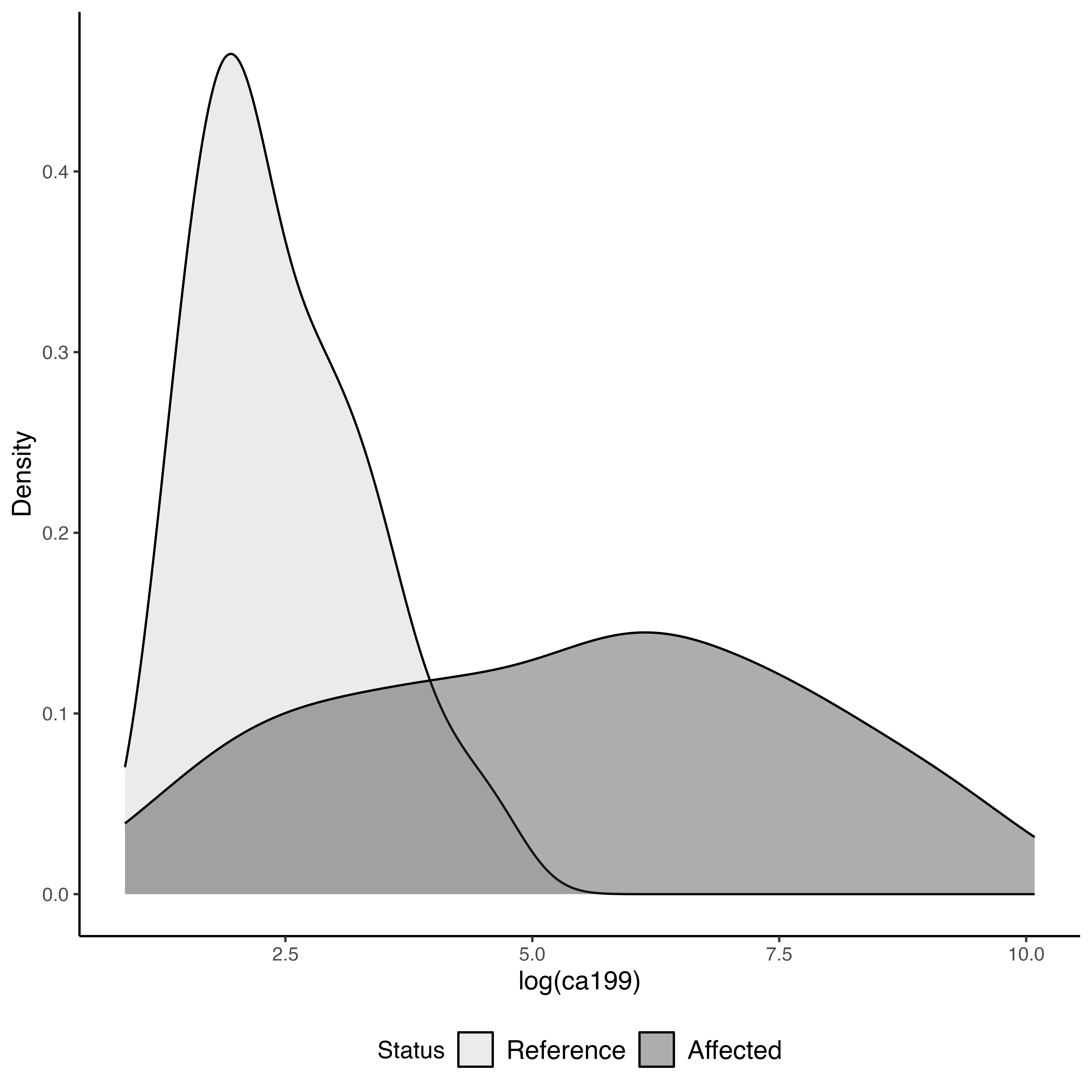}
    \caption{Summary of the pancreatic cancer data %{\color{red} To be consistent, you might want to recreate the figure changing healthy/diseased to reference/affected. Also remove the subtitle of the left panel. Since we describe the distributions in the text, the table on the right panel of Figure 5 can be removed.}
    }
    \label{fig:pancreas_data_summary}
\end{figure}

In the data application, we are only interested in the CA199 marker. Figure~\ref{fig:pancreas_data_summary} has the densities and summary of the log-transformed CA199 marker. Briefly, levels of carbohydrate antigen (CA199) marker of 90 affected (with pancreatic cancer) and 51 reference subjects are used to estimate the ROC curve for discriminating pancreatic cancer. The means (standard deviations) of the logarithm of CA199 marker are 2.472 (0.865) in the reference group and 5.415 (2.342) in the affected group.

\subsection{Data analysis}
\label{Part:Pancreasdata_nocov}
 %We fit BN, BG, PBN, pCN and spCN models. 
 We fit BN, BG, PBN, pCN and spCN models to the pancreatic cancer data. For the parametric pCN model, we take a logarithmic transformation and model $log(Y^0)$ as follows: 
\begin{equation}
log(Y_{i^{'}}^{0}) = \alpha_0+e_{i^{'}}, e_{i^{'}} \sim N(0,\sigma_0^2). \label{Eqn:Data_stage1_NoCov}
\end{equation}
Then the PV $z_i$ are calculated under the parametric approach as
\begin{align*}
z_i = 1-\Phi\left(\frac{log(y_i^1)-\hat{\alpha_0}}{\hat{\sigma}_0} \right).
\end{align*}
For the semiparametric approach, the PV is computed after estimating the distribution of the affected scores using DPM as
\begin{align*}
z_i = 1-\sum_{k=1}^{H}\pi_k\Phi\left(\frac{log(y_i^1)-\hat{\mu_k}}{\hat{\sigma}_0} \right).
\end{align*}
Once we have estimated the PV, we model $z_i$ parametrically as
\begin{align*}
z_i|w_i &\sim U(0, w_i),  \nonumber \\
\eta^{-1}(w_i | \beta_0) &= \beta_0+\epsilon_{1i},  \nonumber \\
\epsilon_{1i} &\sim N(0,\sigma^2), \text{ }i=1,\ldots,N. 
\end{align*}  
The semiparametric concave model follows the same as (\ref{Eqn:Semipar_con}).

\begin{figure}[htbp]
\centering
\begin{subfigure}[t]{.4\textwidth}
\centering
\includegraphics[width=0.8\linewidth]{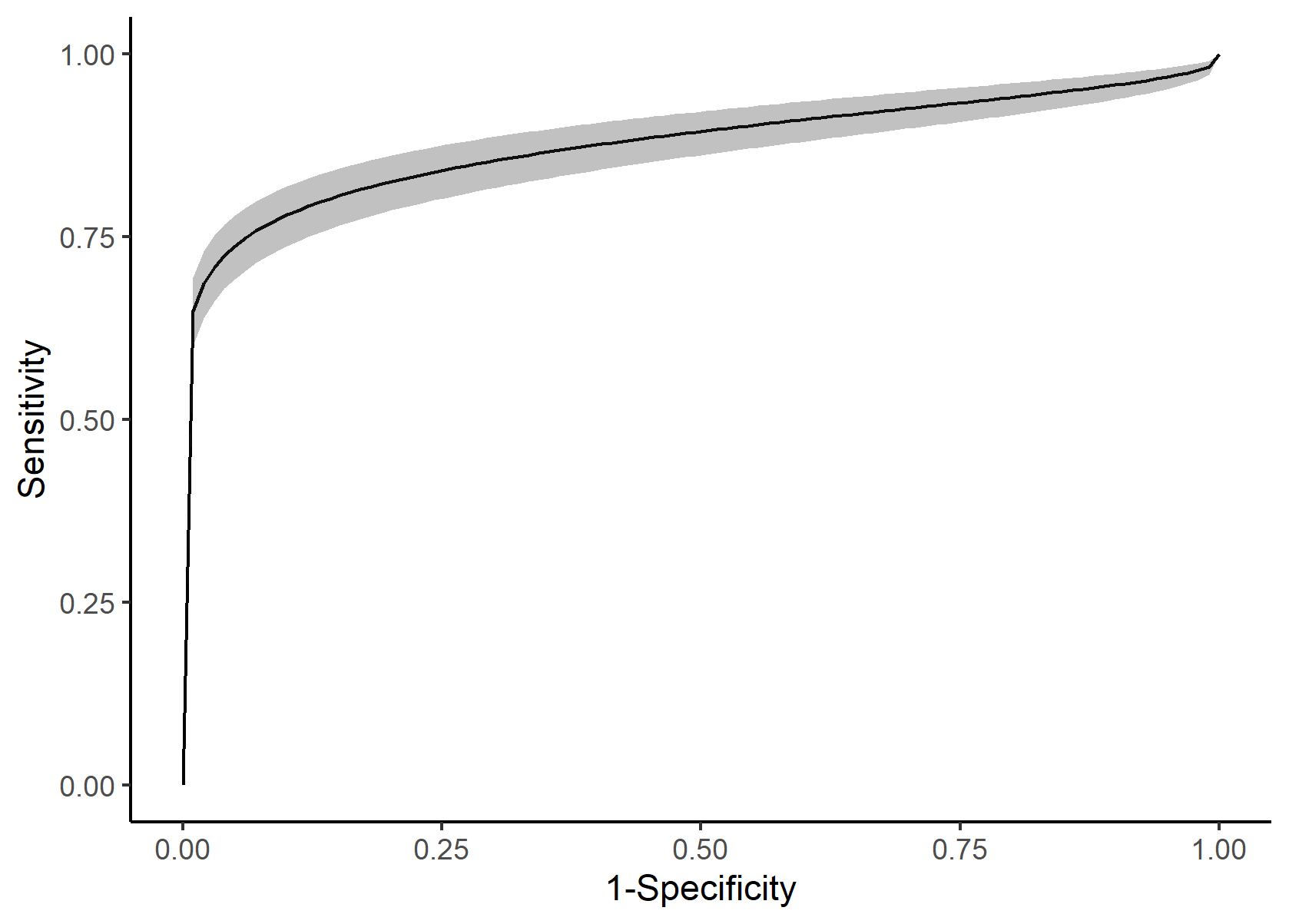}
        \caption{BN}\label{fig:ROC_Pancreas_BN}
\end{subfigure}
\begin{subfigure}[t]{.4\textwidth}
\centering
\includegraphics[width=0.8\linewidth]{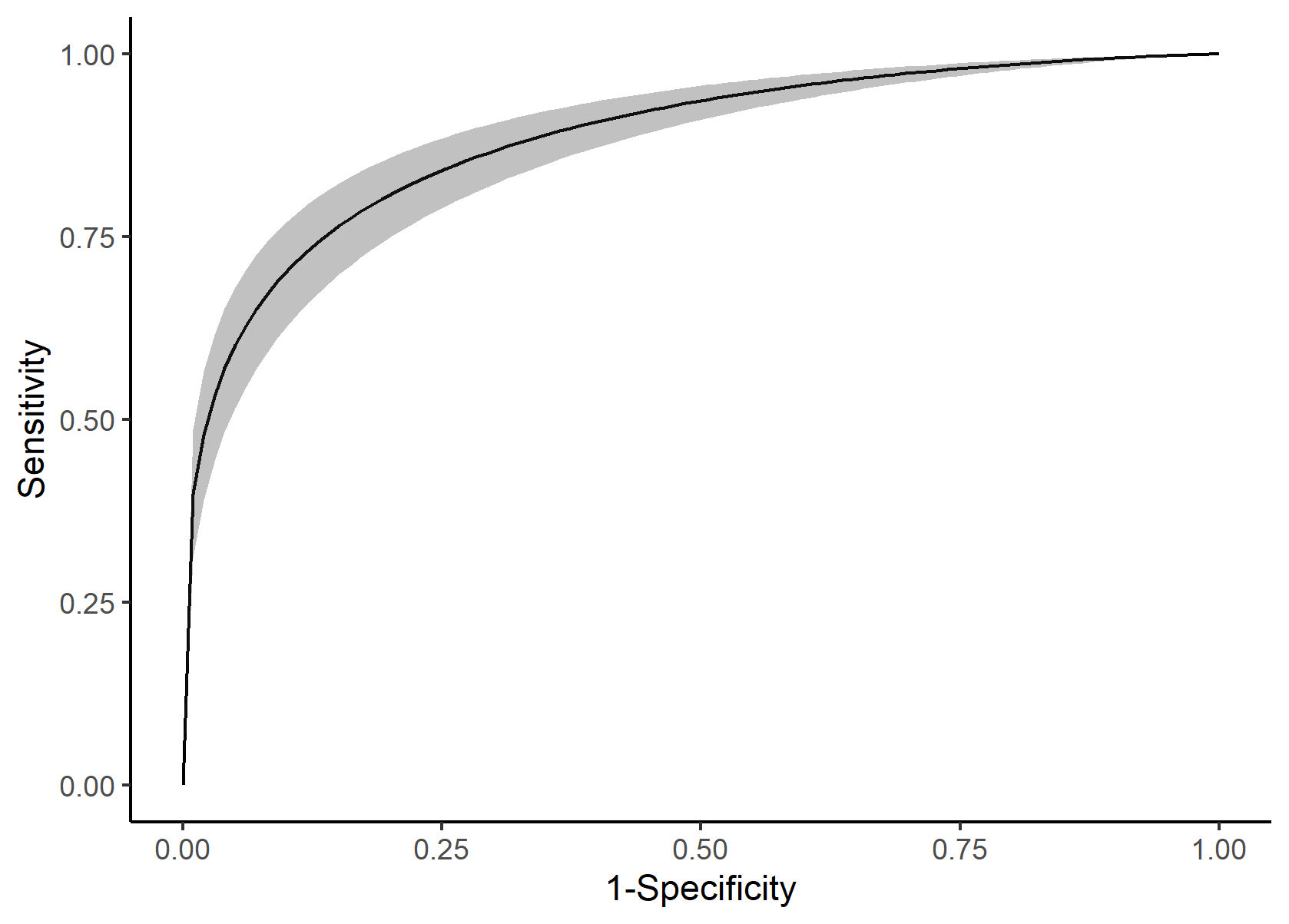}
\caption{BG}\label{fig:ROC_Pancreas_BG}
\end{subfigure}
\medskip
\begin{subfigure}[t]{.4\textwidth}
\centering
\includegraphics[width=0.8\linewidth]{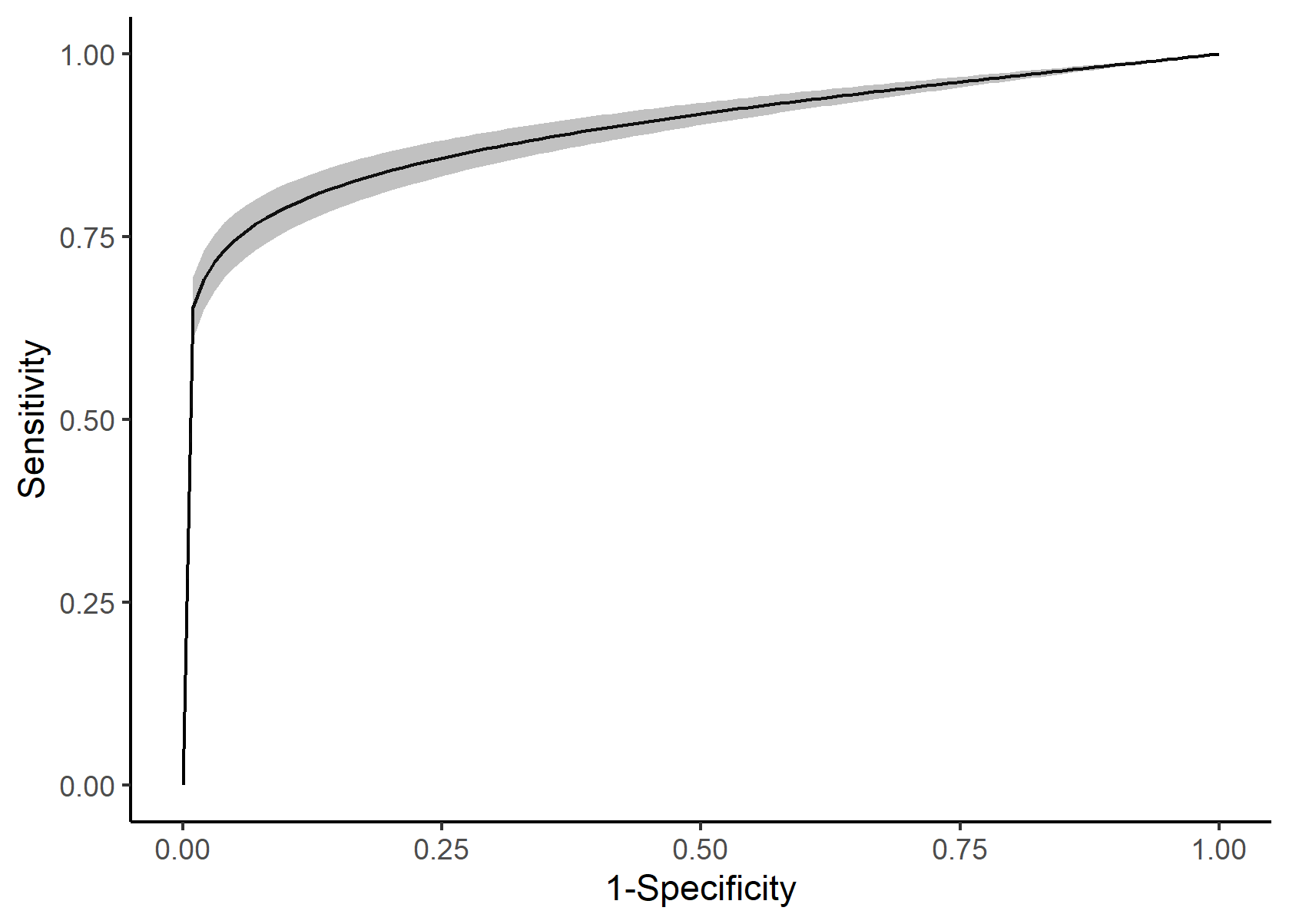}
        \caption{PBN}\label{fig:ROC_Pancreas_PBN}
\end{subfigure}
\begin{subfigure}[t]{.4\textwidth}
\centering
\includegraphics[width=0.8\linewidth]{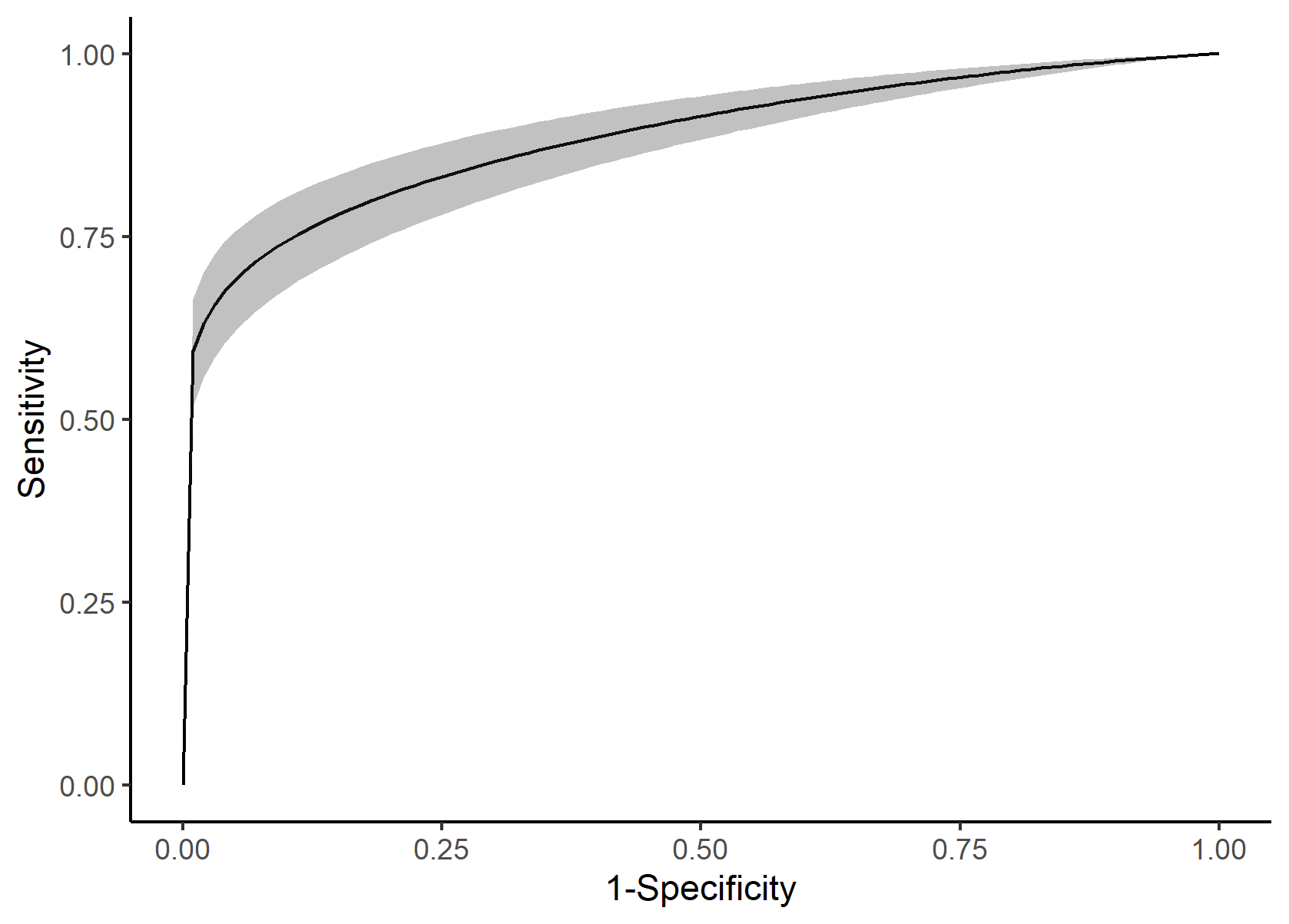}
\caption{pCN}\label{fig:ROC_Pancreas_pCN}
\end{subfigure}
\begin{subfigure}[t]{.4\textwidth}
\centering
\vspace{0pt}% set the real top as the top
\includegraphics[width=0.8\linewidth]{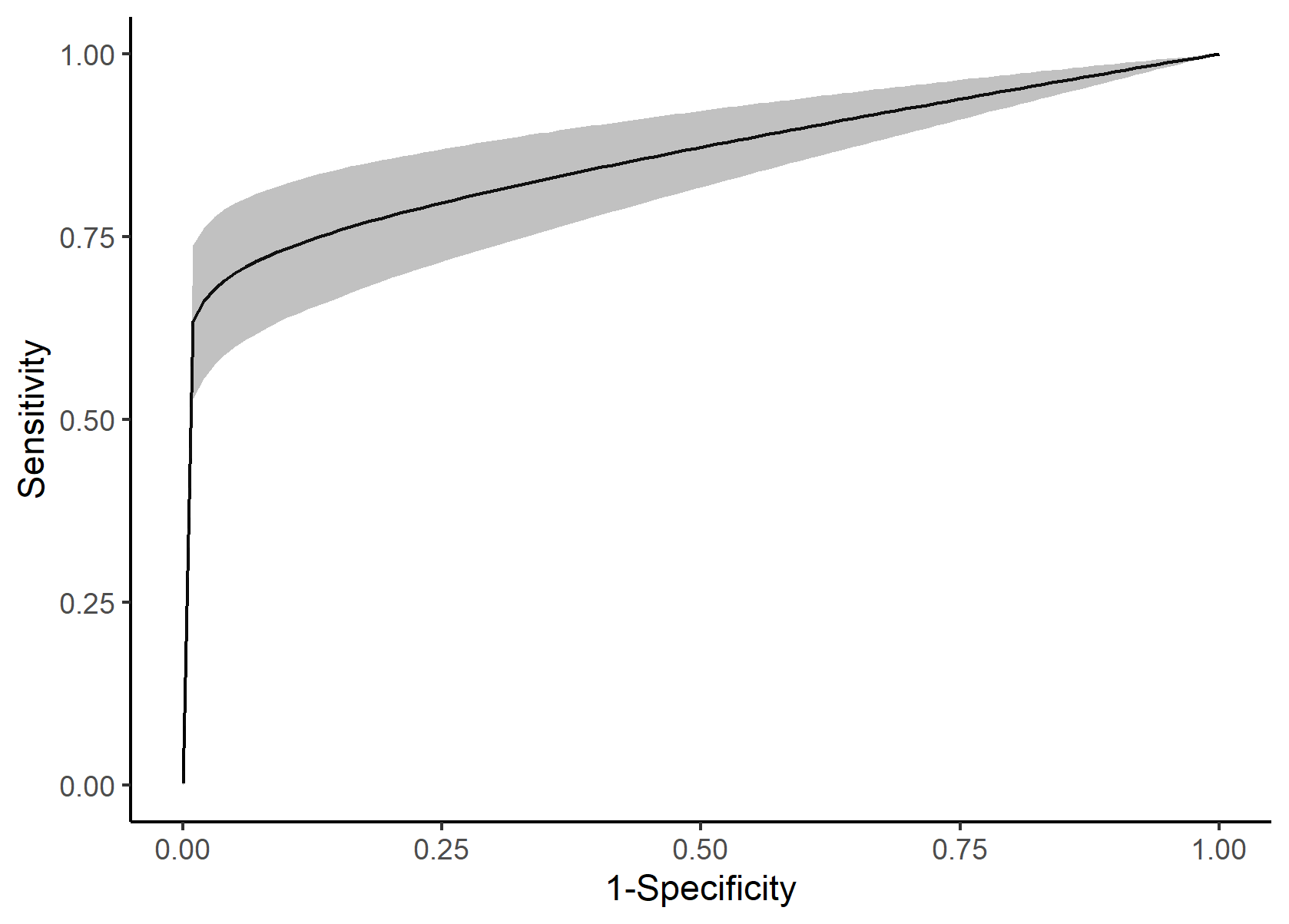}
\caption{spCN}\label{fig:ROC_Pancreas_spCN}
\end{subfigure}
%
%\begin{minipage}[t]{.4\textwidth}
%\caption{Estimated ROC curves for different models in Pancreas data.}\label{fig:ROC_Pancreas_NoCov}
%\end{minipage}
\caption{Estimated ROC curves under different models for the Pancreatic cancer data.}\label{fig:ROC_Pancreas_NoCov}
\end{figure}

We estimate ROC and AUC as discussed in equation (\ref{Eqn:Fest}) and tabulate the posterior mean AUC estimates and 95\% credible interval in Table~\ref{tab:Pancreas_noCov_tab} under different models. The corresponding ROC curves are shown in Figure~\ref{fig:ROC_Pancreas_NoCov}.

\begin{table}[htbp]
\caption{Posterior mean AUC estimates and 95\% credible interval (CI) for Pancreatic cancer data.}
\label{tab:Pancreas_noCov_tab}
\begin{center}
%\adjustbox{max width=0.9\textwidth}{%
\begin{tabular}{lcc}
\hline
\textbf{Model} & \textbf{Mean} & \textbf{95\% CI} \\ \hline
BN              & 0.872         & (0.842, 0.899)   \\
BG              & 0.881         & (0.849, 0.908)   \\
PBN             & 0.900         & (0.883, 0.917)   \\
pCN             & 0.882         & (0.850, 0.910)     \\
spCN           & 0.855         & (0.803, 0.903)   \\ \hline
\end{tabular}%
%}
\end{center}
\end{table}

Based on the AUC estimates Table~\ref{tab:Pancreas_noCov_tab} and ROC curves in Figure~\ref{fig:ROC_Pancreas_NoCov} produced by different models, we infer high diagnostic accuracy for the biomarker CA199 in screening for pancreatic cancer population (mean AUC ranging from 0.86 - 0.90). In Table~\ref{tab:Pancreas_noCov_tab}, we see a reasonable difference in the AUC estimates from different models. Although BN produces an AUC estimate similar to the other four models, estimate of the ROC curve under it is clearly nonconcave (Figure~\ref{fig:ROC_Pancreas_NoCov} (a)). Among the four concave ROC models, PBN produces an AUC estimate with the narrowest 95\% credible interval. It is interesting to see the posterior mean AUC is smallest under SpCN; its wide 95\% credible interval could be a result of small sample size.

\section{Discussion} \label{Part:Discussion}

In this paper, we have proposed a PV-based framework where to impose concavity curvature on ROC curves. This was achieved by utilizing the connection between cumulative distribution function of the placement value of a test score and a ROC curve. Compared to existing concave ROC approaches in the literature, the proposed methodology is more flexible and can better accommodate different distributional features. Our simulation study results suggest that the proposed semiparametric concave approach achieves better accuracy in AUC and ROC estimates in different scenarios, followed by its parametric counterpart. This work provides a new useful tool in diagnostic accuracy analysis, especially when strong parametric distributional assumptions are not supported.

Other forms of constraints have been considered in ROC curve analysis. For example, \citet{ghosal2019discriminatory} introduced a PV-based BN model that allows for multiple ROC curves with AUC ordering constraint. One natural future research direction is to jointly consider shape- and order-constrained multiple ROC curves, in the sense that several ROC curves are ordered (e.g., in AUC) and each individual one is concave. One immediate challenge is the development of efficient computational algorithms, as the multitude of constraints compresses the support of joint posterior distribution which is then difficult to sample from.

As another future direction, we can consider accounting for covariates in the ROC framework which will allow us to estimate the covariate-specific or covariate-adjusted concave ROC curves. Covariate adjustment in the concave ROC framework has never been explored and the PV-based framework potentially provides an attractive platform to account for covariates.

\section*{Acknowledgements}
This research was supported by the Intramural Research Program of {\it Eunice Kennedy Shriver} National Institute of Child Health and Human Development. This work utilized the computational resources of the NIH HPC Biowulf cluster. (\url{http://hpc.nih.gov})

\bibliographystyle{apalike}
%\bibliographystyle{imsart-nameyear}
%\bibliography{Concave_normal}
\input{Concave_normal.bbl}

%\newpage
%\appendix
%\appendixpage
%\addappheadtotoc

\end{document}

%% file: Concave_normal.bbl
 \newcommand{\noop}[1]{}